\documentclass[conference,compsoc]{IEEEtran}
\IEEEoverridecommandlockouts

\usepackage{tikz}
\usepackage{amsmath}
\usepackage{filecontents}

\usepackage{subfigure}
\usepackage{threeparttable}
\usepackage{colortbl}
\usepackage{xcolor}
\usepackage{multirow}
\usepackage{subcaption}

\usepackage{amsmath}
\usepackage{algorithm}
\usepackage{algorithmic}
\usepackage{booktabs}

\usepackage{xurl}

\ifCLASSOPTIONcompsoc
  \usepackage[nocompress]{cite}
\else
  \usepackage{cite}
\fi

\usepackage{eso-pic}

\AddToShipoutPictureFG*{
  \ifnum\value{page}=1
  \AtPageUpperLeft{
    \makebox[\paperwidth]{
      \raisebox{-0.5cm}{
        \footnotesize
        Accepted to the Proceedings of the IEEE Symposium on Security and Privacy (IEEE S\&P) 2026. Please cite the published version when available.
      }
    }
  }
  \fi
}

\begin{document}

\title{SaTor: Exploring Satellite Routing in Tor to Reduce Latency}


\author{\IEEEauthorblockN{Haozhi Li\IEEEauthorrefmark{1}\thanks{\IEEEauthorrefmark{1}Conducted this work during a PhD visit at The University of Edinburgh.}}
\IEEEauthorblockA{School of Cyberspace Science and Technology\\
Beijing Institute of Technology\\
Email: lihaozhi@bit.edu.cn}
\and
\IEEEauthorblockN{Tariq Elahi}
\IEEEauthorblockA{School of Informatics\\
University of Edinburgh\\
Email: t.elahi@ed.ac.uk}
}

\maketitle

\begin{abstract}

High latency is a critical limitation within the Tor network that has a negative impact on web application responsiveness. A key factor exacerbating Tor latency is the creation of lengthy circuits that span across geographically distant regions, causing significant transmission delays. A common solution involves modifying Tor's circuit-building process to reduce the likelihood of selecting lengthy circuits. However, this strategy compromises Tor's routing randomness, increasing the risk of deanonymization. Reducing Tor's latency while minimizing security degradation presents a challenge.

This paper proposes and investigates SaTor, a satellite-assisted routing scheme to reduce Tor latency. By equipping Tor relays with satellite network access, SaTor could accelerate slow circuits via satellite transmission, without biasing the existing path selection process. Our performance evaluation, using a simulator we developed along with real-world measurements, shows that over the long term, SaTor provides an expected speed-up of 21.8 ms for over 40\% of circuits, with only 100 top relays equipped with satellite service. Our research uncovers a viable way to overcome Tor's latency bottleneck, serving as a practical reference for its future enhancement.   

\end{abstract}

\section{Introduction}

Tor is a popular tool for anonymous communication, protecting millions \cite{TorMetrics} of users against surveillance \cite{dingledine2004tor}. Currently, the Tor network is supported by thousands of volunteer relays. The popularity of Tor, a boon for online privacy, places considerable demand on network performance. Improving the network quality of service (QoS) to improve user experience is an important research topic within Tor community \cite{akhoondi2012lastor, wang2012congestion, alsabah2013path, annessi2016navigator, barton2018towards, hogan2022shortor}.

Measuring the time taken for traffic traveling from one endpoint to another, latency has a direct impact on user experience, especially for interactive applications such as web browsing and online calling \cite{arapakis2014impact}. However, Tor often experiences higher latency than regular Internet \cite{dhungel2010waiting,TorMetrics}. A key reason is that Tor traffic is often routed through a series of geographically distributed relays. A Tor client accessing a server may construct a short, fast circuit with all intermediate relays in close proximity, or a longer, slower path spanning the globe. This routing mechanism, while essential for Tor's anonymity, extends the data's journey and results in high latency \cite{akhoondi2012lastor,wacek2013empirical,hogan2022shortor}.


Prior work on reducing Tor latency looks at avoiding slow paths, by preferring relays that are geographically close or have historically demonstrated favorable latency performance~\cite{akhoondi2012lastor,wang2012congestion,wacek2013empirical,annessi2016navigator,barton2018towards,imani2019modified}. However, these methods may compromise Tor's anonymity, as they undermine the randomness of circuit path, making traffic routes predictable to adversaries~\cite{mittal2011stealthy,wan2019guard,karunanayake2021anonymisation,tan2022anonymity}. An alternative approach, ShorTor~\cite{hogan2022shortor}, incorporates extra \emph{via relays} into conventional three-hop circuits, allowing the newly configured circuits to be faster than the standard ones. While ShorTor avoids altering Tor's default path selection, its strategy of recruiting extra relays per circuit still introduces security risks and increases network workload. Currently, it remains an open challenge to reduce Tor's latency without substantially compromising security or increasing network overhead. 

Satellite communication, as a rapidly developing and deployed technology, provides stable and cost-effective network services to millions. The rise of satellite networks has driven research into their performance advantages and novel applications. Some studies suggest that satellite routing may offer faster transmission than terrestrial fiber owing to its higher signal propagation speed. Theoretically, satellite traffic travels at the speed of light in a vacuum via spot-beams, while terrestrial traffic may reach at most two-thirds of the speed of light in an optical fiber due to a physical limit \cite{SMF-28TM2002}. Consider two relays communicating with each other either via terrestrial Internet or a satellite orbiting in low-earth orbit (LEO, 550 kilometers) above the earth's surface. Disregarding node processing delays, the pure propagation latency of satellite traffic becomes lower than that of terrestrial paths when the distance between endpoints exceeds $\approx$1,100 kilometers, as the higher signal speed offsets the longer physical path \cite{chaudhry2022optical, handley2018delay, handley2019using}. 

Another latency advantage of satellite routing lies in its simpler structure, where traffic is relayed through only a few satellites. By contrast, traffic in terrestrial routing often traverses multiple network domains controlled by various service providers, leading to unnecessarily long paths and a greater likelihood of congestion \cite{bozkurt2017internet,hoiland2016measuring,chavula2017insight}. The advantages of satellite routing illuminate a plausible insight to overcoming Tor's latency challenges: by using satellite transmissions between some relays---during the times when terrestrial network is slow---while adhering to Tor's default path selection, it is possible to reduce latency without using extra relays in circuits or biasing circuit path. This paper presents an exploratory study on the latency of satellite-assisted Tor (SaTor for short), aiming to answer:

\textit{Could the introduction of satellite routing technology reduce latency in the Tor network? If so, in what specific ways could Tor take this benefit?}

Answering this question requires comparing satellite and terrestrial routing latencies between all pairs of Tor relays, a challenging task due to the impracticality of equipping every relay with satellite connectivity for direct latency measurements. As a solution, we developed a simulator which could estimate latency between any pair of Tor relays at any given time via both satellite and terrestrial routing. The simulator tracks real-time satellite coordinates, identifies reachable routing paths between relay pairs, and calculates routing latency by dividing the physical path length by a probabilistic distribution of traffic speeds. The speed distribution is derived from public datasets of satellite and terrestrial latencies, specifically LENS \cite{zhao2024lens} and RIPE \cite{RIPE2024}. Compared to existing satellite routing simulations \cite{lai2020starperf, lai2023starrynet, kassing2020exploring}, which use a constant speed of light for all latency calculations, our simulator captures both temporal and geographical characteristics in latency. 

To calibrate potential simulation errors and assess the latency benefits of satellite routing in actual Tor connections, we conducted real-world latency measurements on a subset of relay pairs using a dual-homed testbed with access to both satellite and terrestrial services. The measurements spanned more than a month, characterizing the long-term latency gap between satellite and terrestrial routing. We then compared these measured latencies with the simulated results on the same subset of relay pairs and derived an error distribution to calibrate subsequent simulations.

Both the simulation and measurement show that, while satellite routing may not outperform terrestrial links under normal conditions, more than half of relay pairs achieve notable reductions at tail-percentile ($>$90th) latencies, with improvements averaging up to 1.4 seconds ($\approx$ 50\% lower than terrestrial latency). Since tail latencies significantly affect end-user experience \cite{Alicloud2024}, SaTor presents substantial potential for performance improvement. To this end, we propose a latency reduction scheme for SaTor which supposes a subset of relays to be equipped with satellite network services, enabling traffic to use satellite paths when terrestrial routes are slow. This involves relays installing a satellite dish (similar to a home Wi-Fi) and subscribing to a satellite provider, such as Starlink \cite{StarlinkServicePlans}. These dual-homed relays actively measure latency to other relays through both their satellite and terrestrial interfaces, and instruct Tor to route traffic through the faster interface by adaptively modifying the routing table in the operating system. This process, operating within the system's network stack, requires no modifications to the existing Tor system.

The evaluation of dual-homed SaTor shows that, in a best-case scenario where all relays have satellite access, RTT latency is reduced by an average of 21.7 ms (36.6\%) across all relay pairs and 41.4 ms (41.0\%) across all circuits over a long-term period, with even greater gains at tail latencies. In a more realistic setup with only 100 high-bandwidth relays using satellite, SaTor still accelerates over 40\% of circuits, achieving an average RTT reduction of 21.8 ms, translating to $\approx$400 ms faster page load times for users. Regarding SaTor's practicality, based on our investigation, over 94\% of Tor relays are in regions with current or upcoming satellite service. While $\approx$70\% of relays are cloud-hosted and depend on provider support for satellite access---an area where some providers are already progressing\cite{SpaceCloud2021,HowSatellite2023}---the remaining 30\% of non-cloud relays are sufficient to realize a substantial portion of SaTor's potential. With only a moderate share of relays adopting satellite access, SaTor is unlikely to increase an adversary's network visibility compared to vanilla Tor.

Our contributions are summarized as follows:

\begin{itemize}
    \item We propose the idea of integrating satellite routing into Tor, exploring a novel application of satellite networks and offering a fresh perspective on addressing Tor's long-standing latency challenges.
    \item We present SaTor Simulator, a tool for evaluating real-time satellite and terrestrial latencies. Using globally measured datasets of satellite and terrestrial networks, we assess latency in the current Tor. We also release our simulator publicly\footnote{https://github.com/Reecoach/SaTor}.
    
    \item We measured real-world terrestrial and satellite latencies in Tor to validate the advantages of satellite routing and refine the SaTor simulator.
    \item We propose a practical scheme for deploying satellite routing functionality in Tor and conduct a comprehensive feasibility assessment from both cost-benefit and security perspectives.
\end{itemize}

\section{Background}

\subsection{The Tor Network}

\subsubsection{Onion Routing}

As a key technique in Tor, onion routing encapsulates messages in multiple layers of encryption, and transmits them through a ``circuit'' containing a series of relays\cite{reed1998anonymous, dingledine2004tor}. Each relay decrypts a single layer to uncover the next message destination. This technique maintains the anonymity between communication endpoints, as each intermediary knows only the identity of the immediately preceding and following nodes. 

A typical Tor circuit contains three relays, each selected from a large set of volunteer-operated hosts. The first relay, known as the \emph{entry guard}, is chosen from a list of trustworthy and stable nodes, serving as the entry point for the client's data to the Tor network. The last relay, known as the \emph{exit relay}, is responsible for connecting Tor to the public Internet. The \emph{middle relay} connects the entry guard to the exit to further obscure the source of the traffic. 

\subsubsection{Tor Latency Reduction}

\label{sec:tor-latency-reduction}

The end-to-end latency of a Tor circuit $\mathcal{L}_{\mathrm{cir}}$ arises from two sources: \emph{intra-relay processing} and \emph{inter-relay transmission}, expressed as follows:

\begin{equation}
    \mathcal{L}_{\mathrm{cir}} =
    \mathcal{L}_{\mathrm{guard}}^{\mathrm{P}} +
    \mathcal{L}_{\mathrm{middle}}^{\mathrm{{P}}} +
    \mathcal{L}_{\mathrm{exit}}^{\mathrm{P}} +
    \mathcal{L}_{\mathrm{g} \to \mathrm{m}}^{\mathrm{T}} +
    \mathcal{L}_{\mathrm{m} \to \mathrm{e}}^{\mathrm{T}}
\end{equation}

where $\mathcal{L}_{\mathrm{X}}^{\mathrm{P}}$ is the processing latency within relay $\mathrm{X}$, and $\mathcal{L}_{\mathrm{X} \to \mathrm{Y}}^{\mathrm{T}}$ means the transmission latency during the route from relay $\mathrm{X}$ to $\mathrm{Y}$. This expression excludes latency components external to the Tor network, such as those incurred between client and guard, and between exit and the destination server.

To reduce the \emph{intra-relay processing} latency, Tor maintains minimal hardware requirements for relays and uses a bandwidth-weighted algorithm for load balancing across relays. Further efforts may focus on optimizing the Tor software and enhancing its traffic management mechanism \cite{alsabah2016performance,jansen2012throttling}, thereby reducing processing delays within relays. Moreover, the rapid growth in the number of relays can also contribute to reducing Tor's intra-relay latency.

High \emph{inter-relay transmission} latency is a crucial factor that hampers Tor's performance, which is the primary focus of this paper. Such latency is closely tied to the routes of circuits. By default, Tor selects circuit routes based solely on relay bandwidth and certain security constraints \cite{torproject2023specifications}. This often results in slow circuits over distant geographical or heavily congested areas. A common solution involves modifying Tor's path selection algorithm to reduce the likelihood of choosing distant or congested relays in a single circuit \cite{akhoondi2012lastor, wang2012congestion, alsabah2016performance}. However, this strategy introduces a degree of predictability into which circuits the users are likely to use, which can be exploited to narrow down the anonymity set of users, facilitating more targeted attacks for de-anonymization \cite{backes2014mators, backes2016your, Joshua2015defending}. To address this challenge, this paper explores satellite routing as a potential game-changer for reducing latency in the Tor network.

\subsection{Satellite Communication}

Utilizing orbiting satellites to relay signals, satellite communication technology provides a way for achieving low-latency transmission. This technology has advanced notably in recent years, with an emphasis on deploying vast constellations of small, low Earth orbit (LEO) satellites to provide global, high-speed internet connectivity. 

\subsubsection{Routing Traffic via Satellite}

\begin{figure}
    \centering
    \includegraphics[width=0.8\linewidth]{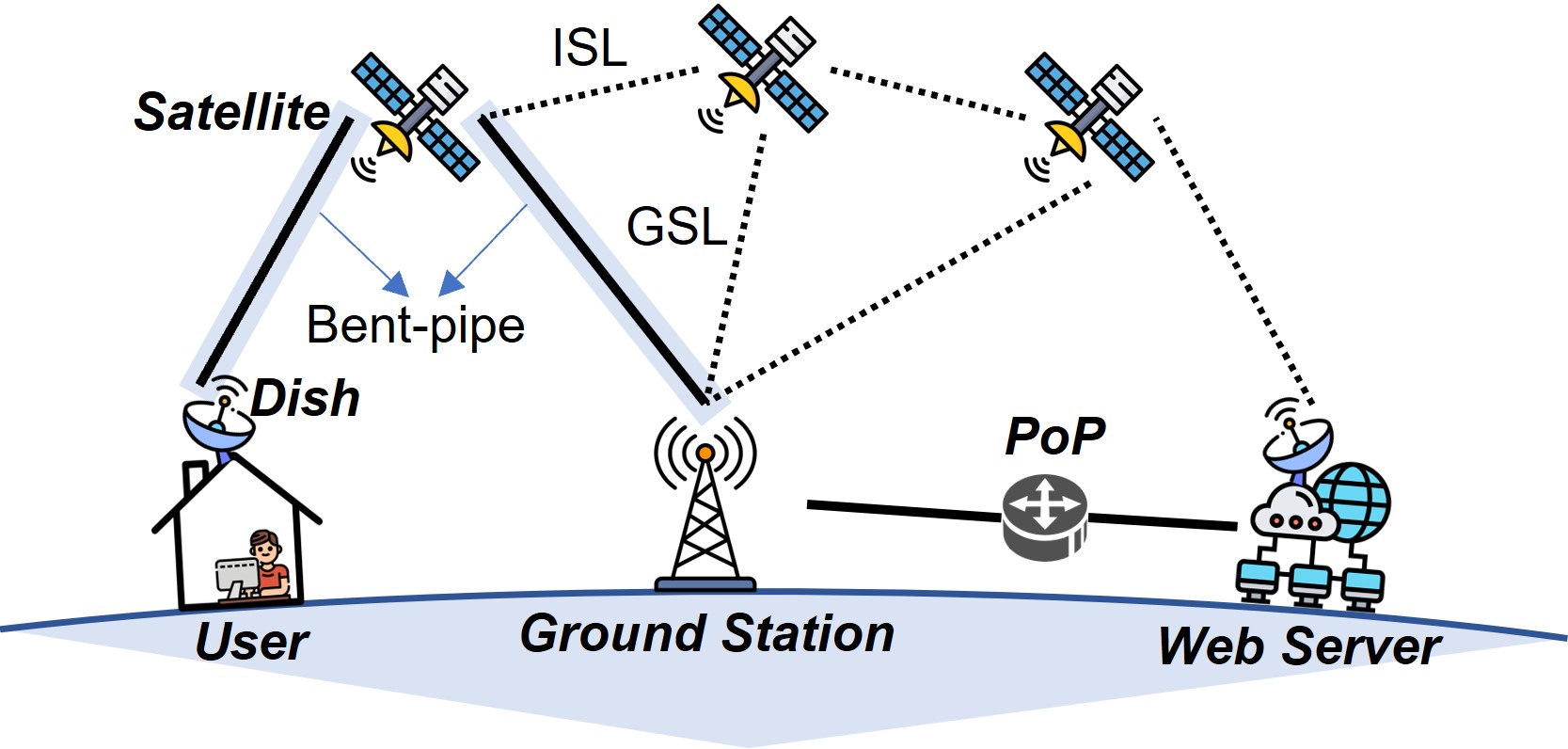}
    \caption{Overview of satellite communications.}
    \label{fig:satellite-communication-tech}
\end{figure}

The principle of satellite data routing is illustrated in Fig. \ref{fig:satellite-communication-tech}. As shown by the solid black line, user data is transmitted from a home dish to a nearby satellite and relayed down to a ground station, forming a ``bent-pipe'' link structure. The data is then routed via a Point of Presence (PoP) to destination server. For the reverse path, the server's data is sent via the PoP to the ground station, and then to the user via satellite connection. This \emph{single bent-pipe routing} strategy is the most fundamental approach in most satellite constellations.

Satellite routing strategies are rapidly evolving. According to Starlink's filings, inter-satellite links (ISLs), which are laser-based connections between satellites, are being incorporated into their satellites\cite{SpaceX2018update}. With ISLs, traffic can be relayed through multiple satellites before being transmitted to a ground station near the destination. This ISL-enabled routing retains most of the transmission path in space, where signals can, in theory, propagate at the speed of light. As a result, it offers lower latency compared to the single bent-pipe method, in which a large portion of the transmission still relies on terrestrial links.

\subsubsection{Commercial LEO Satellite Constellation}

Situated at an altitude ranging from 500 to 2,000 kilometers, LEO satellites enable much lower communication latency compared to traditional geostationary satellites. However, the coverage of a single LEO satellite is limited, so large satellite groups must be arranged in a constellation for reliable and performant global coverage. 

SpaceX's Starlink is currently the largest low Earth orbit (LEO) satellite constellation. As of May 2025, SpaceX has launched over 8,000 satellites into orbits approximately 550 kilometers above Earth, providing service to millions of users \cite{mcdowell2019jonathans, Starlink2023}. SpaceX plans to expand the Starlink constellation to as many as 34,000 satellites in the future. Eutelsat's OneWeb is another major LEO constellation, with 660 satellites launched to an altitude of approximately 1,200 kilometers \cite{OneWeb2023}. Other initiatives, including Kuiper, Yinhe, and Telesat, also have ambitious deployment plans underway. The rapid advancement of commercial LEO constellations signals a promising future for satellite-based Internet, offering potential solutions to the persistent challenges faced by conventional network applications.

\subsubsection{Latency Reduction using Satellite Routing}

The latency $\mathcal{L_{\mathrm{X} \to \mathrm{Y}}^\mathrm{T}}$ from relay $\mathrm{X}$ to relay $\mathrm{Y}$, say the path traverses $n$ intermediate devices $\{d_1, d_2, \dots, d_n\}$, is modeled as:

\begin{equation}
    \mathcal{L}_{\mathrm{X} \to \mathrm{Y}}^\mathrm{T} = 
    \mathcal{L}_{\mathrm{X} \to \mathrm{d_1}}^\mathrm{T} + 
    \sum_{i=1}^n{\mathcal{L}_{d_i}^\mathrm{P}} +
    \sum_{i=1}^{n-1}{\mathcal{L}_{d_i \to d_{i+1}}^\mathrm{T}} +
    \mathcal{L}_{\mathrm{d_n} \to \mathrm{Y}}^\mathrm{T}
\end{equation}

where $\mathcal{L}_{d_i}^\mathrm{P}$ denotes the processing latency at device $d_i$, such as queuing delay in a router, and $\mathcal{L}_{d_i \to d_{i+1}}^\mathrm{T}$ represents the transmission delay between devices $d_i$ and $d_{i+1}$. 

Compared to traditional terrestrial routing, satellite routing offers the potential to reduce both transmission latency $\mathcal{L}^\mathrm{T}$ and processing latency $\mathcal{L}^\mathrm{P}$. Transmission latency can be estimated as the ratio of total link length to signal propagation speed. As illustrated in Fig. \ref{fig:satellite-communication-tech}, satellite routing relays traffic through only a few satellites before reaching a ground station near the destination. In contrast, terrestrial routing often involves complex paths across multiple network domains and devices, leading to longer travel distances and increased processing delays. Additionally, due to the physical properties of the transmission medium (satellite beam in vacuum versus optical signal in fiber), satellite traffic can theoretically propagate at the speed of light, while terrestrial traffic is limited to about two-thirds of that speed. Therefore, prior studies suggested that satellite routing would outperform terrestrial routing, particularly in long-distance transmissions \cite{chaudhry2022optical,chaudhry2022crossover,handley2018delay,handley2019using}.

In addition to the above theoretical analyses, researchers have conducted both simulative and empirical evaluations of satellite routing latency \cite{kassem2022browser,ma2023network,mohan2023multifaceted,lai2020starperf,lai2023starrynet,kassing2020exploring}. These studies demonstrate that satellite routing generally offers latency performance comparable to that of terrestrial routing at the current stage, and even lower latency for transmissions between certain regions. They also indicate that factors such as unstable weather conditions, satellite link reconfiguration, sub-optimal routing path, and the sparse distribution of ground stations can hinder satellite routing from realizing its full potential. However, to the best of our knowledge, no existing work has investigated the potential latency benefits and practical integration strategies of satellite routing in the context of Tor. This gap serves as the primary motivation for our work.



\section{SaTor Latency Evaluation Methodology} 
\label{sec:evaluation-methodology}

\subsection{Evaluation Goals}

This paper aims to explore the potential latency reduction achieved by integrating satellite routing technology into the existing Tor. This evaluation requires estimating both satellite and terrestrial transmission latencies across all Tor relays over an extended period. The evaluation should be scalable across the entire Tor network and flexible enough to accommodate rapidly expanding satellite constellations, including both current developmental stages and future deployments. We conducted both programmatic simulation and real-world measurements, as introduced below.

\subsection{Evaluation Material}
\label{subsec:evaluation-material}

The data and knowledge in evaluation include:

\textbf{Tor Circuits.} A Tor circuit contains two hops across three relays: entry, middle and exit. We collected a dataset containing 100,000 Tor circuits using an instrumented Tor client. The client constructs circuits using a single consensus file at 00:00 AM on December 24, 2024. These circuits include $\approx$ 120,000 unique hops (relay pairs) across 6,964 relays. Our goal is to evaluate the latencies of all circuits in this dataset to assess the latency performance of the entire Tor. Since SaTor focuses exclusively on the latency between relay pairs, the circuits do not include the connections between client and entries, nor between exits and servers.

\textbf{Satellite Constellation.} A constellation consists of satellites, ground stations, and PoPs (Fig. \ref{fig:satellite-communication-tech}). Satellite movements are described by two-line-element (TLE) files---a standard format for recording the trajectory of earth-orbiting objects. The North American Aerospace Defense Command (NORAD) regularly releases the TLE files of LEO constellations, allowing for tracking real-time satellite coordinates (latitude, longitude)\cite{NORAD}. The coordinates of ground stations and PoPs are available on websites managed by astronomy enthusiasts\cite{StarlinkInsider2024,SpaceTrack2024}. While these sites are unofficial, their accuracy is satisfactory for scientific research\cite{lai2020starperf,kassing2020exploring,lai2023starrynet}. The evaluation is performed on Starlink constellation in December 2024, which consists of over 6,700 operational satellites, 25 PoPs, and 277 ground stations.

\textbf{Baseline Latency.} We collect satellite and terrestrial latency measurements from public datasets as an evaluation baseline. The satellite latencies are from LENS dataset \cite{zhao2024lens}, which measures the ICMP RTTs between global-distributed Starlink dishes and PoPs. The LENS data we collected spans a two-month period from April to May, 2024, a total 1.624 billion data points across five geographical locations. The terrestrial latencies are sourced from RIPE Atlas, a global network to measure Internet performance\cite{RIPE2024}. We extract over 104.64 million ICMP RTT latencies from RIPE dataset, covering a one-week period starting on Sep. 2, 2024. Full data information is in Table  \ref{tab:baseline-satellite-latency} and \ref{tab:baseline-terrestrial-latency}, Appendix \ref{appendix:baselin-latency-measurement}. 

\textbf{GeoIP.} The geographical coordinates of each Tor relay are determined based on their IPs using GeoIP service, thus estimating the length of traffic route. The GoeIP service sources information from MaxMind dataset \cite{maxmind2024}, which claims 79\% and 72\% of IPs are within a 100km error-range in Germany and the US, respectively, where the majority of Tor relays are located. The coordinates of RIPE probes are provided in RIPE's official website, deemed accurate.

\subsection{Evaluation Approach}
\label{subsec:evaluation-approach}

The evaluation begins with measuring satellite and terrestrial latency across a subset of Tor circuits using a real-world testbed. Next, we simulated the latencies for the same subset over a custom-built simulator, using the gap between simulation and measurement to improve its accuracy. Finally, the calibrated simulator was employed to estimate the latency of Tor circuits that could not be directly measured.

\subsubsection{Real-world Measurement in Tor}

\begin{figure}
    \centering
    \includegraphics[width=0.85\linewidth]{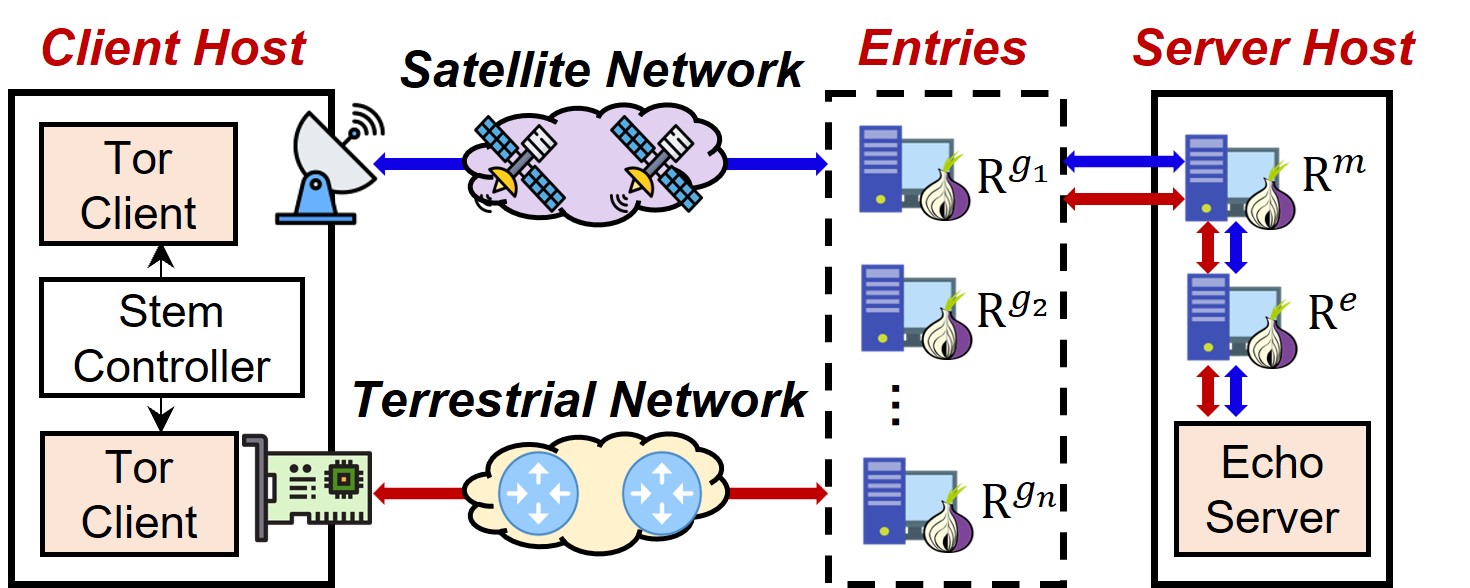}
    \caption{SaTor evaluation over real-world Tor connections.}
    \label{fig:sator-latency-measurement-actual-tor}
\end{figure}

As shown in Fig. \ref{fig:sator-latency-measurement-actual-tor}, two Tor clients, controlled by Stem \cite{Stem}, run in separate Docker containers within a dual-homed host, binding to the satellite interface and the terrestrial interface, respectively. Two Tor relays, denoted as $\mathrm{R}^m$ and $\mathrm{R}^e$, along with an echo server program, run on another host. The client host is in Waterloo, Canada, and the server host is in Los Angeles, US. The two clients, using their bound interfaces, continuously transmit probe packets to the echo server through different circuits $c_i = \{\mathrm{R}^{g_i}, \mathrm{R}^m, \mathrm{R}^e\}$, where the entry $\mathrm{R}^{g_i}$ is sequentially rotated from the Tor consensus. Such measurement constructs numerous circuits that originate from Waterloo, traverse diverse relays worldwide as entries, and terminate in the US. Note that Stem can build circuits with designated entries, not limited to Guard-flagged relays. 

The echo server responds immediately upon receiving each probe packet, enabling RTT calculations at the clients. Packet loss would trigger retransmissions, producing high RTTs. This RTT consists of the latency from the client to $\mathrm{R}^{g_i}$ (via either satellite or terrestrial network) and from $\mathrm{R}^{g_i}$ to $\mathrm{R}^m$ (via terrestrial network), assuming negligible latency from $\mathrm{R}^m$ to $\mathrm{R}^e$. Comparing the RTTs from the two clients yields an assessment of satellite versus terrestrial latency in Tor, since with the same entry $\mathrm{R}^{g_i}$, the $\mathrm{R}^{g_i} \leftrightarrow \mathrm{R}^m$ segment is identical and the observed difference isolates the satellite benefits on the path from client to $\mathrm{R}^{g_i}$.


\subsubsection{Programmatic Simulation} 

We developed a simulator to estimate the satellite and terrestrial latency between each Tor relay pair at any given time. The core concept for calculating latency is to divide the physical length of the traffic route, which is determined using the geographical coordinates of endpoints, by the estimated propagation speeds of traffic. This concept is used in popular satellite routing simulators, such as Hypatia\cite{kassing2020exploring} and StarPerf\cite{lai2020starperf}. As shown in Fig. \ref{fig:sator-latency-evaluation-approach}, the simulator operates in five procedures:

\begin{figure}
    \centering
    \includegraphics[width=1\linewidth]{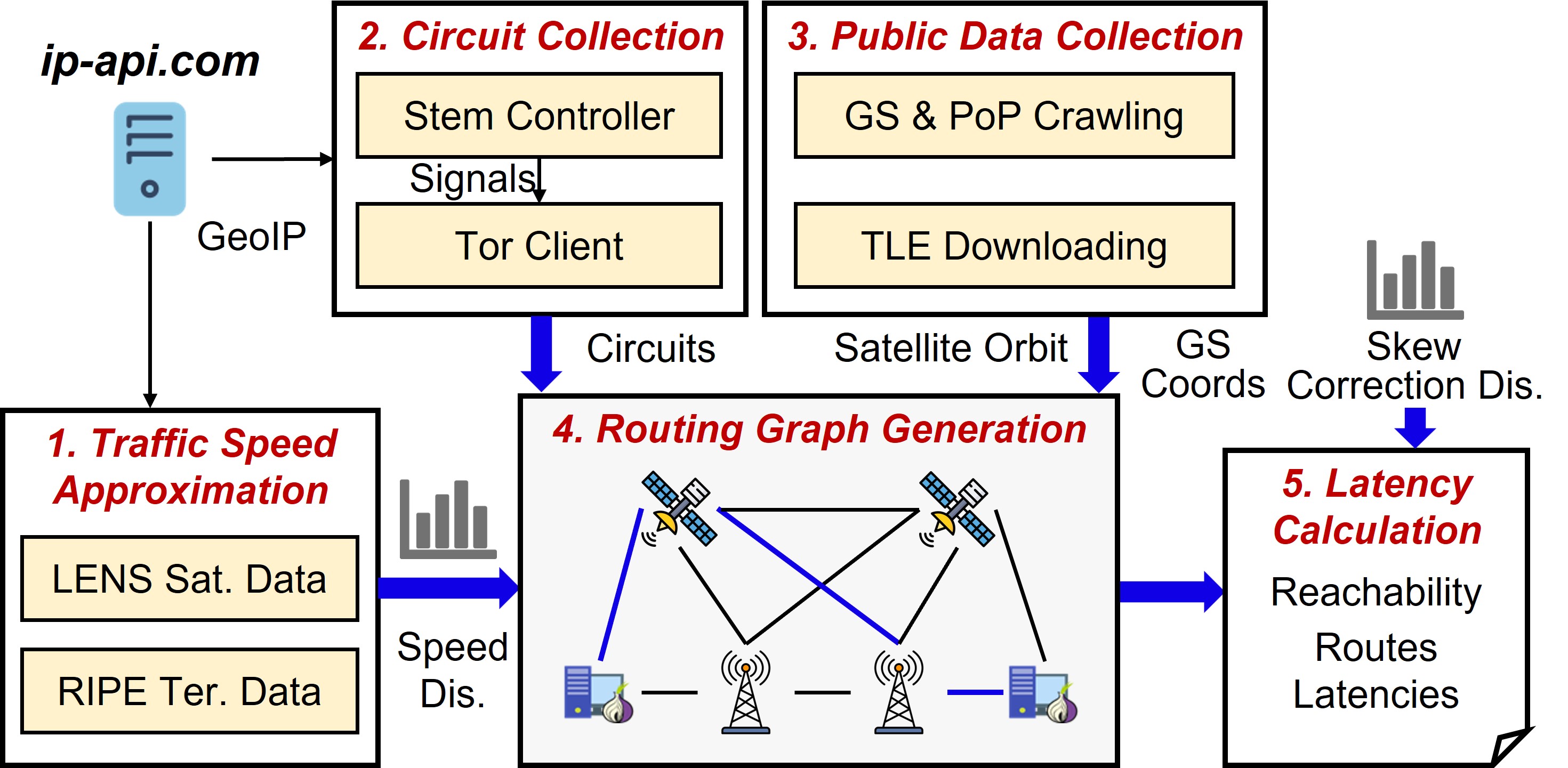}
    \caption{Procedures of simulative evaluation on SaTor.}
    \label{fig:sator-latency-evaluation-approach}
\end{figure}

\textbf{1. Traffic Speed Approximation.} The simulator approximates the probability distribution of satellite and terrestrial traffic speeds based on baseline latencies as follows:

\begin{enumerate}
    \item Calculate the traffic speeds as the ratio of the geographical round-trip length to the RTT latencies.
    \item Partition the range between the maximum and minimum speeds using $n$ delimiters $\delta_1, \dots, \delta_n$.
    \item For each delimiter $\delta_i$, let $f_i$ denote the proportion of speed values lower than $\delta_i$. The collection $\{\delta_i \mapsto f_i\}$ therefore defines an Empirical Cumulative Distribution Function (ECDF).
    \item To yield a speed sample, draw a random number $u \sim \mathcal{U}(0,1)$ and select the delimiter $\delta_i$ such that $f_i$ is the closest cumulative frequency to $u$. The selected $\delta_i$ is then taken as the sampled speed.
\end{enumerate}

For terrestrial traffic, the geographical round-trip length is estimated by doubling the great-circle distance between RIPE Atlas probes, computed using their coordinates obtained from the official Atlas website, which are considered accurate. Note that this estimation accounts only for the great-circle distance (i.e., the shortest path along the Earth's surface) between probe pairs and does not consider actual routing detours through intermediate devices such as routers and switches. Since terrestrial traffic speed may depend on geographic distance, we partitioned probe pairs into groups of 1,000 km intervals. For each group, a separate traffic speed distribution was computed for latency simulation.

\begin{figure}
    \centering
    \includegraphics[width=0.9\linewidth]{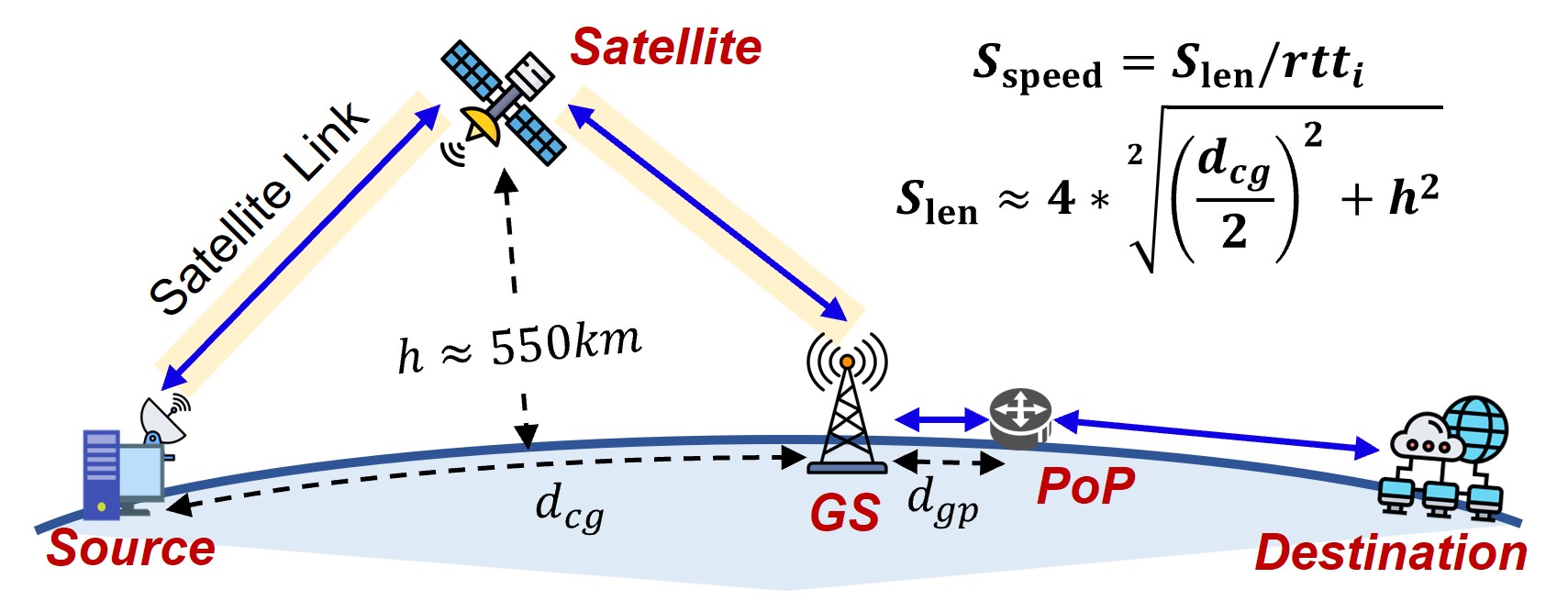}
    \caption{Satellite routing topology to estimate traffic speed.}
    \label{fig:satellite-speed-estimation}
\end{figure}

For satellite traffic, the round-trip length is estimated based on satellite routing topology. 
The LENS latency data are obtained by pinging Starlink's PoPs from multiple client terminals. The traffic path can be represented as: 
\[
\text{Client} \;\longleftrightarrow\; \text{Satellite} \;\longleftrightarrow\; \text{Ground Station} \;\longleftrightarrow\; \text{PoP},
\]
with the segment ``$\text{Client} \leftrightarrow \text{Ground Station}$'' carried over satellite links 
and ``$\text{Ground Station} \leftrightarrow \text{PoP}$'' over terrestrial links. As shown in Fig.~\ref{fig:satellite-speed-estimation}, the satellite path length is:
\[
S_{\text{len}} \approx 4 \cdot \sqrt{\left(\tfrac{d_{cg}}{2}\right)^2 + h^2},
\]
where $h \approx 550 \,\text{km}$ is the orbit altitude of LEO satellites, and $d_{cg}$ is the great-circle distance between the client and the ground station. Since the ground stations and PoPs are often geographically close ($d_{gp} \ll d_{cg}$) with an RTT of $\approx5$ ms~\cite{mohan2023multifaceted}, the satellite traffic speed is given by:

\[
S_{\text{speed}} \approx \frac{S_{\text{len}}}{rtt_i}
\]

where $rtt_i$ is an RTT sampled from the dataset. Note that each $rtt_i$ includes the $\approx 5$ ms ground-station-to-PoP latency, leading to an underestimation of satellite traffic speed and a conservative estimate of satellite latency reduction.


\textbf{2. Tor Circuit Collection.} Circuits are collected using an instrumented Tor client. Stem library~\cite{Stem} is used to send signals to the client, continuously instructing it to construct new circuits. The client is instrumented so that once a circuit is selected and saved, the relevant function will immediately return without actually building the connection. In this way, the simulator can quickly gather a large number of circuits based on up-to-date consensus file, ensuring no impact on the Tor network's burden. Moreover, it fully leverages the most recent algorithms in the Tor system, offering an advantage over classical but outdated circuit generation tools like TorPS\cite{usersrouted-ccs13}. Although the circuits are collected from a single client, given that Tor's path selection relies on the same consensus and algorithm globally, our dataset is representative of worldwide conditions for Tor.


\textbf{3. Public Data Collection.} The TLE files for satellite constellations, along with the coordinates of ground stations and PoPs, are sourced from public websites. The TLE files are acquired from NORAD website\cite{NORAD}, while the coordinates of ground stations and PoPs are taken from the KML file downloaded from Google Maps\cite{StarlinkInsider2024}.

\textbf{4. Routing Graph Generation.} The simulator maintains a routing graph for the entire Tor network. Each node in the graph represents a communication entity, including a relay, satellite, ground station, or PoP. Note that intermediate devices along terrestrial links, such as routers and switches, are not included. Each graph edge represents a data link, with its weight corresponding to the latency between the two connected entities. This latency is estimated using the physical distance between peers and a traffic speed sampled from the distribution obtained in Procedure 1, depending on whether the link is terrestrial (groundstation-to-PoP, PoP-to-relay) or over satellite (relay-to-satellite, satellite-to-groundstation). For terrestrial links, traffic speed is sampled from the distribution within the link's distance interval.

At regular intervals, the simulator updates the routing graph, recalculating the coordinates of each satellite and the physical distance among nodes. It then adjusts the edge weights by resampling a speed for each link to recalculate its latency. The simulator is flexible to simulate various routing strategies of satellite constellation by switching the availability of each type of links, detailed in Appendix \ref{appendix:satellite-routing-strategy}.


\textbf{5. Latency Calculation.} Using the routing graph, the latency between two relays is estimated by summing the weight of each reachable path connecting them. The simulator is designed to output the top-$K$ shortest paths utilizing an efficient algorithm\cite{yen1971finding}. Given that Starlink may not always route traffic through the shortest path, the simulator estimates satellite routing latencies by averaging the top-$K$ paths. The evaluation uses $K=10$ as the baseline, while also examining how varying $K$ affects satellite latency.


\subsubsection{Simulation Calibration}

The simulation is based on latencies measured from global locations over an extended period, which encompass both the transmission latency over terrestrial fiber or satellite links, and the processing latency within intermediate satellites, ground stations, and routers. Continuously sampling from these latencies, our simulation replicates both spatial and temporal variability in satellite and terrestrial transmission. Compared to prior satellite routing simulators \cite{lai2020starperf, kassing2020exploring, lai2023starrynet}, which assume the speed of light for satellite links and two-thirds of the speed of light for terrestrial links, SaTor simulator offers higher fidelity. 

However, simulation errors may arise due to the different latency characteristics between ICMP and TCP. LENS and RIPE latency measurements, relying on ICMP, may not capture Tor's TCP-based latency with multi-layer encapsulation. For example, in ICMP measurements, a lost packet leads to a timeout with no latency recorded, whereas in TCP a lost packet triggers retransmission, resulting in high delay. Unfortunately, to the best of our knowledge, no large-scale TCP-based latency dataset for satellite networks is publicly available. To calibrate protocol-level simulation errors, the relative differences between the simulated and measured latencies are computed over the same circuit path using the same interface, deriving an error distribution. The subsequent simulation is adjusted using error rates sampled repeatedly from this distribution, with the mean of the adjusted values providing a final latency estimate.

\section{SaTor Latency Reduction Scheme}

\label{sec:sator-scheme}

A further question is: If satellite routing can outperform terrestrial routing in latency between certain relays at specific times (it can, as we will show in Section \ref{sec:evaluation-result}), how can Tor take this benefit? A naive idea is to entirely switch certain links from terrestrial to satellite routing.  However, while this may reduce latency at certain times, it could increase it at others, yielding sub-optimal gains. This section presents a latency reduction scheme for SaTor, featuring adaptive dual-homing routing based on active latency measurements.

\subsection{Overview}

\begin{figure}
    \centering
    \includegraphics[width=1\linewidth]{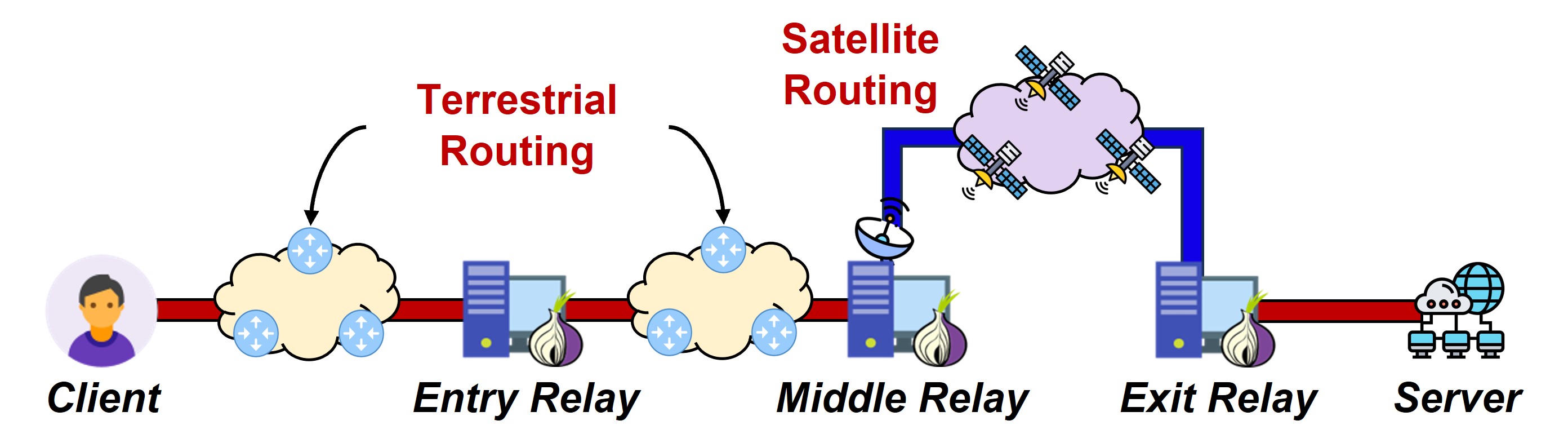}
    \caption{An illustration of circuit in SaTor.}
    \label{fig:SaTor-overview}
\end{figure}

SaTor proposes equipping a subset of Tor relays with satellite network services in addition to their existing terrestrial connections. These dual-homed relays then communicate with other relays via satellite interfaces during periods when terrestrial routing is slow. As shown in Fig. \ref{fig:SaTor-overview}, a Tor client accesses a server through three relays that are selected using Tor's default algorithm. The middle relay chooses to route traffic through satellite network to reach the exit, as it is considered faster than terrestrial link during that period. Thereby, with the circuit path remaining unchanged, SaTor could achieve lower latency compared to the current Tor. This design requires additional hardware and software to be integrated into Tor, as detailed below.

\subsection{SaTor Hardware Configuration}

SaTor assumes that certain Tor relays install a satellite dish and subscribe to a satellite network provider, similar to purchasing a Wi-Fi router and subscribing to traditional Internet service. The new satellite interface creates a dual-homing configuration alongside the existing terrestrial interface. Relays can then route traffic to other relays through the faster interface based on real-time latency performance.

\begin{figure}
    \centering
    \includegraphics[width=0.9\linewidth]{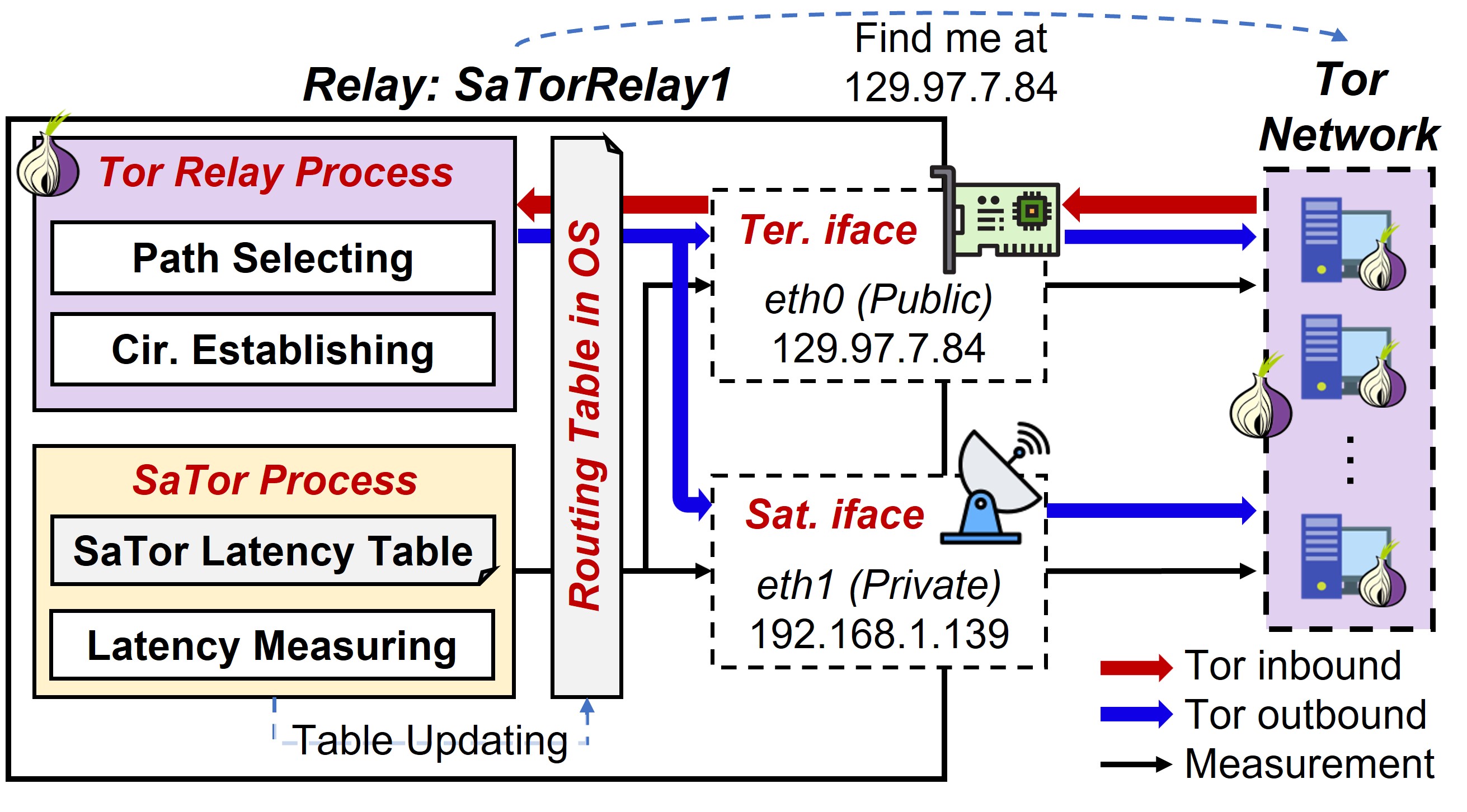}
    \caption{Routing architecture between a dual-homing SaTor relay with the remaining of Tor network.}
    \label{fig:SaTor-relay-overview}
\end{figure}

Starlink, the most widely used satellite network, offers a business plan that includes a public IP address and a personal plan that provides only a private IP \cite{StarlinkServicePlans}. A Tor relay requires a public IP to be uploaded to Tor consensus for handling inbound connections from other relays, but a private IP is sufficient for outbound connections. Therefore, SaTor can choose either plan from Starlink, provided that at least one public IP is available on either the terrestrial or satellite interface. Fig. \ref{fig:SaTor-relay-overview} illustrates a configuration where relay \emph{SaTorRelay1} operates a terrestrial interface (\emph{eth0}, with public IP 129.97.7.84) and a satellite interface (\emph{eth1}, with private IP 192.168.1.139). \emph{SaTorRelay1} publishes its public address by setting the `ORPort' and `Address' fields in Tor's configuration file, allowing inbound connections from other relays to reach \emph{eth0}. When establishing outbound connections to other relays, \emph{SaTorRelay1} can select between \emph{eth0} and \emph{eth1} for optimal latency performance. This functionality is achieved by modifying the routing table in the operating system, without requiring changes to Tor software.

If \emph{SaTorRelay1} subscribes to Starlink's business plan, obtaining a public IP for both satellite and terrestrial interfaces, it could upload both IPs to Tor consensus for inbound connections, similar to running two relays on a single physical host. This would allow relays without satellite access to benefit from satellite acceleration by connecting to the satellite IP of \emph{SaTorRelay1}. However, it would require modifying Tor's default relay selection, as some relays with satellite IPs may have to be prioritized to achieve such acceleration. Therefore, SaTor adopts the ``one inbound, two outbound'' configuration, as shown in Fig. \ref{fig:SaTor-relay-overview}. All changes occur after Tor's current path selection, and relays choose fast interface solely for establishing outbound connections to the next relay. Moreover, to prevent end-to-end traffic correlation by satellite network provider, SaTor operates exclusively within the Tor network, meaning that Tor clients and exit relays do not conduct SaTor's functionality.

\begin{algorithm}[h!]
\caption{SaTor Latency Measuring}
\label{algo:sator-measurement}
\begin{algorithmic}[1]
\STATE \textbf{Define data structures:}
\STATE \hspace{5mm} $relay\_list$: List of all Tor relays
\STATE \hspace{5mm} $latency\_history[relay]$: Historical records of the satellite and terrestrial latencies for each relay
\STATE \hspace{5mm} $last\_time[relay]$: Last measurement time for relays
\STATE \hspace{5mm} $priority[relay]$: Measurement priority for relays

\vspace{0.15cm}

\STATE \textbf{Function} \texttt{GetFasterIfaceEN(relay)}:
\STATE \hspace{5mm} $p_{sat}, p_{ter} \gets$ Probability of satellite/terrestrial interface being faster in $latency\_history[relay]$
\STATE \hspace{5mm} \textbf{return} $-(p_{sat} \log_2 p_{sat} + p_{ter} \log_2 p_{ter})$ \textcolor{blue}{// Entropy}

\vspace{0.15cm}

\STATE \textbf{Function} \texttt{UpdatePriority(time\_now, a)}:
\STATE \hspace{5mm} \textbf{for each} $relay \in relay\_list$:
\STATE \hspace{10mm} $H_r \gets \texttt{GetFasterIfaceEN}(relay)$
\STATE \hspace{10mm} $F_r \gets time\_now - last\_time[relay]$
\STATE \hspace{10mm} $priority[relay] \gets a \cdot H_r + (1 - a) \cdot F_r$
\STATE \hspace{5mm} \textbf{end for}

\vspace{0.15cm}

\STATE \textbf{Function} \texttt{MeasureLatency(relay)}:
\STATE \hspace{5mm} $t_{sat} \gets$ Measured satellite latency to $relay$
\STATE \hspace{5mm} $t_{ter} \gets$ Measured terrestrial latency to $relay$
\STATE \hspace{5mm} $latency\_history[relay].append(t_{sat}, t_{ter})$

\vspace{0.15cm}

\STATE \textbf{Function} \texttt{MainLoop(T, N, a)}: 
\newline \textcolor{blue}{// T: interval time to pause before next measurement}
\newline \textcolor{blue}{// N: Maximum of relays to be measured each time}
\newline \textcolor{blue}{// a: Tunable, for calculating relay measurement priority}
\STATE \hspace{5mm} \textbf{while} True:
\STATE \hspace{10mm} $time\_now \gets$ Get current time
\STATE \hspace{10mm} \texttt{UpdatePriority}(time\_now, a)
\STATE \hspace{10mm} $top\_relays \gets$ Select top $n$ relays by priority
\STATE \hspace{10mm} \textbf{for each} $relay \in top\_relays$:
\STATE \hspace{15mm} \texttt{MeasureLatency}(relay)
\STATE \hspace{15mm} $last\_time[relay] \gets time\_now$
\STATE \hspace{10mm} \textbf{end for}
\STATE \hspace{5mm} \texttt{sleep}(t)
\end{algorithmic}
\end{algorithm}

\subsection{SaTor Software Functionality}

\begin{figure}
    \centering
    \includegraphics[width=1\linewidth]{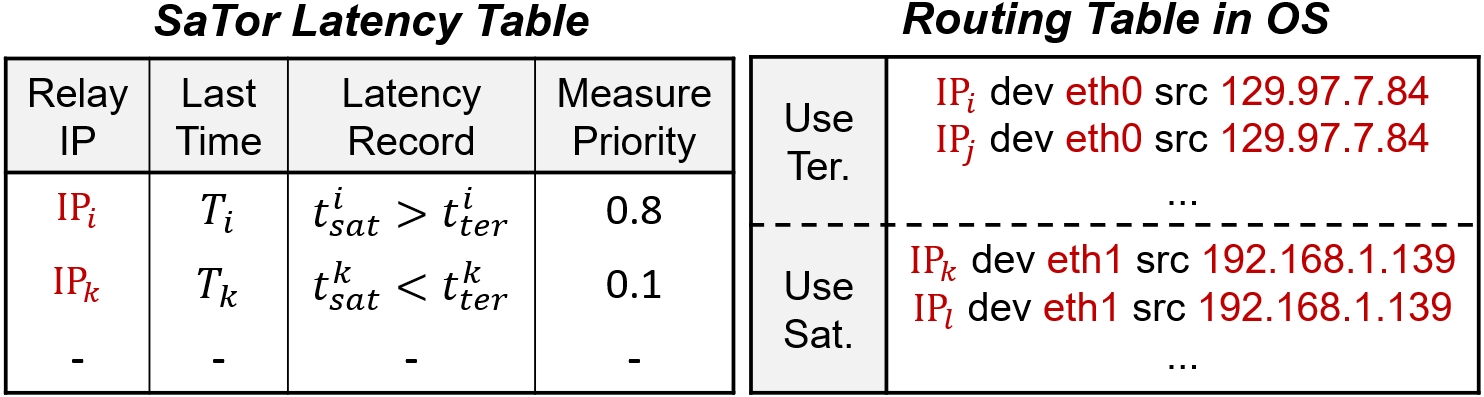}
    \caption{SaTor maintains a latency Table to record the measured latencies, and updates the system routing table to control relay's routing behavior for reducing latency.}
    \label{fig:SaTor-table-overview}
\end{figure}

SaTor works as an independent program separate from the Tor system. As shown in Fig. \ref{fig:SaTor-relay-overview}, in \emph{SaTorRelay1}, SaTor process periodically measures latency to a set of other relays through both satellite and terrestrial interfaces, maintaining a table of the measured latencies. Latency can be measured using hping3 tool\cite{kali-hping3}, by attempting to establish TCP connections to an open port of relays, as recorded in Tor's consensus. Based on the measured latencies, the SaTor process updates the routing table in operating system, connecting to some relays via the satellite interface, and others via the terrestrial. As a result, the Tor relay software, operating at the application layer, will establish new connections using the network interface specified by the system's routing table, without requiring any code changes.

SaTor's latency measurement, as outlined in Algorithm \ref{algo:sator-measurement}, is conducted at regular intervals $T$ and measures up to $N$ relays each time. Relays are prioritized for measurement based on a score calculated by the time since the last measurement and the uncertainty of the faster interface to them. This uncertainty is quantified as the entropy of the random variable for each interface being faster, based on historical records. Simply put, if the terrestrial interface is highly likely to be faster than satellite for reaching a relay, it is continuously used for communication to that relay without the need for frequent measurements. After each interval $T$, SaTor updates the priority scores, selects the top $N$ relays with the highest priority, and sends probe packets through both interfaces to measure latency. Subsequently, SaTor updates its latency table and the system's routing table, as shown in Fig. \ref{fig:SaTor-table-overview}. Until the next measurement cycle, Tor follows the routing table for connecting to other relays. Adjusting $T$ and $N$ could balance SaTor's efficiency and effectiveness, as well as the extra workload imposed on Tor.

\section{SaTor Evaluation Result}
\label{sec:evaluation-result}
\subsection{Setup Overview}

The simulation was conducted on 100k Tor circuits (over 120k relay pairs), as described in Section \ref{subsec:evaluation-material}. Each relay pair was simulated every 5 minutes over a 24-hour period, covering multiple LEO satellite orbits ($\approx$100 minutes each) \cite{handley2018delay} to capture the periodic latency variations of satellite routing. The simulation adopted \emph{ISL-enabled} routing strategy (Fig. \ref{fig:routing-strategy-ISL}), as currently used in Starlink. The SaTor latency measurement, as introduced in Section \ref{subsec:evaluation-approach}, encompasses 7,280 circuits with distinct entry relays, conducted over $\approx$30 rounds during August 2024. 
In each round, every circuit is measured 10 times at 1-second intervals. Ultimately, 6,897 circuits yielded over 300 valid latency data points on both satellite and terrestrial interfaces, which were selected to comprise the measurement dataset.

The evaluation employs various percentiles of the latency data. For each relay pair, we obtain a simulated set and a measured set, each containing $\approx$300 data points. The 50th percentile of these latencies reflects link performance under typical conditions, indicating that latency is lower than this value half of the time and higher the other half. Values above the 95th percentile indicate performance under congested scenarios, and the 90th percentile represents the latency experienced by the majority of users \cite{statistic-latency}.

The evaluation is organized as follows. First, baseline traffic speeds are analyzed to offer a preliminary view of satellite and terrestrial latencies. Second, simulation and measurement are compared to derive an error distribution to correct subsequent simulations. Third, corrected simulations are performed across all circuits. Fourth, a dual-homing scenario, as proposed in Section \ref{sec:sator-scheme}, is simulated to evaluate SaTor's best-case potential. Finally, several practical challenges in SaTor deployment are investigated.

\subsection{Baseline Traffic Speed}

The ECDFs for the speed of satellite and terrestrial traffic are derived from LENS and RIPE datasets. Satellite traffic speed is calculated along actual routing paths, from the testbed, through satellites to a ground station. Terrestrial speeds are estimated using straight-line distances between RIPE probes, due to the difficulty of predicting real-world Internet routing paths, as introduced in Section \ref{subsec:evaluation-approach}. 

As shown in Fig. \ref{fig:baseline-speed} (Appendix \ref{appendix:baselin-latency-measurement}), when the distance between endpoints is 0-2k kilometers (0-4k km round-trip), satellite traffic often transmits faster than terrestrial traffic. Beyond this range, terrestrial routing generally achieves higher speeds, while satellites still retain a non-negligible chance of being faster. This is likely due to the fact that long-haul terrestrial links often follow optimized backbone routes, while short paths may suffer from inefficient inter-domain routing, leading to higher-than-necessary delays \cite{bozkurt2017internet,hoiland2016measuring}. However, the higher speed of satellite traffic does not necessarily equate to lower latency, since the routing path is often longer due to bent-pipe detours via satellites and ground stations. Satellite routing yields a latency advantage only when the speed gain outweighs the additional path length, which will be examined by SaTor simulator.


\subsection{Simulation Error Evaluation}

\begin{figure}[h]
    \centering
	\subfigure[Satellite (Starlink)]{		
		\centering
		\includegraphics[width=0.46\linewidth]{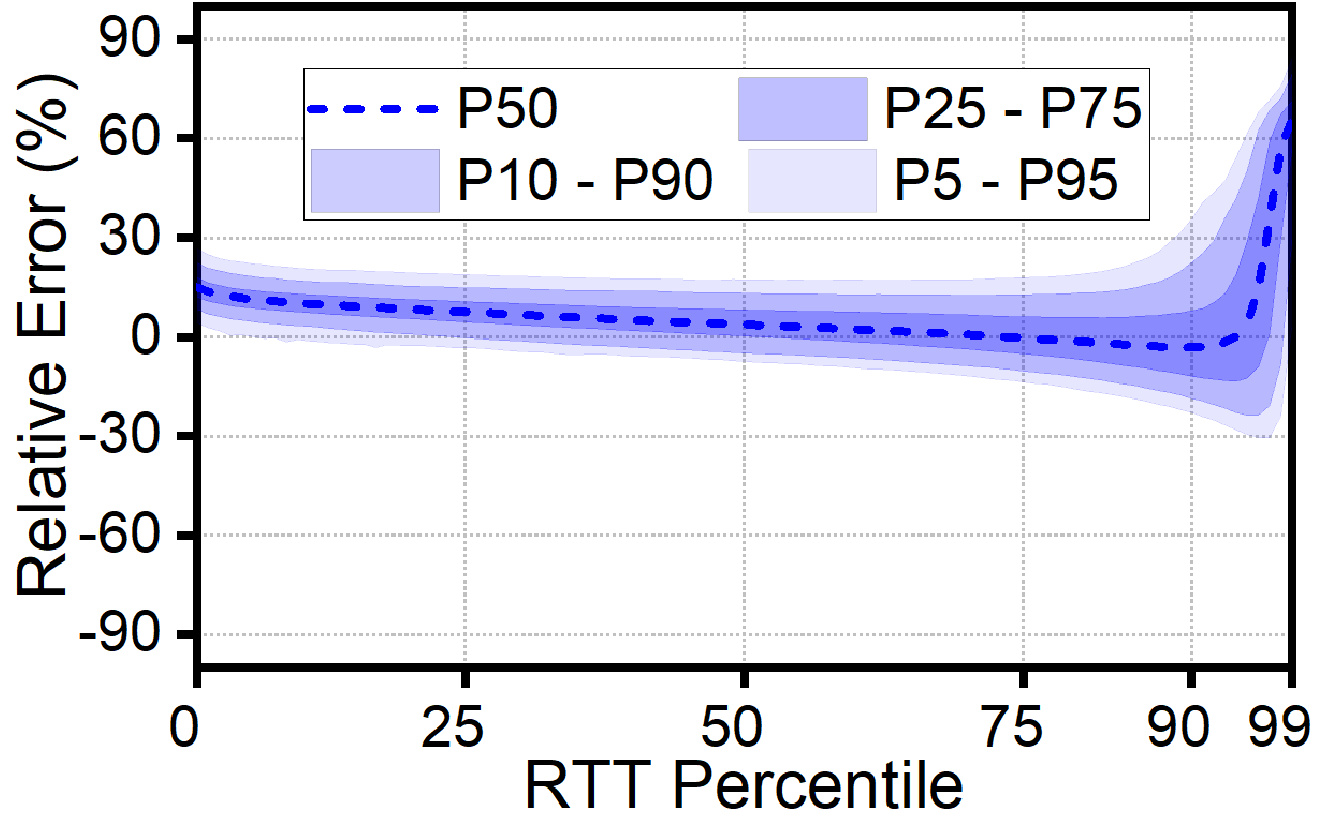}
		\label{fig:relative-error-waterloo-percentile-sat}
	}
	\subfigure[Terrestrial]{
		\centering
		\includegraphics[width=0.46\linewidth]{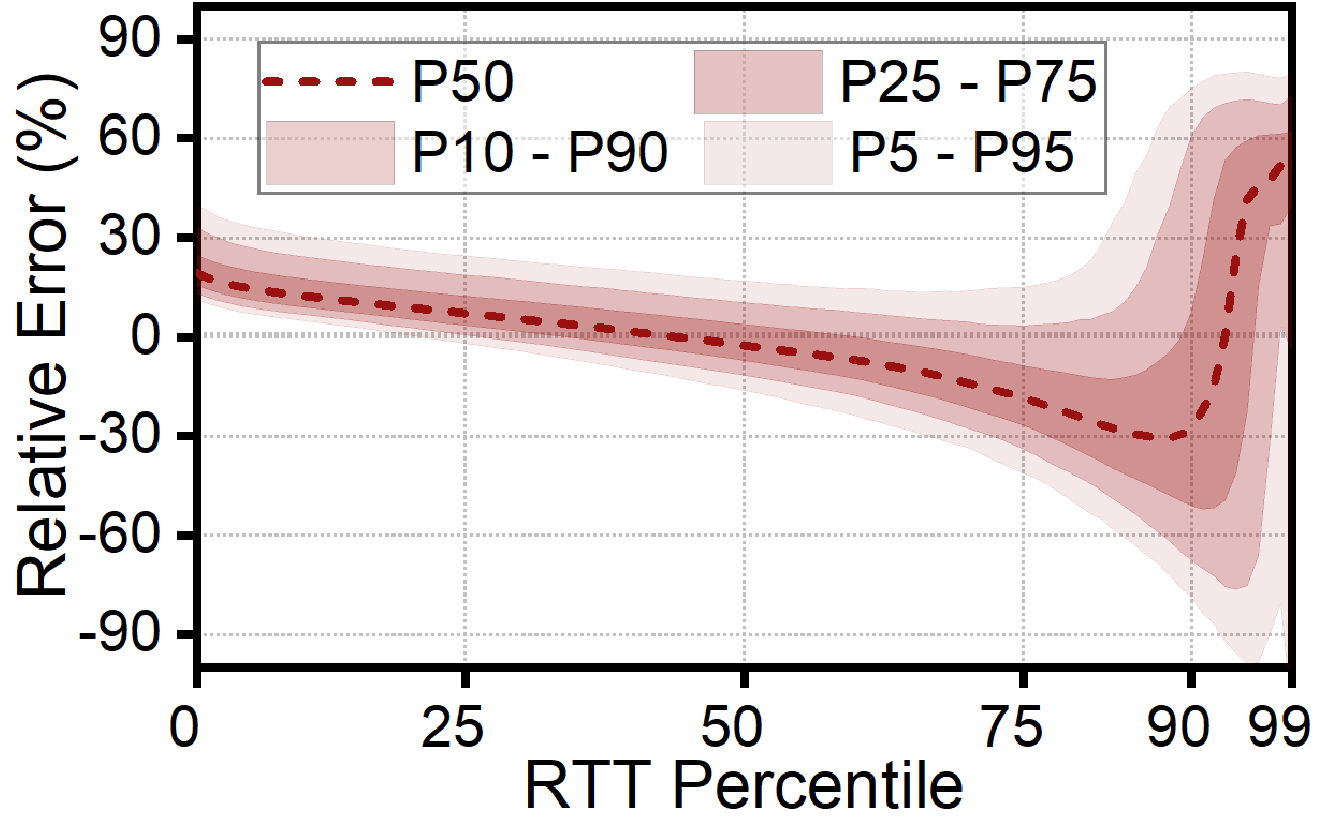}
		\label{fig:relative-error-waterloo-percentile-ter}
	}
	\caption{Relative error between simulated and measured latencies across Waterloo circuits at varying percentiles. }
	\label{fig:relative-error-waterloo}
\end{figure}

The traffic speed CDF enables latency estimation between two relays using their geographical distance. However, the ICMP speeds may deviate from Tor's TCP-based, onion-routed traffic. For a given circuit, let $n$ simulated latencies be $\mathcal{L}_s=\{l_s^1, l_s^2,\dots,l_s^n\}$ and $n$ measured latencies be $\mathcal{L}_m=\{l_m^1, l_m^2,\dots,l_m^n\}$. The simulation-measurement relative error at the $i$-th percentile is calculated as $e_i=(P_i(\mathcal{L}_m)-P_i(\mathcal{L}_s))/P_i(\mathcal{L}_m)$, where $P_i(\cdot)$ denotes the operation that extracts the $i$-th percentile from a latency set. We examine 5,280 circuits with both simulated and measured data. Some circuits are excluded because their relays either exited the network or changed fingerprints during the time gap between the simulation and measurement. 

Fig. \ref{fig:relative-error-waterloo} shows the relative errors between simulated and measured latencies. The x-axis is the evaluated latency percentiles, and the y-axis displays the relative error in percentage. The dashed line shows the median error across all circuits, while the shaded regions indicate error ranges. For instance, the lightest shaded band represents the range between the 5th and 95th percentile of errors across circuits. As shown in Fig. \ref{fig:relative-error-waterloo-percentile-sat}, satellite simulations generally exhibit high fidelity. Below the 70th percentile, 90\% of circuits show relative errors between $\approx$-10\% and 25\%. Beyond the 75th, the error range widens, reaching from -30\% to 60\% at the 95th. A further trend is that beyond the 95th many circuits show positive errors, indicating that simulations underestimate latency compared to real measurements.

As shown in Fig. \ref{fig:relative-error-waterloo-percentile-ter}, below the 50th percentile, terrestrial simulations yield a narrow error range of roughly -15\% to 40\% for 90\% of circuits. At the 75th, this 90\% error range shifts to between -40\% and 15\%, while beyond the 90th, it further broadens to -90\% to 70\%. Across all percentiles, over 50\% of circuits stay within an error range between roughly -45\% to 60\%, supporting the overall effectiveness of simulation. Another notable trend is that the simulation tends to overestimate real-world measurements between the 50th and 95th latency percentiles, but underestimates them beyond the 95th percentile. Such underestimation is likely because tail latency reflects packet loss and retransmission in TCP, which ICMP-based simulations cannot capture.

Tail latency typically indicates link congestion with increased variability and unpredictability, leading to higher simulation errors. Terrestrial simulation exhibits higher error levels than satellite routing, mainly due to its coarse-grained approach, which estimates latency using straight-line distances and overlooks actual circuitous routing paths. In contrast, the satellite simulation tracks the precise coordinates of intermediate nodes to replicate fine-grained routing paths. Simulation discrepancies may also arise from the gap between the globally collected latency dataset and the specific latency characteristics at the measurement testbed. To mitigate this, the simulator is calibrated using the observed errors before being applied to the entire Tor network.

\subsection{Simulation Error Calibration}

\begin{figure}[]
    \centering
        \subfigure[Sat. $P_{50}$]{
		\centering
		\includegraphics[width=0.29\linewidth]{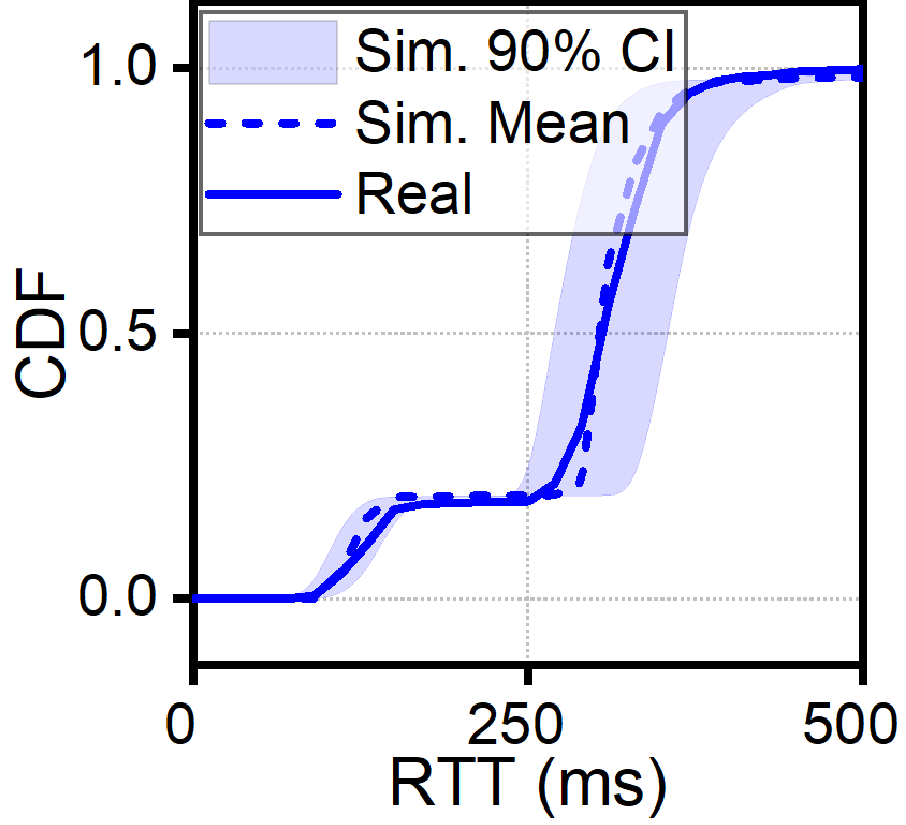}
		\label{fig:waterloo-latency-sat-50}
	}
        \subfigure[Sat. $P_{90}$]{
		\centering
		\includegraphics[width=0.29\linewidth]{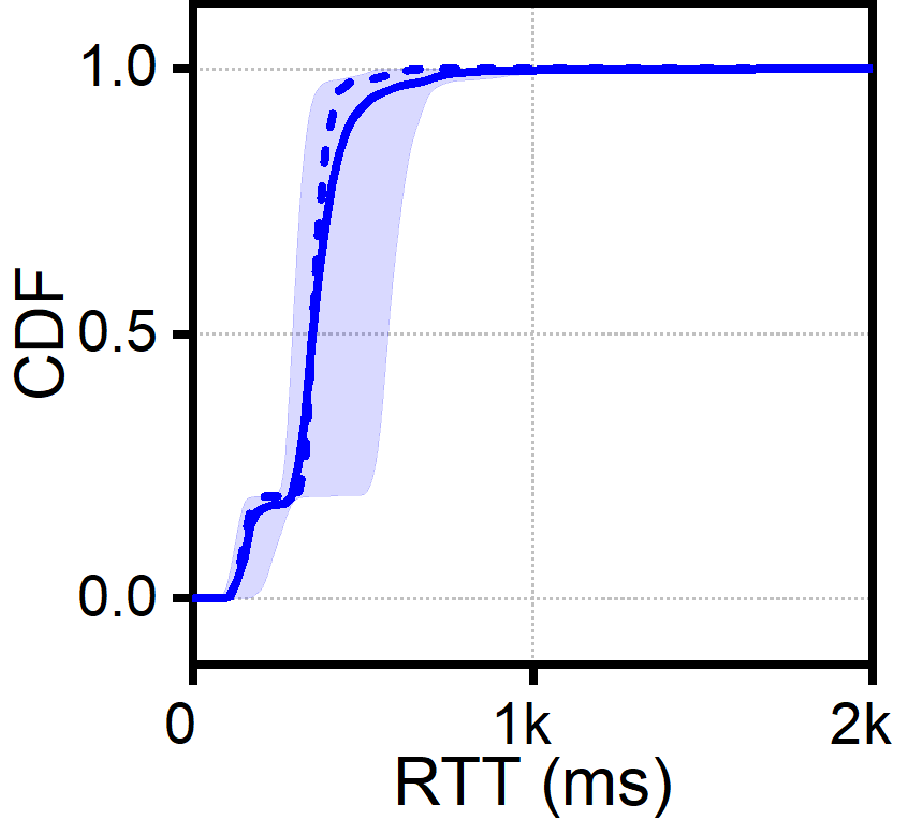}
		\label{fig:waterloo-latency-sat-90}
	}
        \subfigure[Sat. $P_{95}$]{		
		\centering
		\includegraphics[width=0.29\linewidth]{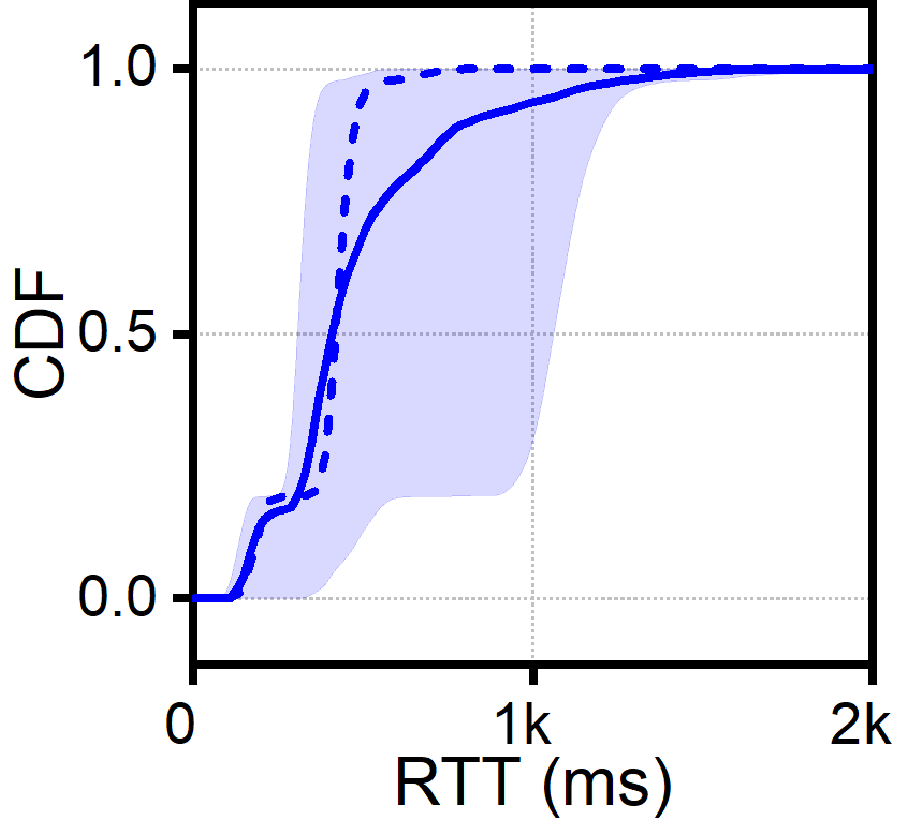}
		\label{fig:waterloo-latency-sat-95}
	}
	\subfigure[Ter. $P_{50}$]{		
		\centering
		\includegraphics[width=0.29\linewidth]{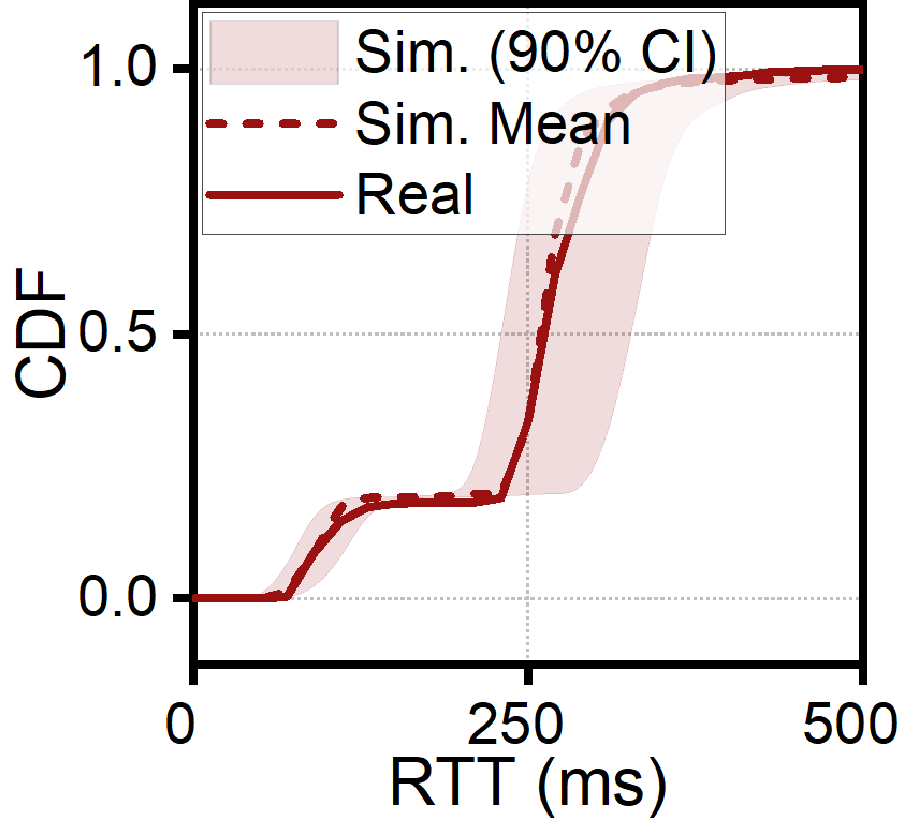}
		\label{fig:waterloo-latency-ter-50}
	}
        \subfigure[Ter. $P_{90}$]{
		\centering
		\includegraphics[width=0.29\linewidth]{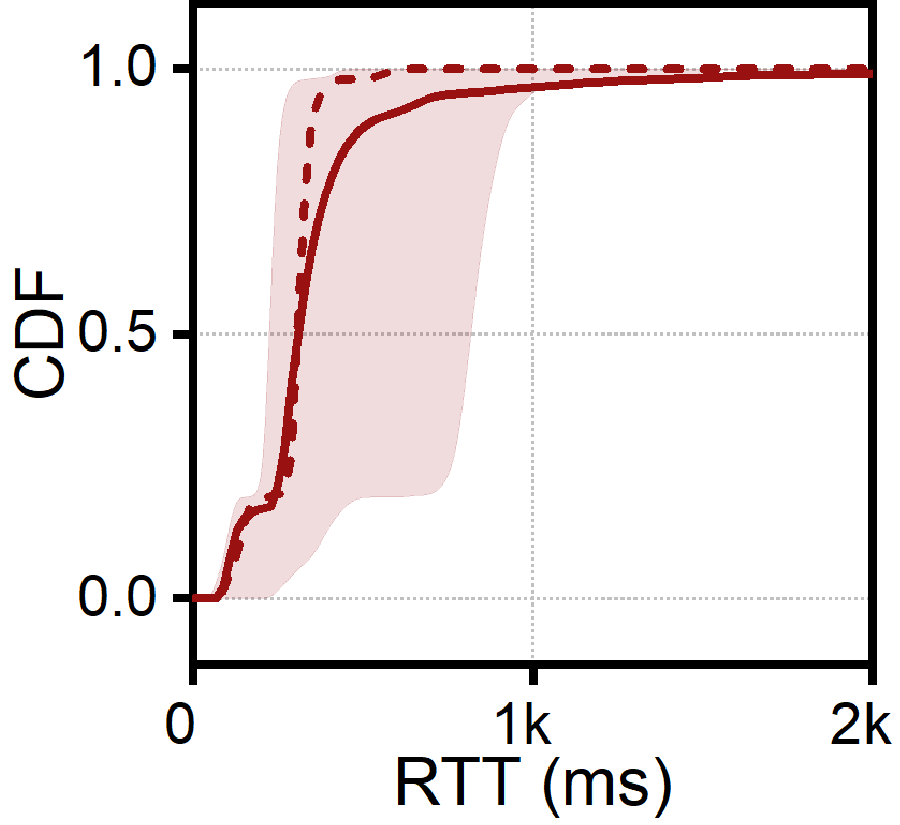}
		\label{fig:waterloo-latency-ter-90}
	}
	\subfigure[Ter. $P_{95}$]{
		\centering
		\includegraphics[width=0.29\linewidth]{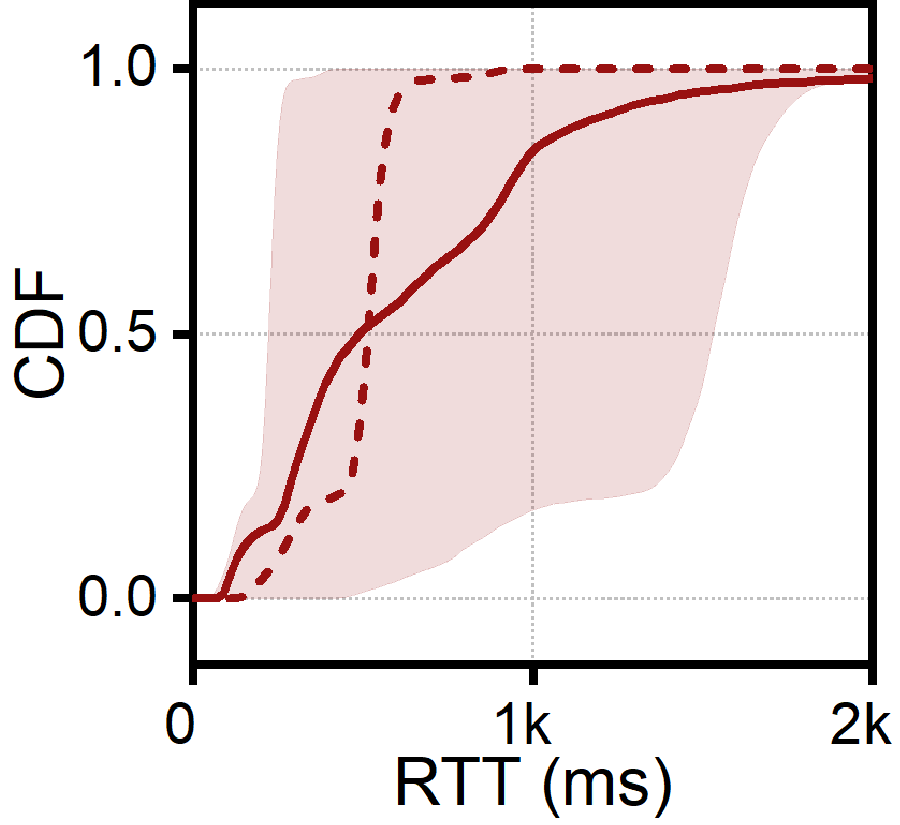}
		\label{fig:waterloo-latency-ter-95}
	}
	\caption{Latency CDFs from corrected simulations and real measurements for Waterloo-originating circuits. }
	\label{fig:simulation-measurement}
\end{figure}

\begin{table*}[ht!]
  \centering
  \caption{Simulated and Measured Latency Reduction After Using Satellite Routing for Waterloo-originating Circuits}
  \label{tab:relative-latency-reduction-waterloo}
  \setlength{\tabcolsep}{6pt}
  \begin{threeparttable}
    \begin{tabular}{lccccccc}
      \toprule
      & \multicolumn{6}{c}{\textbf{Relative Reduction (\%)}} & \\
      \cmidrule(lr){2-7}
      \multirow{-2}{*}{\textbf{Percentile}}
        & \textbf{$<$20} & \textbf{20--40} & \textbf{40--60} & \textbf{60--80} & \textbf{80--100} & \textbf{Sum.\ $>$20} & \multirow{-2}{*}{\textbf{Avg.\ Reduction}} \\
      \midrule
      \textbf{50th}
        & {\color[HTML]{9A0000} 100.00\% (99.89)\tnote{1}}
        & {\color[HTML]{3531FF} 0.00\% (0.03)}
        & {\color[HTML]{3531FF} 0.00\% (0.04)}
        & {\color[HTML]{3531FF} 0.00\% (0.04)}
        & {\color[HTML]{000000} 0.00\% (0.00)}
        & {\color[HTML]{3531FF} 0.00\% (0.11)}
        & {\color[HTML]{3531FF} 0.00\% (48.95)} \\
      \textbf{90th}
        & {\color[HTML]{9A0000} 97.06\% (79.72)}
        & {\color[HTML]{3531FF} 2.35\% (5.88)}
        & {\color[HTML]{3531FF} 0.00\% (4.12)}
        & {\color[HTML]{3531FF} 0.00\% (5.90)}
        & {\color[HTML]{3531FF} 0.59\% (4.38)}
        & {\color[HTML]{3531FF} 2.37\% (20.28)}
        & {\color[HTML]{3531FF} 25.06\% (57.89)\tnote{2}} \\
      \textbf{95th}
        & {\color[HTML]{3531FF} 19.01\% (39.30)}
        & {\color[HTML]{9A0000} 71.38\% (10.96)}
        & {\color[HTML]{3531FF} 8.55\% (18.84)}
        & {\color[HTML]{3531FF} 0.47\% (24.94)}
        & {\color[HTML]{3531FF} 0.59\% (5.96)}
        & {\color[HTML]{9A0000} 80.65\% (60.70)}
        & {\color[HTML]{3531FF} 29.93\% (58.08)} \\
      \bottomrule
    \end{tabular}
    \begin{tablenotes}
      \footnotesize
      \item[1] Indicating that for the 50th-percentile latency over the simulation period, 99.41\% of circuits show $<20\%$ reduction after using satellite, while the corresponding fraction in real measurements is 99.89\%.
      \item[2] Among relay pairs with significant latency reduction ($>20\%$), satellite routing is, on average, 25.06\% faster than terrestrial routing in simulation; the corresponding figure from real measurements is 57.89\%.
    \end{tablenotes}
  \end{threeparttable}
\end{table*}

At each latency percentile $i$, the relative errors between simulation and measurement across all circuits form a distribution $\mathcal{E}_i$. Then, the simulator estimates the $i$-th percentile latency by sampling $\mathcal{E}_i$ for $R$ times ($R=$10k in the following evaluation) to adjust the baseline simulation, yielding $R$ latency candidates. The mean of these candidates is taken as the representative estimate, while the 5th and 95th percentiles of them define a 90\% confidence interval (CI). 

Fig. \ref{fig:simulation-measurement} shows the CDFs of the calibrated simulation and measurement for the 50th, 90th, and 95th latency percentiles. The solid line represents the measured data, the dashed line indicates the representative latency estimate, and the shaded area denotes the 90\% simulation CI. At the 50th percentile (Fig. \ref{fig:waterloo-latency-sat-50} and \ref{fig:waterloo-latency-ter-50}), the representative estimate matches the measurement. At the 90th percentile, the satellite simulation still demonstrates high fidelity (Fig. \ref{fig:waterloo-latency-sat-90}), whereas the terrestrial simulation exhibits slight underestimation, as some simulated latencies fall below the measured values (Fig. \ref{fig:waterloo-latency-ter-90}). This trend becomes more pronounced at the 95th percentile (Fig. \ref{fig:waterloo-latency-sat-95} and \ref{fig:waterloo-latency-ter-95}). Nevertheless, in all cases, the 90\% CI encompasses the actual measurement curve.

\textbf{Calibration Limitations.} Overall, simulation calibration consists of three steps. First, the testbed measures TCP-based latencies in real Tor on a subset of circuits, each comprising two segments: the client–entry segment, which traverses either satellite or terrestrial links, and the entry–middle segment, which always traverses terrestrial links. Second, the simulator reproduces these circuit paths and settings to generate ICMP-based latency simulations. Third, an error distribution is derived by comparing the simulated and measured results to calibrate subsequent simulations. However, such calibration settings have two limitations.

First, the measurements do not directly capture the latency of individual relay pairs but instead reflect circuit-level latency over two consecutive pairs. The measured satellite latencies are also biased, as the entry–middle segments always traverse terrestrial networks. These issues may introduce skew when the calibration is applied to pair-level latency simulations. This limitation arises because the satellite testbed is research-oriented and does not permit running public proxy services such as Tor relays. With only Tor clients deployable, prior relay-pair-level latency measurement tools such as Ting~\cite{cangialosi2015ting} cannot be used.

Second, while the operation of repeated sampling and averaging can capture overall error trends across circuits, it may smooth out deviations at the individual level, making calibration less effective for circuits whose errors deviate significantly from the average, especially for tail latencies with greater variability and a wider error range. 

Despite these limitations, recall that the calibration targets two goals. The first is to capture protocol-level differences between ICMP and TCP latencies, which the current setup could capture. The second is to assess the benefits of satellite routing relative to terrestrial routing. This assessment remains valid as long as the latency benefits of satellite routing are not overestimated, regardless of moderate errors in their individual estimates. Table~\ref{tab:relative-latency-reduction-waterloo} presents the relative reduction between satellite latency ($l_{\text{sat}}$) and terrestrial latency ($l_{\text{ter}}$), computed as $(l_{\text{ter}} -l_{\text{sat}})/l_{\text{ter}} \times 100\%$, based on both measurement and calibrated simulation. Blue text indicates underestimation by the calibrated simulation (more circuits actually experience such level of reduction), while red indicates overestimation. 

At the 50th percentile, the measurement shows that 0.11\% of circuits achieve a significant reduction ($>$20\%), whereas the simulation reports none. At the 95th, the simulation overestimates this figure at 80.65\%, compared to the measured 60.70\%. However, this overestimation only occurs in the 20–40\% reduction range, as the simulation underestimates the number of circuits achieving larger reductions. Overall, the calibrated simulator is expected to \textit{conservatively} estimate satellite latency reduction, providing cautious assertions regarding the benefits of satellite routing.

\subsection{Satellite and Terrestrial Latency in Tor}

\begin{figure}[h!]
    \centering
        \subfigure[Sat. $P_{90}$]{
		\centering
		\includegraphics[width=0.29\linewidth]{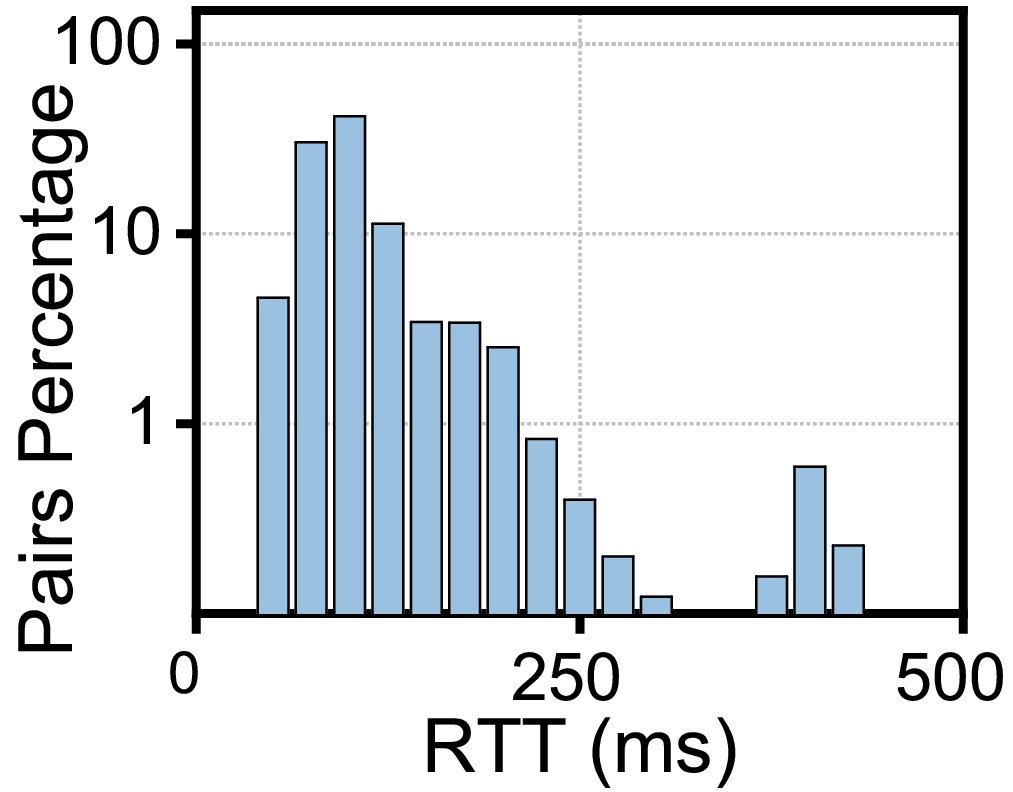}
		\label{fig:simulation-latency-sat-90}
	}
        \subfigure[Sat. $P_{95}$]{
		\centering
		\includegraphics[width=0.29\linewidth]{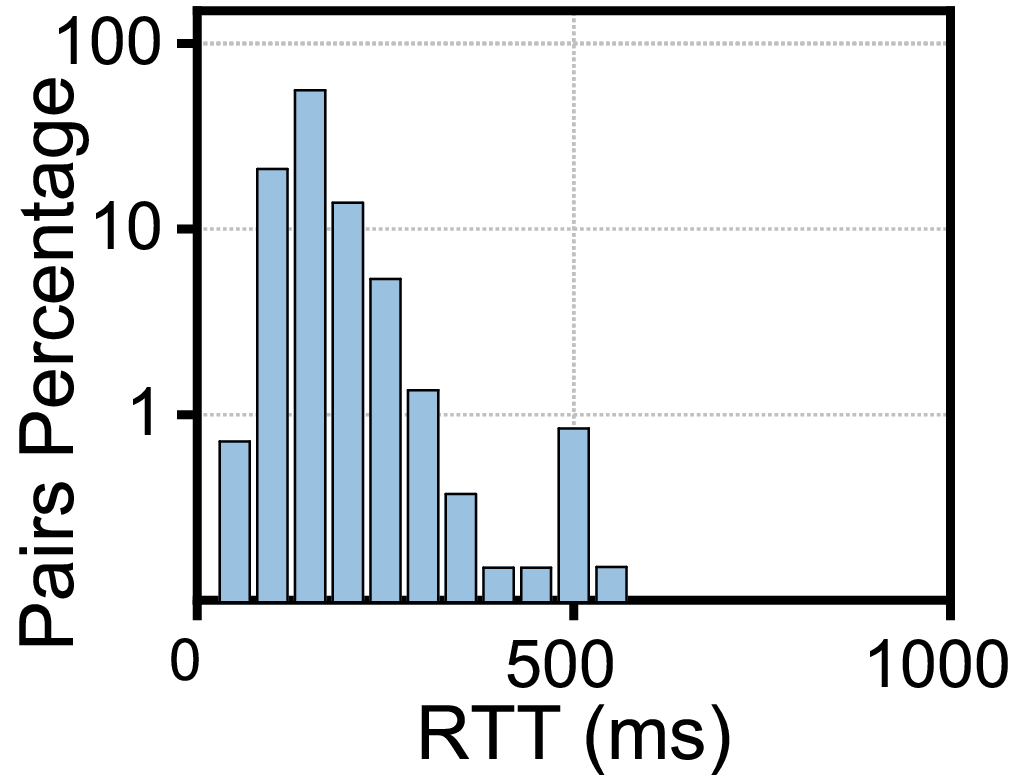}
		\label{fig:simulation-latency-sat-95}
	}
        \subfigure[Sat. $P_{98}$]{		
		\centering
		\includegraphics[width=0.29\linewidth]{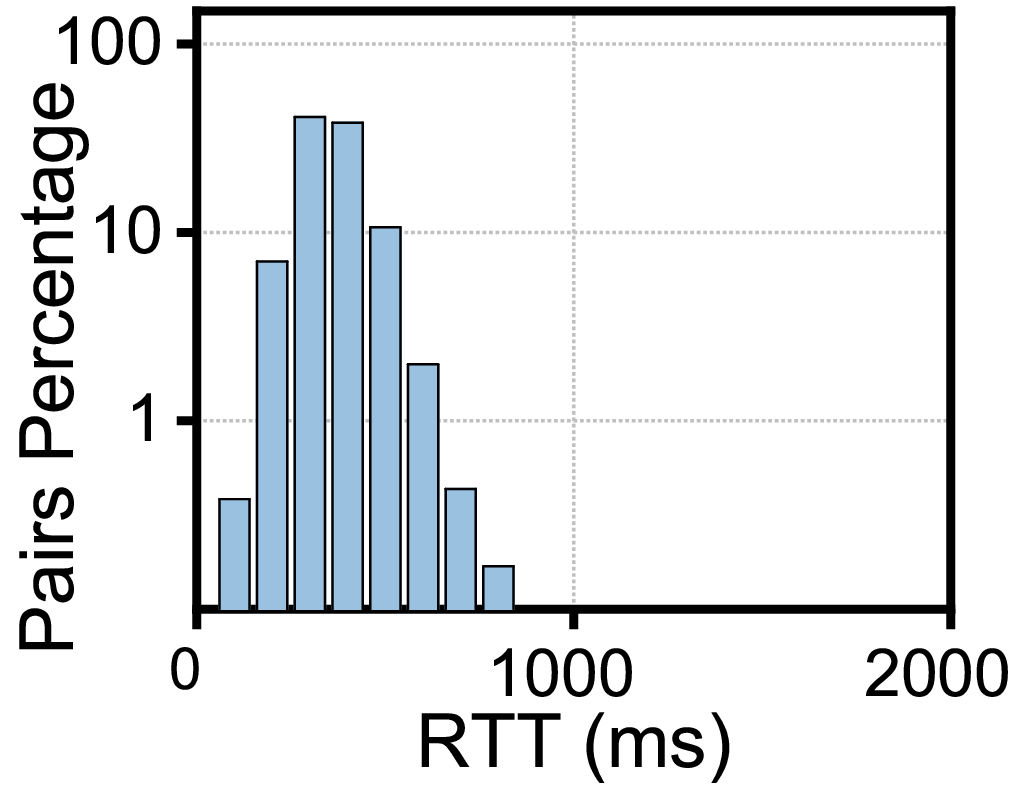}
		\label{fig:simulation-latency-sat-98}
	}
	\subfigure[Ter. $P_{90}$]{		
		\centering
		\includegraphics[width=0.29\linewidth]{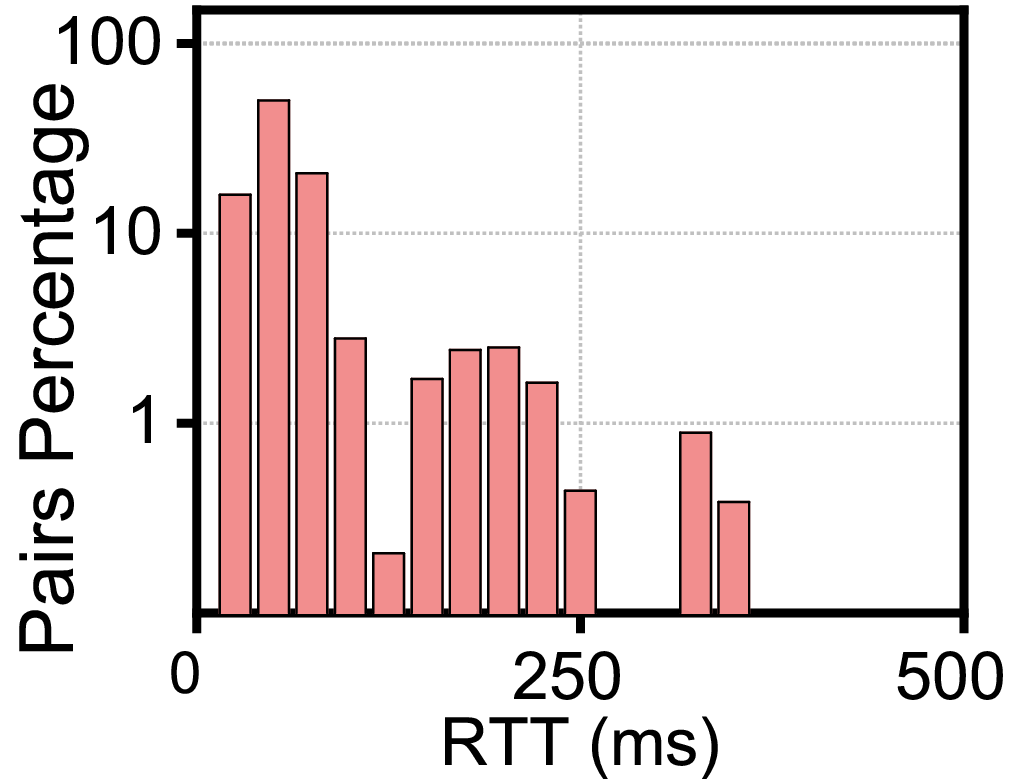}
		\label{fig:simulation-latency-ter-90}
	}
        \subfigure[Ter. $P_{95}$]{
		\centering
		\includegraphics[width=0.29\linewidth]{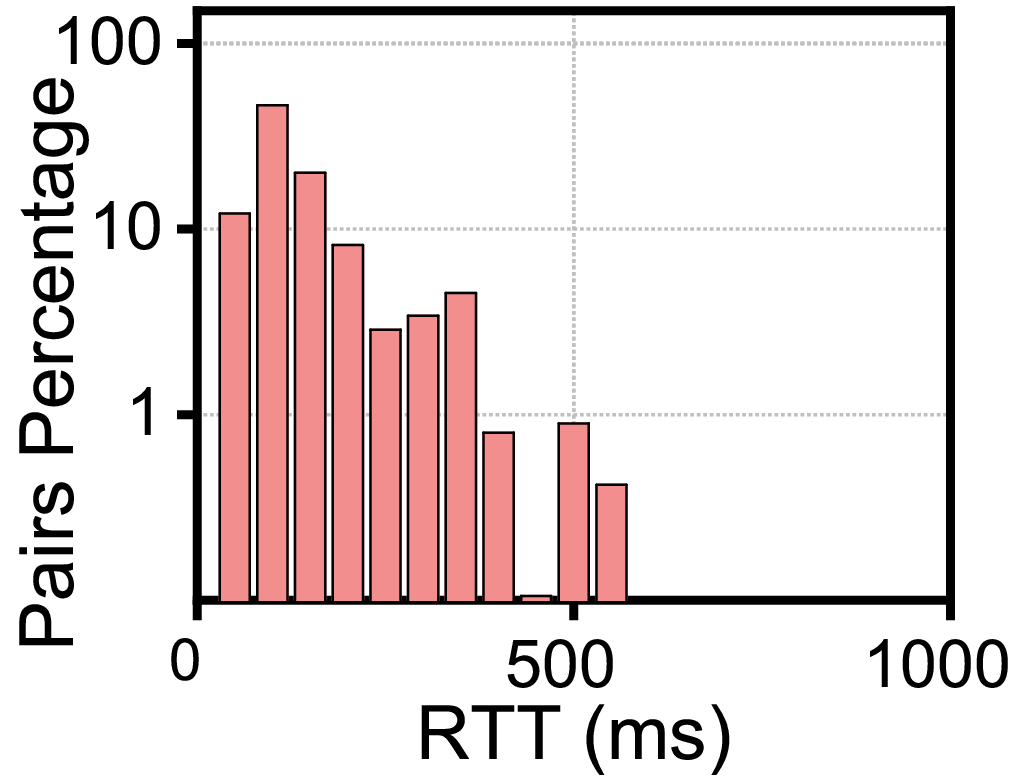}
		\label{fig:simulation-latency-ter-95}
	}
	\subfigure[Ter. $P_{98}$]{
		\centering
		\includegraphics[width=0.29\linewidth]{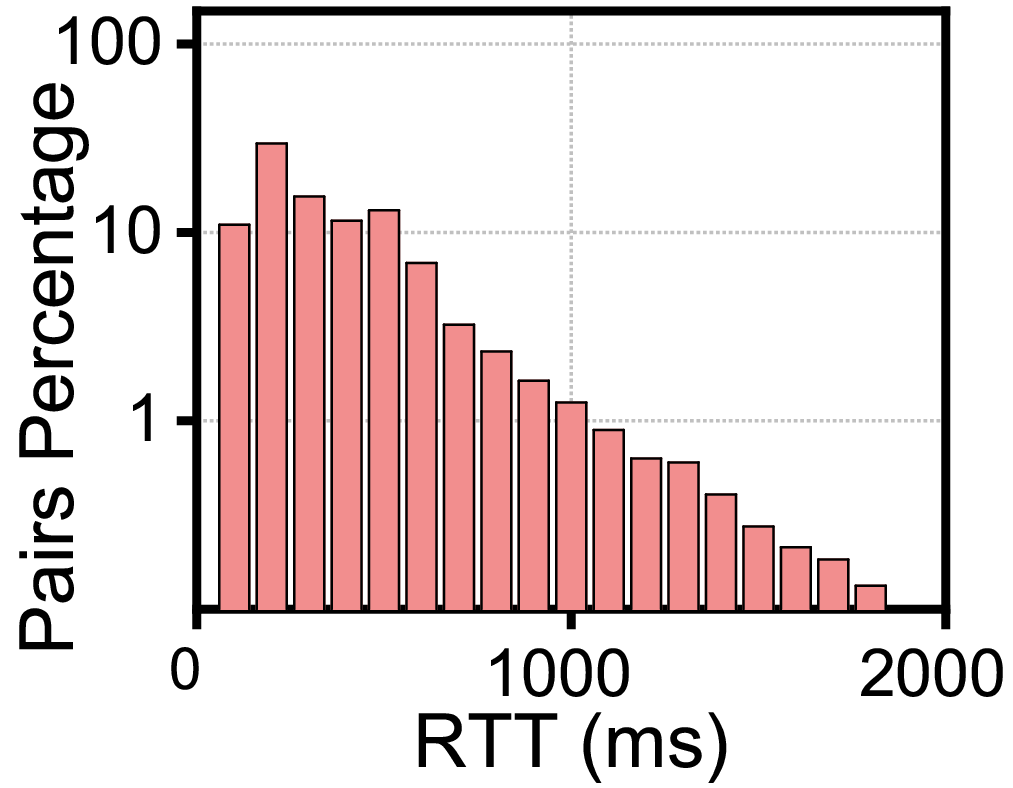}
		\label{fig:simulation-latency-ter-98}
	}
	\caption{Simulated latency histogram for pairs using satellite (a-c) and terrestrial routing (d-f). }
	\label{fig:simulation-latency-his}
\end{figure}


\begin{figure}[h!]
    \centering
        \subfigure[Lat-Varied Pairs (\%)]{
		\centering
		\includegraphics[width=0.293\linewidth]{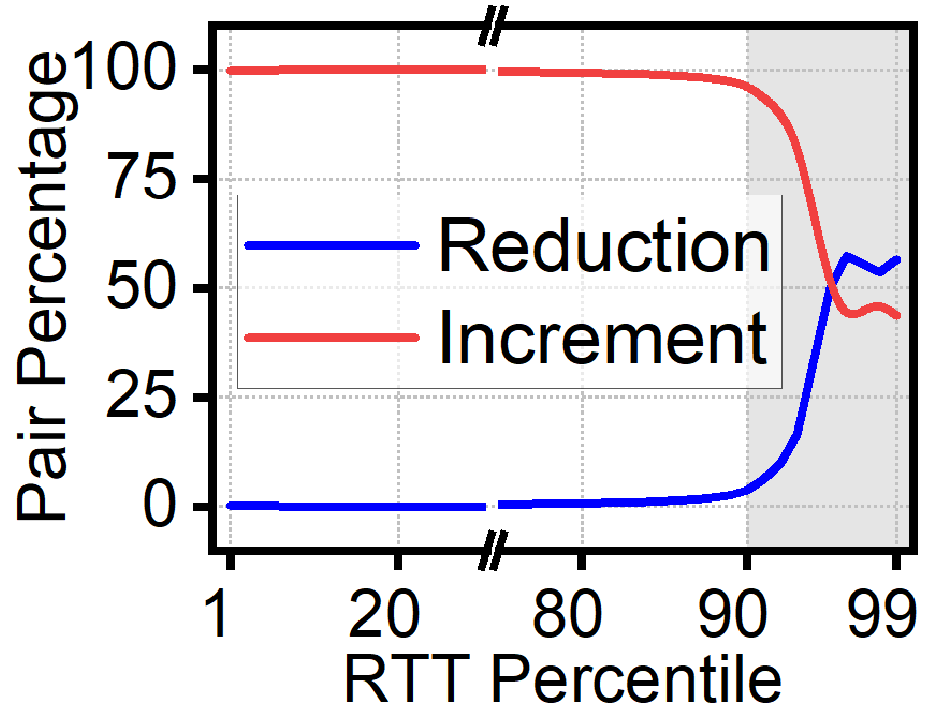}
		\label{fig:simulation-latency-reduced-pair-percent}
	}
        \subfigure[Variation (ms)]{
		\centering
		\includegraphics[width=0.293\linewidth]{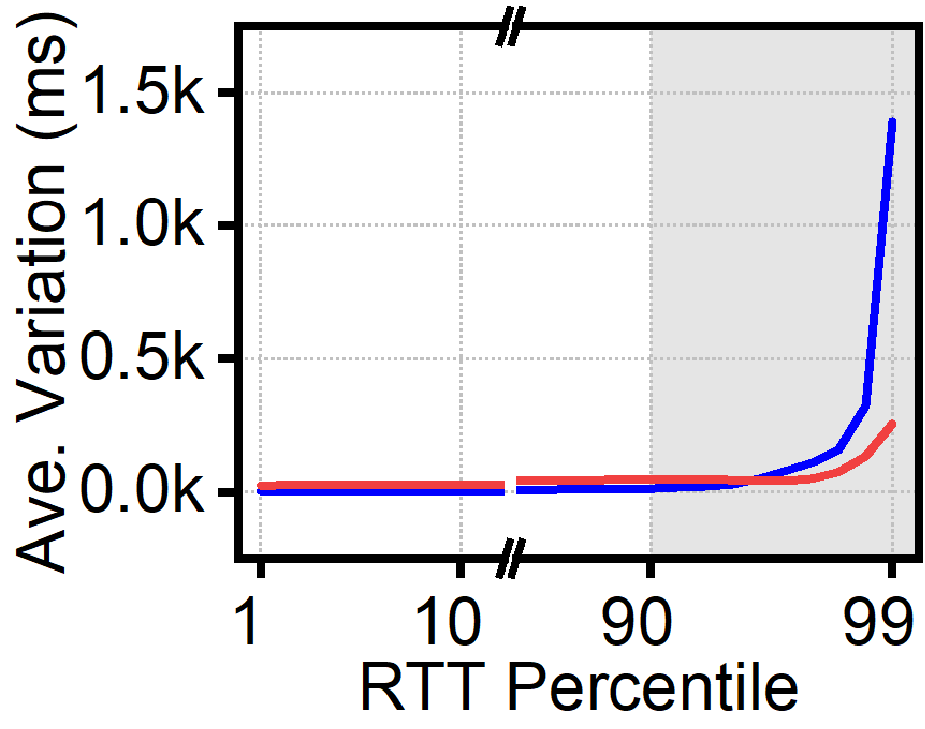}
		\label{fig:simulation-latency-average-reduction}
	}
        \subfigure[Variation (\%)]{		
		\centering
		\includegraphics[width=0.29\linewidth]{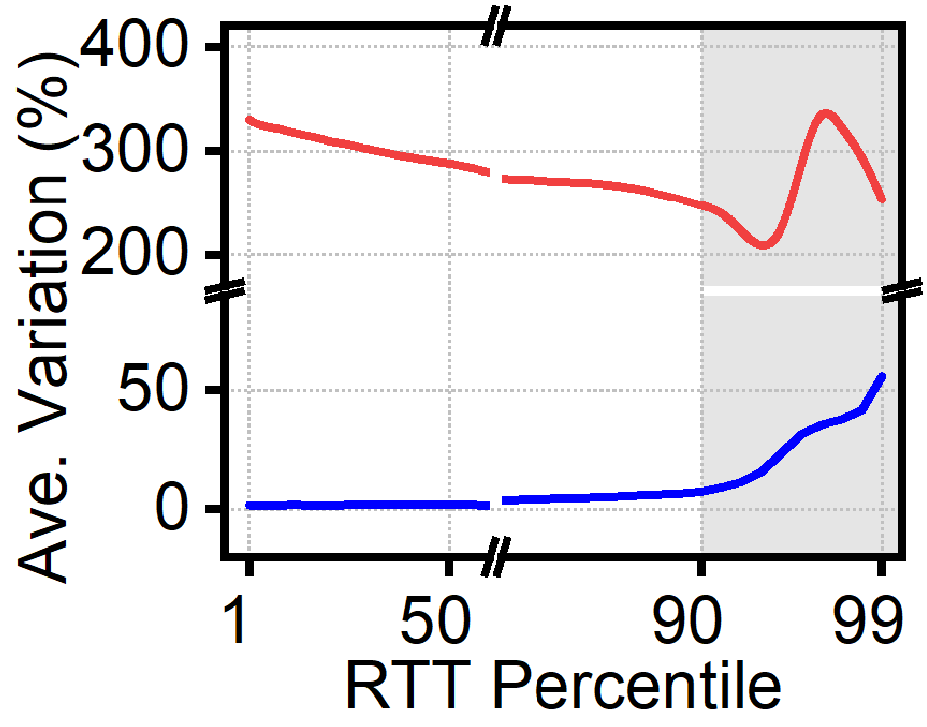}
		\label{fig:simulation-latency-average-reduction-relative}
	}
	\caption{Simulated latency variation using satellite routing across the 1-99th percentile throughout simulation. }
	\label{fig:simulation-latency-varying-percentile}
\end{figure}

Fig. \ref{fig:simulation-latency-his} presents the latency histograms for all 120k relay pairs, focusing exclusively on tail latencies at the 90th, 95th, and 98th percentiles, as satellite routing shows limited latency reduction at lower percentiles in the measurements. At the 90th percentile, satellite routing shows similar latency distributions as terrestrial. At the 95th, satellite routing achieves latency reduction efficacy, presenting a slightly smaller portion of slow relay pairs compared to terrestrial. At the 98th percentile, its effectiveness becomes pronounced, eliminating all tail pairs with latencies $>$1k ms.

\begin{table}[ht]
  \centering
  \caption{Simulated Latency Reduction of Satellite Routing}
  \label{tab:relative-latency-reduction-simulation}
  \setlength{\tabcolsep}{3pt}
  \begin{tabular}{lccccccc}
    \toprule
    \multirow{2}{*}{\textbf{P}} & \multicolumn{6}{c}{\textbf{Relative Reduction (\%)}} & \multirow{2}{*}{\textbf{Avg.}} \\
    \cmidrule(lr){2-7}
      & \textbf{$<$20} & \textbf{20--40} & \textbf{40--60} & \textbf{60--80} & \textbf{80--100} & \textbf{$>$20} & \\
    \midrule
    \textbf{90th} & 99.47\% & 0.53\% & 0.00\% & 0.00\% & 0.00\% & 0.53\% & 24.39\% \\
    \textbf{95th} & 81.95\% & 13.28\% & 4.48\% & 0.29\% & 0.00\% & 18.06\% & 34.26\% \\
    \textbf{98th} & 69.66\% & 10.84\% & 9.62\% & 8.10\% & 1.78\% & 30.35\% & 49.85\% \\
    \bottomrule
  \end{tabular}
\end{table}

\begin{figure*}[h!]
    \centering
    \subfigure[Latency-Varied Pairs (\%)]{
		\centering
		\includegraphics[width=0.222\linewidth]{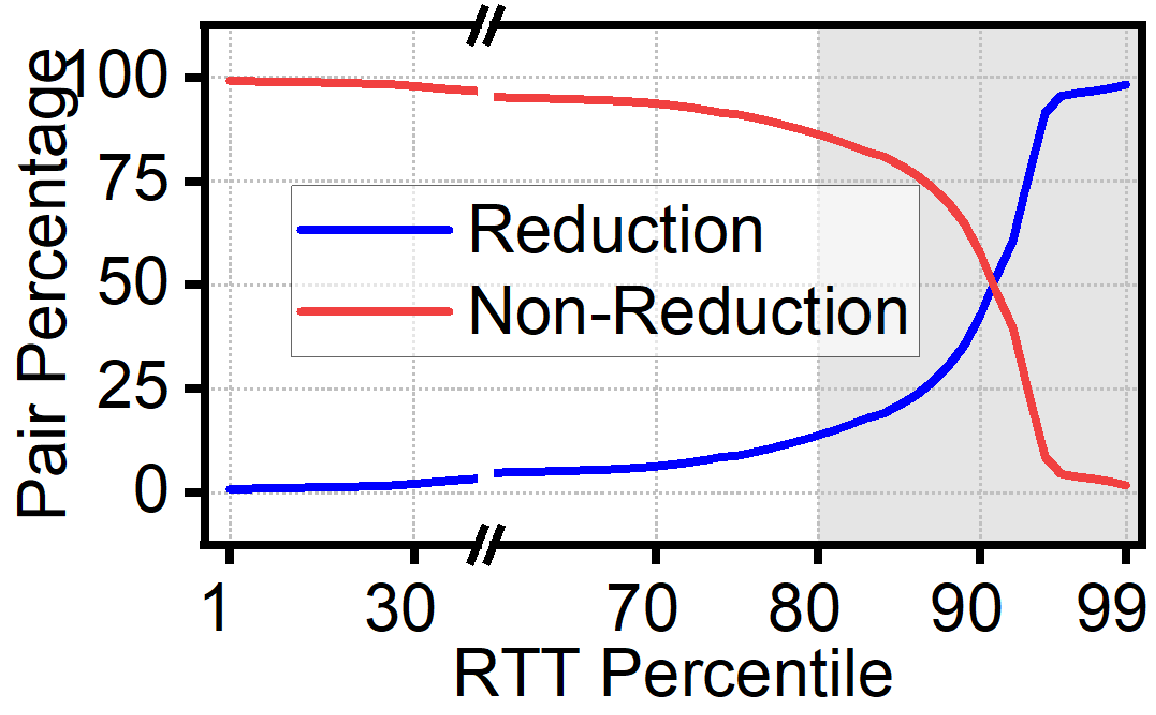}
		\label{fig:simulation-latency-reduced-pair-percent-dual}
	}
        \subfigure[Latency-Varied Circuits (\%)]{
		\centering
		\includegraphics[width=0.222\linewidth]{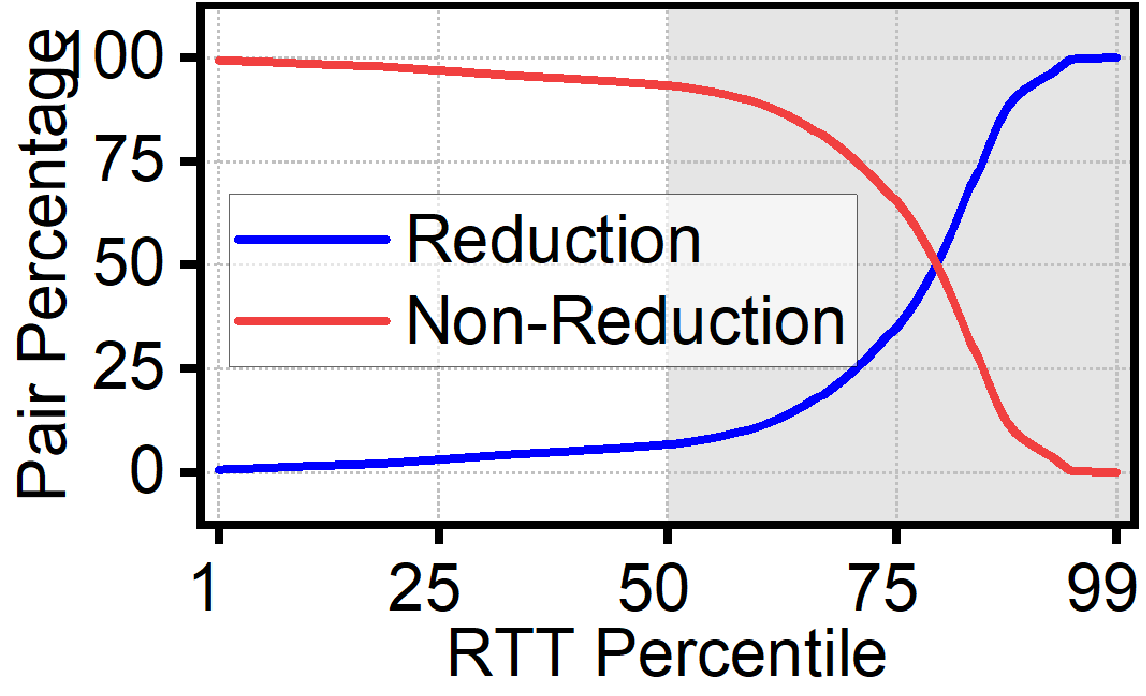}
		\label{fig:simulation-latency-reduced-cir-percent-dual}
	}
        \subfigure[Latency Variation (ms)]{		
		\centering
		\includegraphics[width=0.231\linewidth]{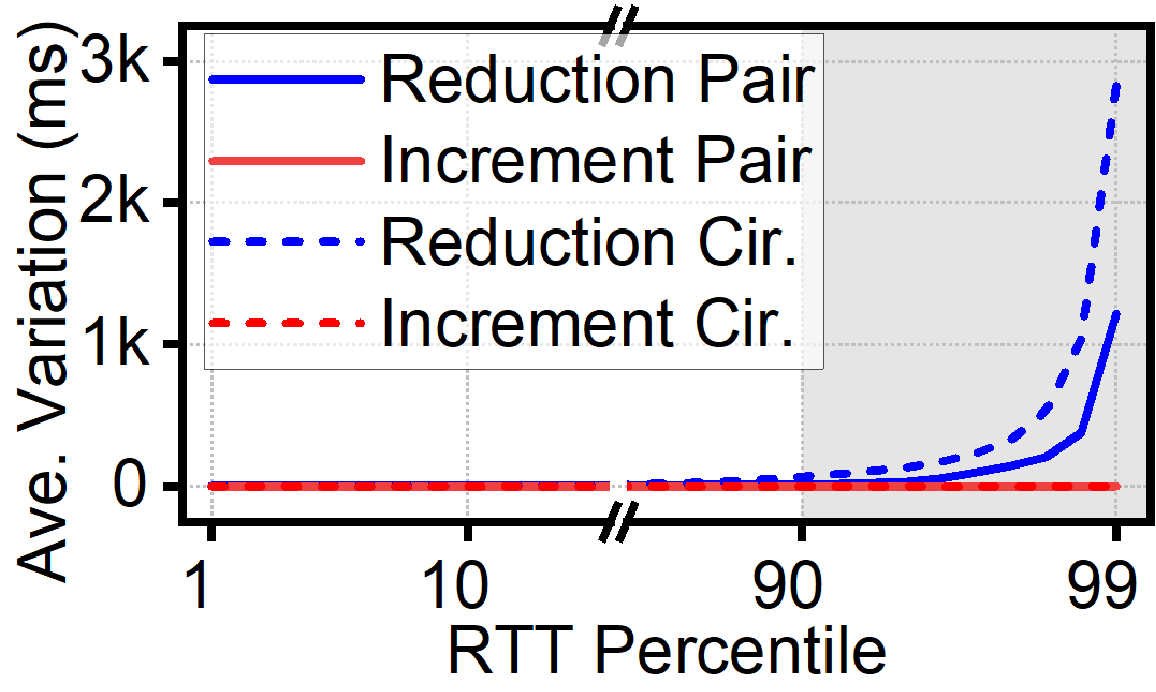}
		\label{fig:simulation-latency-average-reduction-abs-dual}
	}
        \subfigure[Latency Variation (\%)]{
		\centering
		\includegraphics[width=0.231\linewidth]{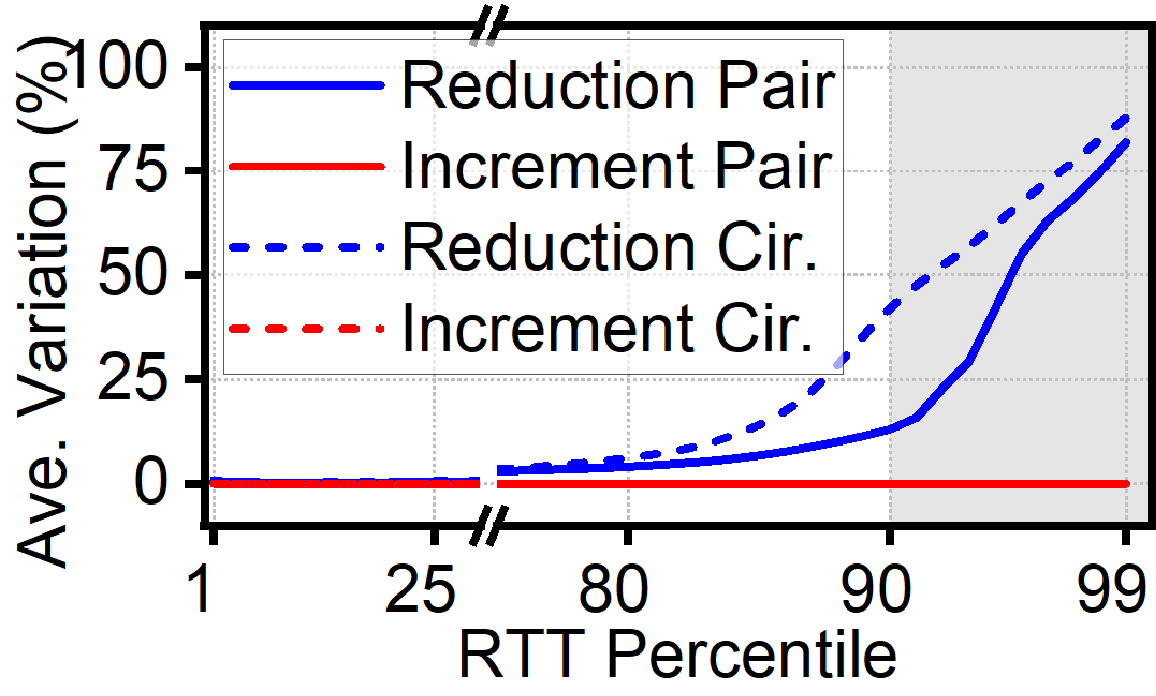}
		\label{fig:simulation-latency-average-reduction-percent-dual}
	}
	\caption{Simulated latency in dual-homed SaTor, including the percentage of pairs (a) and circuits (b) experiencing latency reduction, and the average latency reduction or increment in milliseconds (c) and percentage (d).}
	\label{fig:simulation-latency-varying-percentile-dual}
\end{figure*}

\begin{figure*}[ht!]
    \centering
    \subfigure[Reduced Circuits (\%)]{		
		\centering
		\includegraphics[width=0.179\linewidth]{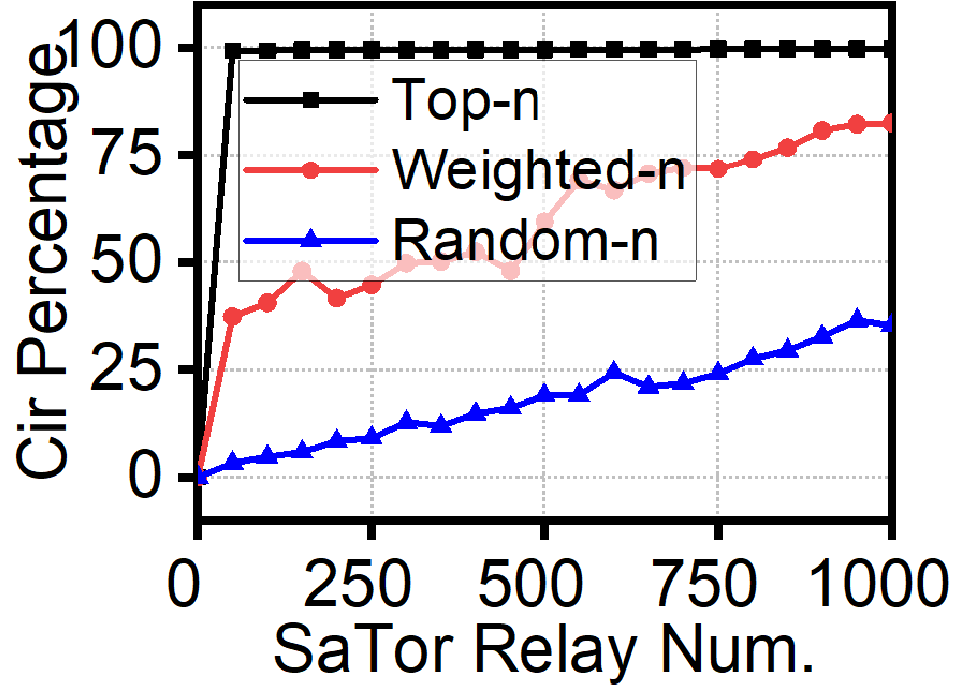}
		\label{fig:simulation-latency-reduced-cir-percent-incremental}
	}
    \subfigure[Ave. Reduction (ms)]{
		\centering
		\includegraphics[width=0.179\linewidth]{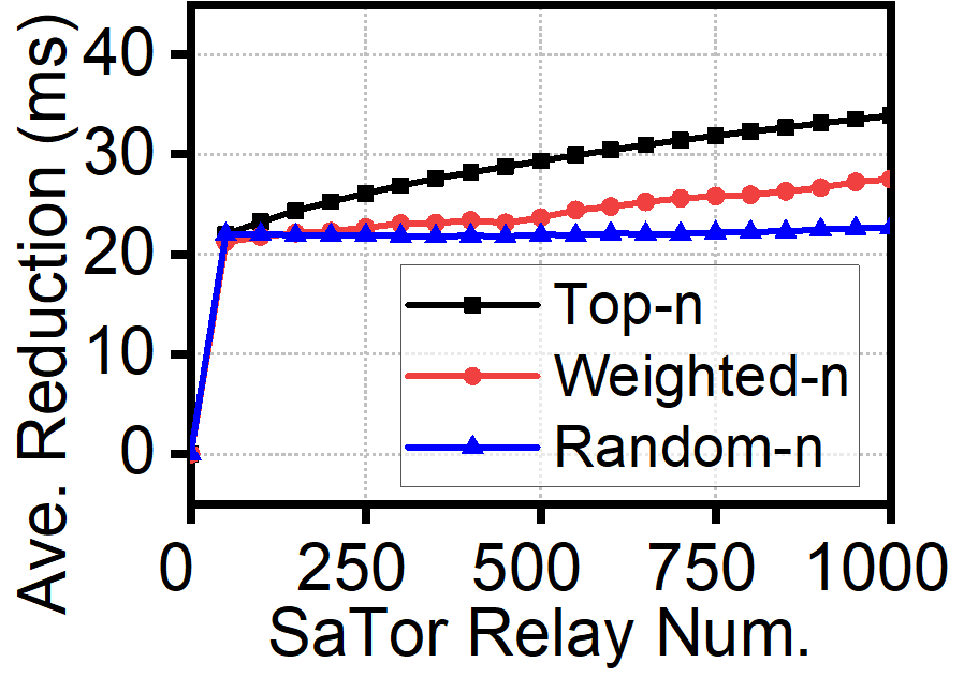}
		\label{fig:simulation-latency-average-cir-reduction-dual-incremental}
	}
    \subfigure[Ave. Reduction (\%)]{		
		\centering
		\includegraphics[width=0.179\linewidth]{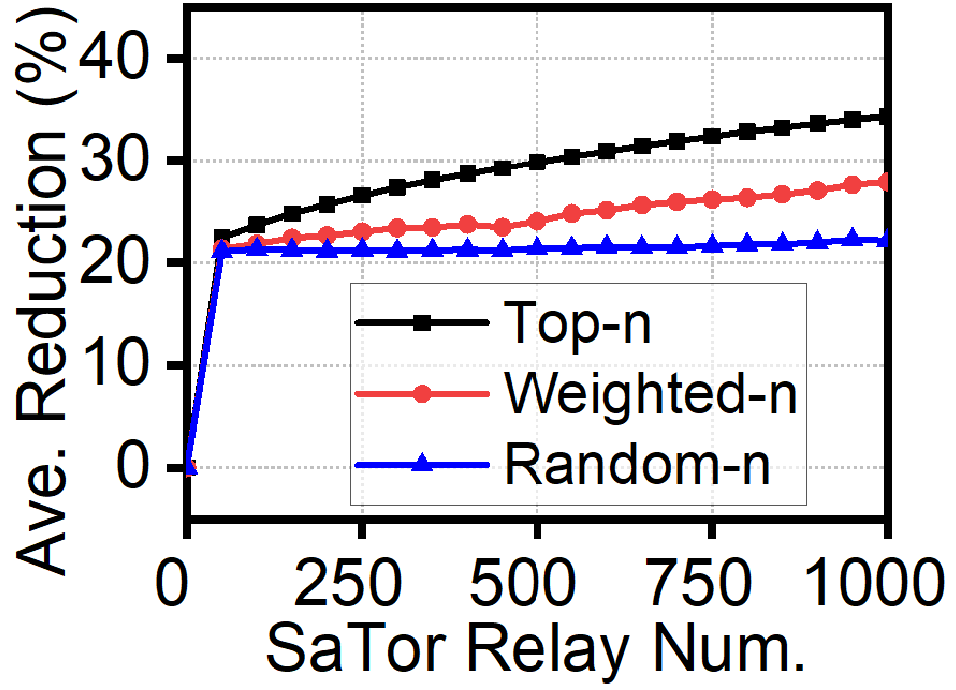}
		\label{fig:simulation-latency-average-cir-reduction-relative-dual-incremental}
    }
    \subfigure[Ave. Reduction CDF (ms)]{
            \centering
            \includegraphics[width=0.179\linewidth]{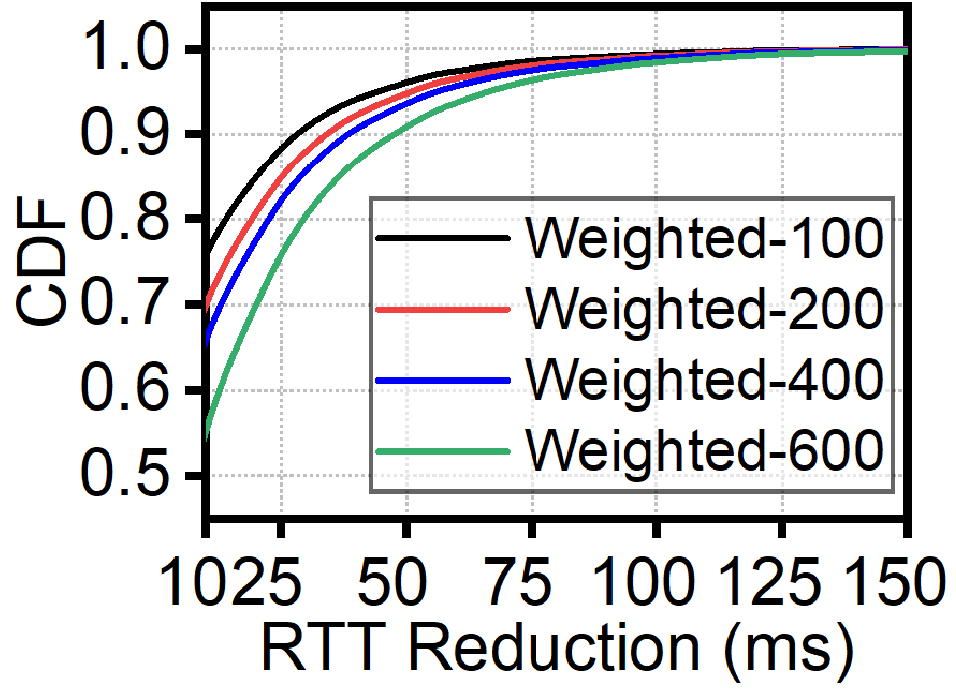}
            \label{fig:simulation-latency-cdf-cir-reduction-dual-incremental}
    }
    \subfigure[Ave. Reduction CDF (\%)]{
            \centering
            \includegraphics[width=0.179\linewidth]{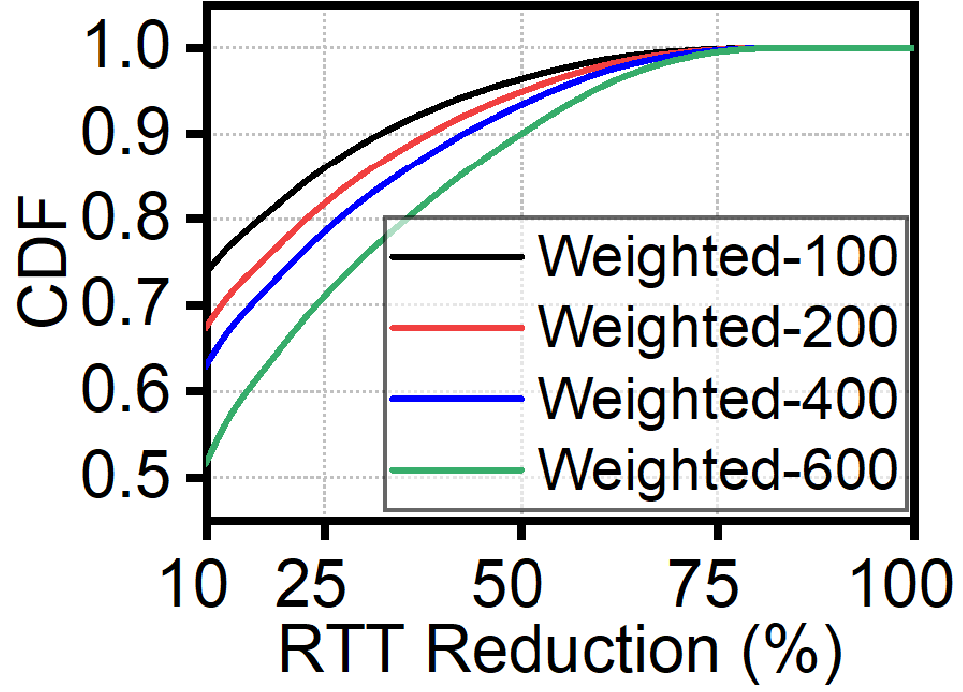}
            \label{fig:simulation-latency-cdf-cir-reduction-relative-dual-incremental}
    }
    \caption{Circuit latency reduction averaged across all percentiles of dual-homed SaTor in the three deployment scenarios.}
    \label{fig:simulation-latency-cir-dual-incremental}
\end{figure*}

\definecolor{darkred}{HTML}{9A0000}

\begin{table*}[ht]
  \centering
  \caption{Concurrent Tail Latency Probability Between Dual-Terrestrial and Satellite–Terrestrial Links}
  \label{tab:source-destination-reduction}
  \setlength{\tabcolsep}{2pt}
  \begin{threeparttable}
  \begin{tabular}{lccccccccccc}
    \toprule
    & \multicolumn{11}{c}{\textbf{Country of Destination Relays}} \\
    \cmidrule(lr){2-12}
    \textbf{Measurement Source} &
    \textbf{Germany} & \textbf{Japan} & \textbf{New Zealand} & \textbf{Singapore} &
    \textbf{Brazil} & \textbf{Australia} & \textbf{UAE} &
    \textbf{US} & \textbf{Canada} & \textbf{Indonesia} & \textbf{Argentina} \\
    \midrule
    London (Dual Terrestrial)      
      & 2.81\%
      & 6.38\% 
      & 8.80\% 
      & \textcolor{darkred}{18.05\%} 
      & 6.25\% 
      & 4.07\% 
      & \textcolor{darkred}{17.78\%} 
      & 3.57\% 
      & 7.40\% 
      & \textcolor{darkred}{15.07\%} 
      & \textcolor{darkred}{25.63\%} \\
    Waterloo (Sat. and Ter.)   
      & 6.62\% 
      & 4.01\% 
      & 5.83\% 
      & 5.38\%  
      & 5.23\% 
      & 5.50\% 
      & 6.67\%  
      & \textcolor{darkred}{40.80\%} 
      & \textcolor{darkred}{49.76\%} 
      & / & / \\
    \bottomrule
  \end{tabular}
  \begin{tablenotes}
    \footnotesize
    \item Values denote the conditional probability that the second link is above the 95th percentile latency, given that the first link is also above this threshold.
  \end{tablenotes}
  \end{threeparttable}
\end{table*}

Table \ref{tab:relative-latency-reduction-simulation} presents the percentage relative latency reduction of satellite routing compared to terrestrial routing. At the 90th percentile, approximately 0.53\% of relay pairs show significant reductions ($>$20\%), with an average reduction of 24.39\%. This share increases to 18.06\% at the 95th and 30.35\% at the 98th. Fig. \ref{fig:simulation-latency-varying-percentile} illustrates the overall latency reduction of satellite routing across 1-99th percentiles. Satellite routing begins to show notable latency reduction above the 90th percentile, and at the 99th, it accelerates over 55\% of relay pairs by an average of 1.4s (or 50\%). However, it is worth noting that a comparable portion of relay pairs ($\approx$45\% at the 99th) experience no latency reduction, and in fact suffer from an average latency increase of nearly 3x when switching from terrestrial to satellite routing, as shown by the red line in Fig. \ref{fig:simulation-latency-varying-percentile}. 

\textbf{Impact of Sub-optimal Satellite Routing.} Prior studies suggest that satellite networks may not always route traffic along the shortest path \cite{mohan2023multifaceted,izhikevich2024democratizing}, so the satellite simulations above report the average latency of the top $K=10$ shortest routes. To examine the impact of different $K$ values, we re-simulated satellite latencies for all 100k circuits over a one-hour testing period, using $K$ values up to 200. Compared to $K=10$, the average satellite latency across all circuits increases by 4.56\% for $K=50$, 7.24\% for $K=100$, and 10.68\% for $K=200$. Nonetheless, even with $K=10$, the simulation tends to underestimate the benefits of satellite routing compared to real measurements, as reported in Table~\ref{tab:relative-latency-reduction-waterloo}. The community is actively refining routing algorithms in satellite networks, which will make latency converge toward the shortest paths \cite{pan2025stableroute,izhikevich2024democratizing}. Therefore, choosing $K=10$ is a reasonable basis for simulation.


\subsection{Dual-Homing Latency in SaTor}

The preceding evaluation confirms SaTor's latency advantages, particularly in tail latencies that critically impact user experience. The dual-homing scheme proposed in Section \ref{sec:sator-scheme} suggests equipping relays with satellite access alongside their existing terrestrial connectivity, enabling real-time interface switching for latency optimization. The evaluation begins by envisioning an ideal scenario in which all relays are dual-homed, providing a best-case estimate.


Fig. \ref{fig:simulation-latency-varying-percentile-dual} presents the overall latency reduction achieved by integrating the dual-homing routing scheme in SaTor, at both relay-pair and circuit levels. In this setup, the SaTor software periodically probes latency to 50 other relays at 5-minute intervals. It is found that nearly all links exhibit latency reduction at tail percentiles (Fig. \ref{fig:simulation-latency-reduced-pair-percent-dual} and Fig. \ref{fig:simulation-latency-reduced-cir-percent-dual}). This aligns with expectations, as the relay can keep selecting the faster interface throughout the simulation. Combined with the results in Fig. \ref{fig:simulation-latency-average-reduction-abs-dual} and Fig. \ref{fig:simulation-latency-average-reduction-percent-dual}, at the 99th percentile, over 98\% of pairs show an expected latency reduction of over $\approx$1.2 seconds (82\%), with almost all circuits showing $\approx$2.8 seconds (87.9\%). Using the average across all percentiles as a metric, SaTor scheme achieves an expected latency reduction of $\approx$21.7 ms (36.6\%) at relay-pair level and $\approx$41.4 ms (41.0\%) at circuit level.

\subsection{Incremental Deployment of SaTor}

\label{sec:incremental-sator}

Equipping all relays with satellite service, as assumed in the previous evaluation, is a costly endeavor. This subsection evaluates whether SaTor's dual-homing scheme, deployed only to a small subset of relays, can still deliver meaningful benefits. The evaluation considers three deployment scenarios: \emph{top-n}, \emph{weighted-n}, and \emph{random-n}. The \emph{top-n} scenario assumes that the $n$ relays with the highest likelihood of being selected adopt satellite connectivity, achieving the highest efficacy with minimal investment. The \emph{weighted-n} assumes that relays adopt satellite with a probability proportional to their normalized bandwidth---higher-bandwidth relays have a greater incentive and resources to invest in SaTor. This scenario differs from the \emph{top-n} in that satellite adoption is probabilistic rather than deterministic. In the \emph{random-n} scenario, relays adopt satellite independently at random. 

As shown in Fig. \ref{fig:simulation-latency-reduced-cir-percent-incremental}–\ref{fig:simulation-latency-average-cir-reduction-relative-dual-incremental}, in the \emph{top-n} scenario, equipping just 50 relays with SaTor accelerates over 99.3\% of circuits, achieving an expected latency reduction of $\approx$21.9 ms (22.4\%). This reduction is based on the average latency across all percentiles. Expanding to top 1k relays yields an expected reduction of $\approx$33.9 ms (34.3\%) across all circuits. In the \emph{weighted-n} scenario, $\approx$50 SaTor relays achieve an expected reduction of $\approx$21.2 ms (21.4\%) for 37.3\% of circuits, while 1k relays benefit 82.5\% of circuits with an average of 27.5 ms (27.8\%). The \emph{random-n} shows the least efficacy: with 1k SaTor relays, only 35.2\% of circuits are accelerated with an average of 22.6 ms (22.1\%).

Fig. \ref{fig:simulation-latency-cdf-cir-reduction-dual-incremental} shows the CDF of absolute latency reduction across all circuits under the \emph{weighted-n} scenario---the most reasonable assumption in practice. With 100 SaTor relays, $\approx$24\% of circuits show a latency reduction greater than 10 ms, 3.95\% exceed 50 ms, and 0.07\% exceed 100 ms. When using 600 SaTor relays, 44.5\% of circuits show a reduction greater than 10 ms, 9.1\% exceed 50 ms, and 1.2\% exceed 100 ms. In terms of relative reduction (Fig. \ref{fig:simulation-latency-cdf-cir-reduction-relative-dual-incremental}), deploying 600 SaTor relays yields a $>$10\% latency reduction for 47.2\% of circuits, $>$25\% for 29.9\%, and $>$50\% for 10.1\%.

\textbf{Practical Impact on User Experience.} The evaluation uses RTT latency, which may not directly reflect SaTor's benefits from an end-user perspective. Based on \cite{hogan2022shortor}, a 50 ms RTT reduction roughly corresponds to a 1 second ($\approx$20x) page-load-time (PLT) improvement during web browsing. With 100 SaTor relays in the weighted-n scenario, $\approx$40.5\% of circuits see an average RTT reduction of 21.8 ms, translating to an expected $\approx$400 ms PLT minimization. Also, 3.95\% of circuits see $>$50 ms RTT reduction and 0.7\% exceed 100 ms, implying $>$1 s and $>$2 s PLT improvements, respectively. Notably, these figures are averaged across all percentiles. SaTor's impact at tail latencies, those most affecting user experience, would be more significant.

\subsection{Practicality of SaTor}

SaTor requires the \emph{financial investment} for equipping relays with satellite service, and the \emph{extra bandwidth} required for real-time latency measurements. Currently, satellite networks are widely deployed by several commercial companies, providing services to millions with comparable cost to traditional terrestrial networks. For example, Starlink offers a minimum monthly subscription fee of 80 USD\cite{Starlink2023}, while fiber subscriptions may range from 50 to 100 USD per month\cite{FiberPrice2025}. With Starlink's rapid expansion, satellite costs are expected to decrease further. Regarding additional bandwidth costs, assuming each SaTor relay sends a probe packet via both satellite and terrestrial interfaces to 50 other relays every 5 minutes, the total bandwidth consumption is roughly 3MB per relay per day, a minimal overhead compared to Tor's recommended relay bandwidth \cite{TorRequiement}. 

Currently, over 94\% of relays are located in regions where Starlink service is available. Based on the Tor metrics \cite{TorMetrics} and the IPinfo database \cite{ipinfo2024}, cloud platforms host $\approx$70\% of relays, followed by general ISPs (20\%), business (3\%), and education networks (7\%), as discussed in Appendix \ref{appendix:sator-in-the-real-world}. While deploying SaTor on cloud-hosted relays may require collaboration with cloud providers, major cloud companies are partnering with satellite services, making SaTor deployment hopeful \cite{SpaceCloud2021,HowSatellite2023}. On non-cloud relays (30\% of all relays), direct access to physical machines makes satellite installation simpler. In the weighted-n setting, equipping 50 non-cloud relays accelerates 14.5\% of circuits with an average of 22.8 ms, and removing the non-cloud constraint increases these figures to 37.3\% and 21.2 ms, respectively. 

The adoption of the SaTor may require additional configurations, such as installing satellite dishes (similar to home Wi-Fi). While this may present challenges for resource-constrained relay operators, several factors encourage its practicality. First, $\approx$20\% of top relays are hosted by universities and companies that can afford, or already operate, satellite terminals. Each terminal can serve multiple relays, making pilot SaTor deployment feasible \cite{oneterminalmulti}. Second, Tor allocates most of its budget to program services, indicating the capacity to support infrastructure improvements \cite{torbudget}. Third, SaTor can achieve full functionality without requiring any changes to the Tor application itself, while assistance from non-profit organizations may help relay volunteers overcome deployment and configuration challenges \cite{torserver}.

\subsection{Broader Dual-homing Spectrums}

Dual-homing schemes can be extended to broader spectrums, i.e., dual terrestrial links from different providers. However, dual-homing effectiveness depends on the independence of latency variations between the two links. If both links go slow under similar congestion, benefits are limited. To assess this, we evaluate latency correlations across different types of dual-homed links under two scenarios.

First, we deploy two virtual private servers in London, each hosted by a distinct cloud provider. The two servers simultaneously probe global Tor relays using hping3 \cite{kali-hping3}, recording RTTs for roughly one day per relay. Second, we analyze measurements from the dual-homed Waterloo testbed, which comprises one satellite and one terrestrial link. In this setup, traffic originates from Waterloo, traverses global relays as entries, and terminates in Los Angeles. In these measurements, the geographic location of each destination relay is identified only at the country level. Tail latency correlation between the two links is quantified as the conditional probability that one link experiences tail latency (above the 95th percentile) given that the other also does.

As shown in Table \ref{tab:source-destination-reduction}, dual-terrestrial links from London exhibit strong tail-latency correlation toward Singapore, the UAE, Indonesia, and Argentina, with conditional probabilities $>$15\%, whereas other destinations show moderate correlation. This reflects the routing topology of the terrestrial Internet. When destinations are close, network connectivity is often rich, allowing traffic from different providers to traverse largely disjoint paths. For distant destinations, especially those far from the global Internet core, dual-terrestrial routes often converge at the same exchange points or transit backbones. Congestion at these shared bottlenecks can lead to simultaneous latency spikes across both links, as observed in prior studies \cite{fanou2017investigating,fontugne2020persistent}. Exceptions exist, however, such as on Europe-US routes, where correlation remains low despite long distances, likely due to multiple high-capacity submarine cables and extensive exchange infrastructure.

Dual-homed satellite–terrestrial links (the second row of Table \ref{tab:source-destination-reduction}) exhibit weaker tail-latency correlation than dual-terrestrial links in most cases. Note that the measurements in the Waterloo testbed are circuit-level, where the dual-homed links share a terrestrial hop from the entry relay to the fixed endpoint in the US. Thus, the observed correlation reflects (i) the relationship between the satellite and terrestrial paths on the client–entry segment, and (ii) the inherent correlation of the common terrestrial segment from entry to destination. Even with this overlap, the overall correlation remains low, supporting the view that satellite and terrestrial paths behave largely independently in most cases.

An exception arises when the entry relays are located in the US or Canada, which exhibit a conditional probability $>$40\%. Two factors may account for this. First, in these cases, the circuit length is short, so processing delays within relays and the testbed become more significant. Second, in satellite routing, when endpoint distance is small, the proportion of the path that traverses satellite links from the client to the PoP decreases (Fig. \ref{fig:satellite-communication-tech}). Once traffic reaches the PoP, the remaining route becomes terrestrial, leading to strong latency correlation between the two links. In contrast, for distant destinations, a large fraction of the path traverses inter-satellite links, and traffic rejoins the terrestrial Internet only near the endpoint. This results in orthogonal paths to terrestrial ones, keeping latency correlation low.

Overall, SaTor's satellite–terrestrial dual-homing fits Tor. From a cost perspective, satellite and terrestrial network services are now comparable, and satellite costs are expected to decline further. In terms of performance, the satellite–terrestrial combination offers greater benefits for distant transmissions, aligning with Tor's characteristic of building cross-country circuits to maintain anonymity. While dual-terrestrial links may benefit short hops, these already-fast connections are not major latency bottlenecks, where single-homing is often enough. We further find that satellite acceleration tends to be pronounced for pairs with geographic distances between 5–6k km, detailed in Appendix \ref{Appendix:satellite-routing-across-varying-distance}. Operating on an independent physical layer, satellite routing is a compelling complement to terrestrial links in Tor.

\section{SaTor Security Analysis}

\subsection{Adversary Model}

SaTor considers two adversaries: \emph{relay-level}, referring to malicious relays, and \emph{network-level}, denoting adversarial network service providers. A \emph{relay-level} adversary may actively manipulate traffic to disrupt Tor functionality \cite{sendnermirageflow} or passively analyze metadata, such as packet size and timing, for deanonymization through website fingerprinting techniques \cite{mathews2023sok}. An adversary controlling both the entry and exit relay of a circuit can perform end-to-end correlation attacks to deanonymize users \cite{oh2022deepcoffea}. A \emph{network-level} adversary typically conducts passive surveillance and may attempt AS-level correlation by observing both ends of circuits from different vantage points \cite{sun2017counter}. In restrictive scenarios, state-level actors may drop connections to known relays or block Tor protocol signatures, limiting user access to Tor \cite{master2023worldwide}. Multiple network-level adversaries may also collude, increasing their traffic visibility and attack effectiveness.

An adversary's advantage is measured by the proportion of connections it observes, a prerequisite for most attacks. SaTor's security implications are assessed using \emph{differential advantage}, which measures the additional benefit an adversary gains after introducing satellite routing in Tor.

\subsection{Malicious Relays}

SaTor preserves Tor's default path selection and only opportunistically employs satellite on certain hops within the default circuit. The presence of satellite interfaces does not confer any selection priority. Hence, malicious relays in SaTor appear in circuits with the same probability as in vanilla Tor and gain no additional visibility. While a malicious SaTor relay may attempt to undermine latency benefits by consistently avoiding, or exclusively using, satellite interfaces, such misbehavior is not unique to SaTor. Comparable actions, such as deliberate throttling, can also be carried out by relays in vanilla Tor. Therefore, SaTor is unlikely to increase the capabilities of relay-level adversaries.

\subsection{Adversarial Satellite Service Provider}

\begin{figure}[h]
    \centering
	\subfigure[Relay-pair Observation]{		
		\centering
		\includegraphics[width=0.46\linewidth]{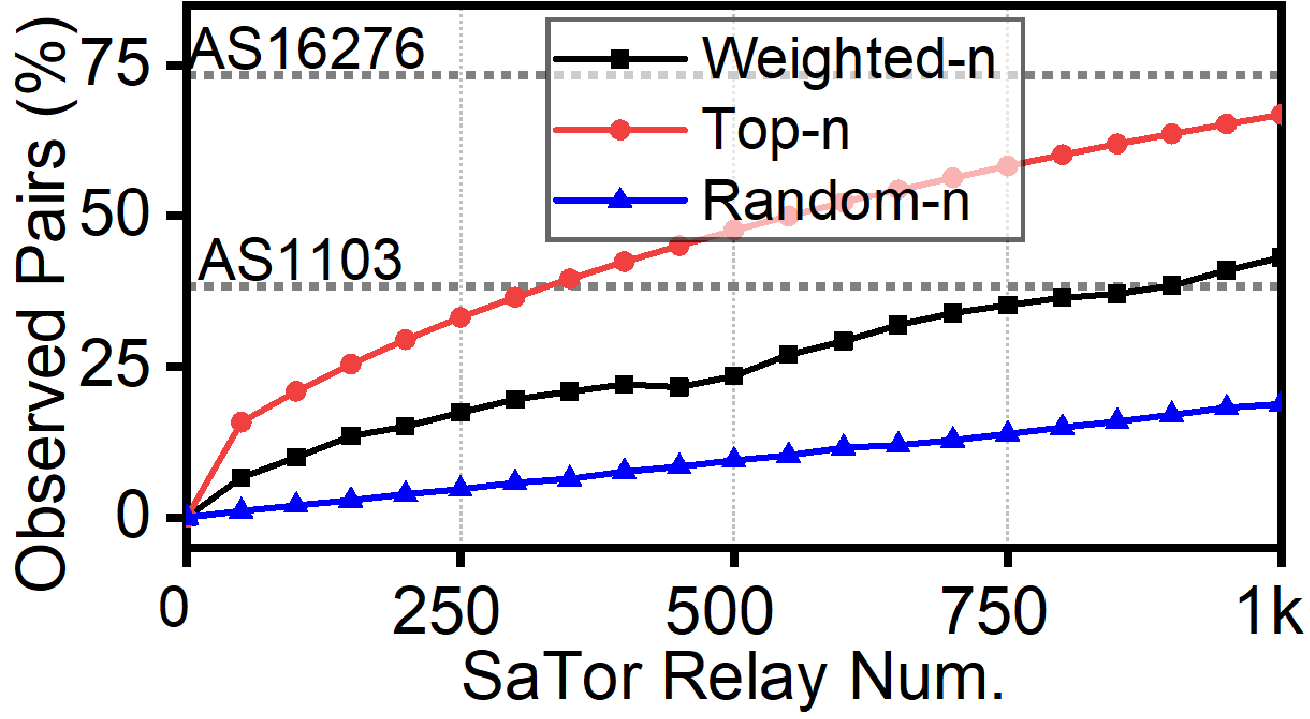}
		\label{fig:advantage-satellite-company-relay-pair}
	}
	\subfigure[Circuit Observation]{
		\centering
		\includegraphics[width=0.46\linewidth]{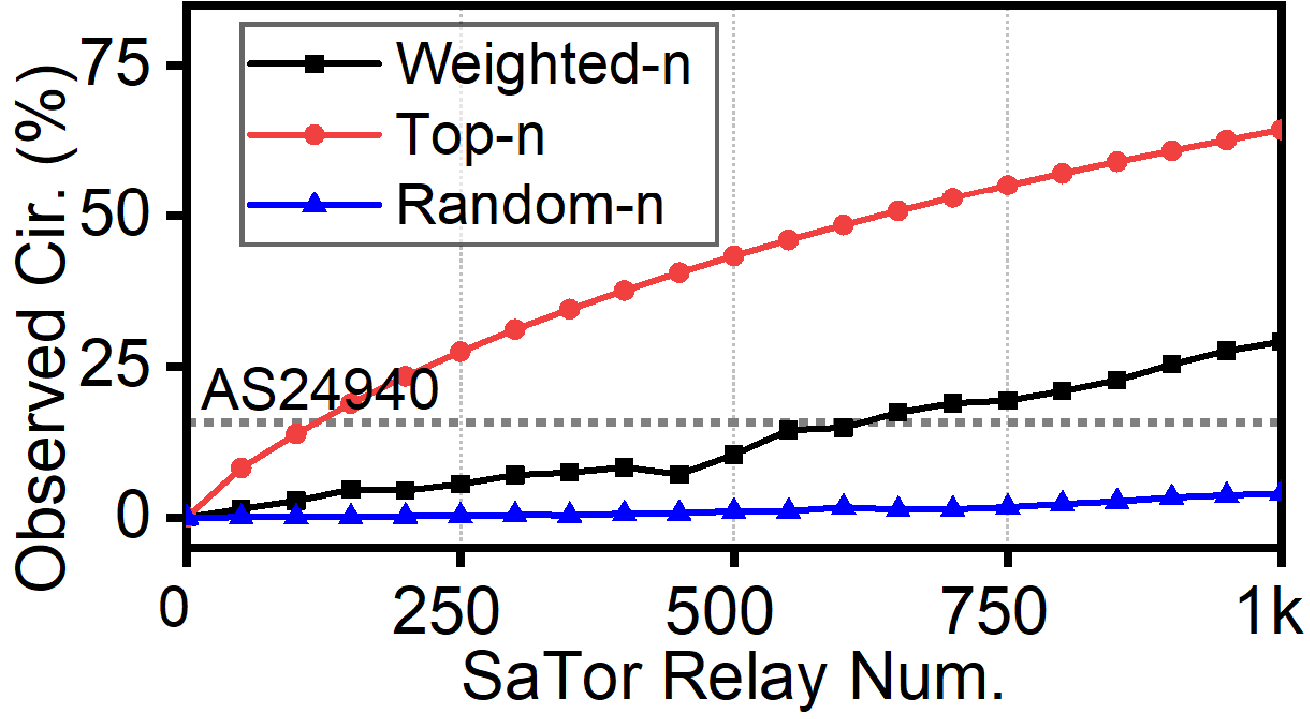}
		\label{fig:advantage-satellite-company-circuit}
	}
	\caption{Observable relay pairs and circuits from satellite providers and from major terrestrial providers. }
	\label{fig:advantage-satellite-company}
\end{figure}

Satellite services are dominated by a few providers, which may act as new network-level adversaries introduced by SaTor. However, in terms of threat model, these satellite providers are fundamentally similar to traditional fiber-based providers. Currently, Tor is unevenly distributed, with many relays hosted by a small number of providers \cite{TorMetrics}. In SaTor, as long as satellite deployment remains moderate, satellite providers do not gain undue advantage. Fig. \ref{fig:advantage-satellite-company-relay-pair} shows the percentage of relay pairs observable by satellite providers under varying numbers of SaTor relays, alongside the visibility of the two most dominant terrestrial providers. To match AS1103's visibility (40\% of pairs), the weighted-n scenario requires 900 satellite relays, while the top-n scenario requires 350. To reach AS16276's visibility (70\%), both scenarios require over 1k satellite relays.

Circuit observation is defined as the case where the entire entry-to-exit path is visible to the adversary. A terrestrial network operator achieves this simply by hosting a middle relay and observing its ingress and egress traffic. In contrast, a satellite provider must host both the entry and middle, since SaTor uses satellite only for outbound connections. Exits are excluded as Tor disallows one operator from hosting both entry and exit, and SaTor prevents satellite usage at exits. As shown in Fig. \ref{fig:advantage-satellite-company-circuit}, matching AS24940's visibility (15.68\% of circuits) requires $>$600 satellite relays in the weighted-n scenario and $\approx$150 in the top-n.

SaTor could achieve reasonable benefits with $\approx$100 dual-homed relays, without posing significant advantages to satellite providers. Moreover, SaTor can distribute dual-homed relays across different providers, such as Starlink\cite{Starlink2023}, OneWeb\cite{OneWeb2023}, and Amazon Kuiper\cite{AmazonKuiper2025}, thereby avoiding excessive visibility by any single entity. Moreover, recall that SaTor only modifies the data forwarding process between entry and exit relays, leaving the connections between client and entry, as well as between exit and destination server, unchanged. Therefore, SaTor does not increase the risk of end-to-end deanonymization attacks, which require a global adversary observing both ends of a circuit. To this end, \emph{collusion} with satellite providers, whether by other network operators or malicious relays, is similar to collusion with fiber-based providers. Introducing satellite routing into Tor simply implies a small subset of relays migrating to a few new providers---whether satellite or terrestrial-based---without altering Tor's overall trust model.


\subsection{Fingerprinting Attacks in SaTor}

Attackers can infer the websites users visit by analyzing traffic at entry relays, exploiting the fact that traffic metadata varies across websites. Since satellite transmissions may introduce distinguishable traffic metadata compared to terrestrial links, it is important to reassess SaTor's vulnerability to website fingerprinting. Encouragingly, a recent study \cite{singh2024connecting} finds that Tor over Starlink is comparably susceptible to website fingerprinting as Tor over fiber. By collecting traffic datasets from both terrestrial and Starlink connections to the Tor network, the study observed that state-of-the-art fingerprinting attacks show a $\approx$2–8\% drop in accuracy over satellite links. Furthermore, fingerprinting defenses are similarly, or even more effective on Starlink, reducing attack accuracy by up to 12\% more than on fiber.


Satellite traffic may be fingerprinted for \emph{relay discovery attacks}. For example, an exit relay could infer the use of satellite transmission between the guard and middle by identifying satellite-specific metadata, such as distinctive packet sizes or timing patterns compared to terrestrial routing. If the exit determines that the guard-middle link is on satellite, it may narrow down the guard's location, since satellite acceleration is often tied to geographical factors, implying that the guard is likely situated far from the middle relay. One possible mitigation is to broaden satellite adoption across more links, allowing relays to use satellite even when it is up to $t$\% slower than terrestrial routing, thereby trading a small performance cost for improved anonymity, which could be a future direction to explore.


\section{Related Work}

Tor's latency reduction efforts can be categorized as \emph{biased} or \emph{non-biased}. The \emph{biased} approaches prioritize links with lower latency, based on either ad-hoc measuring prior to link selection or historical latency records. Instead of Tor's default relay-by-relay selection, these approaches often perform circuit-level selection \cite{annessi2016navigator, barton2018towards, imani2019modified}, evaluating the overall performance of candidate circuits and assigning higher selection weights to those with lower latency. Such active manipulation on circuit selection often yields notable latency improvements, e.g., PredicTor\cite{barton2018towards} accelerates 10\% of circuits by $>$1.5 s in PLT, while SaTor achieves only a $>$600 ms reduction for 10\% using 100 weighted-n relays.

However, the biased approaches impose additional predictability to users' data paths, offering advantages to adversaries. Algorithms favoring geographically short circuits might exacerbate the node-placement attack, wherein adversaries position malicious nodes close to users to increase their chances of being selected \cite{wan2019guard, tan2022anonymity}. Passive attackers, who deanonymize users by monitoring the latency or speed of a circuit, now only need to observe fewer circuits to make accurate inferences \cite{nasr2018deepcorr, rochet2018dropping}. For Tor, which relies on a large and diverse user base to maintain anonymity, reduction in the randomness of user behavior could heighten deanonymization risk \cite{backes2014mators, backes2016your, Joshua2015defending}.

The \emph{non-biased} approaches retain Tor's default path selection algorithm, while modifying the subsequent data routing process. Multi-path routing approaches split traffic across multiple semi-disjoint circuits to avoid single-path congestion, based on the fact that due to the heterogeneity of relay bandwidths, some relays are heavily used while others are less traveled \cite{alsabah2013path,yang2015mtor}. However, simultaneously using multiple circuits increases vulnerability to end-to-end correlation attacks by broadening the exposure to network-level adversaries and malicious relays, which may deanonymize users by observing just one sub-circuit. In contrast, SaTor does not introduce additional relays or alter the end segments of circuits, thereby avoiding extra correlation risks.

Another non-biased approach, \emph{ShorTor}, adds extra \emph{via relays} into standard circuits, to make the newly built circuit faster \cite{hogan2022shortor}. Although ShorTor reduces Tor latency without significant security compromises, its effectiveness is largely concentrated in a small fraction of circuits. Concretely, it reduces latency by $>$100 ms for 1.1\% of circuits, $>$25 ms for 8\%, and  $>$10 ms for 11\%, while the remaining 90\% show no significant reduction. In comparison, SaTor under the weighted-100 scenario improves latency by $>$100 ms for 0.7\% of circuits, $>$25 ms for 14\%, and $>$10 ms for 24\%, with $\approx$40\% of circuits experiencing improvements. Moreover, ShorTor may impose a burden on Tor by requiring some relays to manage traffic that would not ordinarily be routed through them. It consumes $\approx$100 MB of additional bandwidth per relay per day, while SaTor requires 3 MB.

\section{Conclusion}

High latency remains an essential challenge for Tor. Traditional latency reducing strategies often bias shorter circuits, compromising traffic path randomness and degrading security. This paper explores satellite routing in Tor, outfitting relays with satellite connections, coupled with a dual-homing mechanism to adaptively arrange the usage of satellite and terrestrial interfaces for optimal latency. Our evaluation shows that over a long-term observation, SaTor in general accelerates over 40\% circuits by an average of 21.8 ms, with the top 100 most popular relays accessed to satellite network. This study offers a first perspective on utilizing satellite technology to address Tor's challenges, serving as a reference for Tor's future improvements.

\section*{Acknowledgement}

We thank the anonymous reviewers and shepherd for their insightful feedback. We are grateful to Diogo Barradas at the University of Waterloo for his valuable suggestions and for coordinating the satellite testbed. We further thank Professor Tian Song at the Beijing Institute of Technology, Andrew Ferguson and Mohamad Kassem at the University of Edinburgh for their assistance in improving this work.

\bibliographystyle{ieeetr}
\bibliography{citations}

\begin{thebibliography}{10}

\bibitem{TorMetrics}
{The Tor Project}, ``{Tor Metrics}.'' \url{https://metrics.torproject.org/}, 2025.
\newblock Online.

\bibitem{dingledine2004tor}
R.~Dingledine, N.~Mathewson, P.~F. Syverson, {\em et~al.}, ``Tor: The second-generation onion router,'' in {\em USENIX security symposium}, vol.~4, pp.~303--320, 2004.

\bibitem{akhoondi2012lastor}
M.~Akhoondi, C.~Yu, and H.~V. Madhyastha, ``{LASTor}: A low-latency {AS-aware} {Tor} client,'' in {\em 2012 IEEE Symposium on Security and Privacy}, pp.~476--490, IEEE, 2012.

\bibitem{wang2012congestion}
T.~Wang, K.~Bauer, C.~Forero, and I.~Goldberg, ``Congestion-aware path selection for {Tor},'' in {\em Financial Cryptography and Data Security: 16th International Conference, FC 2012, Kralendijk, Bonaire, Februray 27-March 2, 2012, Revised Selected Papers 16}, pp.~98--113, Springer, 2012.

\bibitem{alsabah2013path}
M.~AlSabah, K.~Bauer, T.~Elahi, and I.~Goldberg, ``The path less travelled: Overcoming {Tor’s} bottlenecks with traffic splitting,'' in {\em Privacy Enhancing Technologies: 13th International Symposium, PETS 2013, Bloomington, IN, USA, July 10-12, 2013. Proceedings 13}, pp.~143--163, Springer, 2013.

\bibitem{annessi2016navigator}
R.~Annessi and M.~Schmiedecker, ``{NavigaTor}: Finding faster paths to anonymity,'' in {\em 2016 IEEE European Symposium on Security and Privacy (EuroS\&P)}, pp.~214--226, IEEE, 2016.

\bibitem{barton2018towards}
A.~Barton, M.~Wright, J.~Ming, and M.~Imani, ``Towards predicting efficient and anonymous {Tor} circuits,'' in {\em 27th USENIX Security Symposium (USENIX Security 18)}, pp.~429--444, 2018.

\bibitem{hogan2022shortor}
K.~Hogan, S.~Servan-Schreiber, Z.~Newman, B.~Weintraub, C.~Nita-Rotaru, and S.~Devadas, ``{ShorTor}: Improving {Tor} network latency via multi-hop overlay routing,'' in {\em 2022 IEEE Symposium on Security and Privacy (SP)}, pp.~1933--1952, IEEE, 2022.

\bibitem{arapakis2014impact}
I.~Arapakis, X.~Bai, and B.~B. Cambazoglu, ``Impact of response latency on user behavior in {Web} search,'' in {\em Proceedings of the international ACM SIGIR conference on Research \& development in information retrieval}, pp.~103--112, 2014.

\bibitem{dhungel2010waiting}
P.~Dhungel, M.~Steiner, I.~Rimac, V.~Hilt, and K.~W. Ross, ``Waiting for anonymity: Understanding delays in the {Tor} overlay,'' in {\em 2010 IEEE Tenth International Conference on Peer-to-Peer Computing (P2P)}, pp.~1--4, IEEE, 2010.

\bibitem{wacek2013empirical}
C.~Wacek, H.~Tan, K.~S. Bauer, and M.~Sherr, ``An empirical evaluation of relay selection in {Tor},'' in {\em Proceedings of the Network and Distributed System Security Symposium}, 2013.

\bibitem{imani2019modified}
M.~Imani, M.~Amirabadi, and M.~Wright, ``Modified relay selection and circuit selection for faster {Tor},'' {\em IET Communications}, vol.~13, no.~17, pp.~2723--2734, 2019.

\bibitem{mittal2011stealthy}
P.~Mittal, A.~Khurshid, J.~Juen, M.~Caesar, and N.~Borisov, ``Stealthy traffic analysis of low-latency anonymous communication using throughput fingerprinting,'' in {\em Proceedings of the 18th ACM conference on Computer and Communications Security}, pp.~215--226, 2011.

\bibitem{wan2019guard}
G.~Wan, A.~Johnson, R.~Wails, S.~Wagh, and P.~Mittal, ``Guard placement attacks on path selection algorithms for {Tor},'' {\em Proceedings on Privacy Enhancing Technologies}, vol.~2019, no.~4, 2019.

\bibitem{karunanayake2021anonymisation}
I.~Karunanayake, N.~Ahmed, R.~Malaney, R.~Islam, and S.~K. Jha, ``De-anonymisation attacks on {Tor}: A survey,'' {\em IEEE Communications Surveys \& Tutorials}, vol.~23, no.~4, pp.~2324--2350, 2021.

\bibitem{tan2022anonymity}
Q.~Tan, X.~Wang, W.~Shi, J.~Tang, and Z.~Tian, ``An anonymity vulnerability in {Tor},'' {\em IEEE/ACM Transactions on Networking}, vol.~30, no.~6, pp.~2574--2587, 2022.

\bibitem{SMF-28TM2002}
C.~Incorporated, ``Smf-28tm optical fiber product information,'' 2002.

\bibitem{chaudhry2022optical}
A.~U. Chaudhry and H.~Yanikomeroglu, ``Optical wireless satellite networks versus optical fiber terrestrial networks: The latency perspective: Invited chapter,'' in {\em 30th Biennial Symposium on Communications 2021}, pp.~225--234, Springer, 2022.

\bibitem{handley2018delay}
M.~Handley, ``Delay is not an option: Low latency routing in space,'' in {\em Proceedings of the 17th ACM Workshop on Hot Topics in Networks}, pp.~85--91, 2018.

\bibitem{handley2019using}
M.~Handley, ``Using ground relays for low-latency wide-area routing in megaconstellations,'' in {\em Proceedings of the 18th ACM Workshop on Hot Topics in Networks}, pp.~125--132, 2019.

\bibitem{bozkurt2017internet}
I.~N. Bozkurt, A.~Aguirre, B.~Chandrasekaran, P.~B. Godfrey, G.~Laughlin, B.~Maggs, and A.~Singla, ``Why is the {Internet} so slow?!,'' in {\em Passive and Active Measurement: 18th International Conference, PAM 2017, Sydney, NSW, Australia, March 30-31, 2017, Proceedings 18}, pp.~173--187, Springer, 2017.

\bibitem{hoiland2016measuring}
T.~H{\o}iland-J{\o}rgensen, B.~Ahlgren, P.~Hurtig, and A.~Brunstrom, ``Measuring latency variation in the {Internet},'' in {\em Proceedings of the 12th International on Conference on emerging Networking EXperiments and Technologies}, pp.~473--480, 2016.

\bibitem{chavula2017insight}
J.~Chavula, A.~Phokeer, A.~Formoso, and N.~Feamster, ``Insight into {Africa's} country-level latencies,'' in {\em IEEE AFRICON 2017}, pp.~938--944, Institute of Electrical and Electronics Engineers Inc., 2017.

\bibitem{zhao2024lens}
J.~Zhao and J.~Pan, ``{LENS}: A {LEO} satellite network measurement dataset,'' in {\em Proceedings of the 15th ACM Multimedia Systems Conference}, pp.~278--284, 2024.

\bibitem{RIPE2024}
RIPE, ``{RIPE Atlas}.'' \url{https://atlas.ripe.net/}, 2024.

\bibitem{lai2020starperf}
Z.~Lai, H.~Li, and J.~Li, ``{StarPerf}: Characterizing network performance for emerging mega-constellations,'' in {\em 2020 IEEE 28th International Conference on Network Protocols (ICNP)}, pp.~1--11, IEEE, 2020.

\bibitem{lai2023starrynet}
Z.~Lai, H.~Li, Y.~Deng, Q.~Wu, J.~Liu, Y.~Li, J.~Li, L.~Liu, W.~Liu, and J.~Wu, ``{StarryNet}: Empowering researchers to evaluate futuristic integrated space and terrestrial networks,'' in {\em 20th USENIX Symposium on Networked Systems Design and Implementation (NSDI 23)}, pp.~1309--1324, 2023.

\bibitem{kassing2020exploring}
S.~Kassing, D.~Bhattacherjee, A.~B. {\'A}guas, J.~E. Saethre, and A.~Singla, ``Exploring the {Internet} from space with {Hypatia},'' in {\em Proceedings of the ACM Internet Measurement conference}, pp.~214--229, 2020.

\bibitem{Alicloud2024}
A.~Cloud, ``Performance monitoring metrics.'' \url{https://www.alibabacloud.com/help/en/well-architected/latest/performance-monitoring-indicators/}, 2025.
\newblock Online.

\bibitem{StarlinkServicePlans}
Starlink, ``Starlink service plans.'' \url{https://www.starlink.com/service-plans/}, 2025.
\newblock Online.

\bibitem{SpaceCloud2021}
D.~Mohney, ``The space cloud: Satellite strategies for {AWS, Google and Microsoft}.'' \url{https://www.datacenterfrontier.com/cloud/article/11428161/the-space-cloud-satellite-strategies-for-aws-google-and-microsoft}, 2021.

\bibitem{HowSatellite2023}
A.~Raj, ``How satellites are becoming a game changer for cloud service providers.'' \url{https://techwireasia.com/03/2023/heres-how-satellites-are-enhancing-connectivity-for-cloud-service-providers/}, 2023.

\bibitem{reed1998anonymous}
M.~G. Reed, P.~F. Syverson, and D.~M. Goldschlag, ``Anonymous connections and onion routing,'' {\em IEEE Journal on Selected areas in Communications}, vol.~16, no.~4, pp.~482--494, 1998.

\bibitem{alsabah2016performance}
M.~AlSabah and I.~Goldberg, ``Performance and security improvements for {Tor}: A survey,'' {\em ACM Computing Surveys (CSUR)}, vol.~49, no.~2, pp.~1--36, 2016.

\bibitem{jansen2012throttling}
R.~Jansen, P.~Syverson, and N.~Hopper, ``Throttling {Tor} bandwidth parasites,'' in {\em 21st USENIX Security Symposium (USENIX Security 12)}, pp.~349--363, 2012.

\bibitem{torproject2023specifications}
R.~Dingledine and N.~Mathewson, ``Tor specifications,'' 2023.

\bibitem{backes2014mators}
M.~Backes, A.~Kate, S.~Meiser, and E.~Mohammadi, ``{(Nothing else) {MAT}or(s): Monitoring the Anonymity of Tor's Path Selection},'' in {\em Proceedings of the 21th ACM conference on Computer and Communications Security (CCS 2014)}, November 2014.

\bibitem{backes2016your}
M.~Backes, S.~Meiser, and M.~Slowik, ``{Your Choice MATor(s)}: Large-scale quantitative anonymity assessment of {Tor} path selection algorithms against structural attacks,'' {\em Proceedings on Privacy Enhancing Technologies}, vol.~2016, April 2016.

\bibitem{Joshua2015defending}
J.~Juen, A.~Johnson, A.~Das, N.~Borisov, and M.~Caesar, ``Defending {Tor} from network adversaries: A case study of network path prediction,'' {\em Proceedings on Privacy Enhancing Technologies}, vol.~2015, no.~2, pp.~1--17, 2015.

\bibitem{SpaceX2018update}
{SpaceX FCC update}, ``{SPACEX NON-GEOSTATIONARY SATELLITE SYSTEM}.'' \url{https://licensing.fcc.gov/myibfs/download.do?attachment_key=1569860}, 2018.
\newblock Online.

\bibitem{mcdowell2019jonathans}
J.~McDowell, ``{Jonathan's Space Report}.'' \url{https://planet4589.org/}, 2025.
\newblock Online.

\bibitem{Starlink2023}
SpaceX, ``Starlink.'' \url{https://www.starlink.com/}, 2025.
\newblock Online.

\bibitem{OneWeb2023}
OneWeb, ``Oneweb.'' \url{https://oneweb.net/}, 2025.
\newblock Online.

\bibitem{chaudhry2022crossover}
A.~U. Chaudhry and H.~Yanikomeroglu, ``On crossover distance for optical wireless satellite networks and optical fiber terrestrial networks,'' in {\em 2022 IEEE Future Networks World Forum (FNWF)}, pp.~480--485, IEEE, 2022.

\bibitem{kassem2022browser}
M.~M. Kassem, A.~Raman, D.~Perino, and N.~Sastry, ``A browser-side view of {Starlink} connectivity,'' in {\em Proceedings of the 22nd ACM Internet Measurement Conference}, pp.~151--158, 2022.

\bibitem{ma2023network}
S.~Ma, Y.~C. Chou, H.~Zhao, L.~Chen, X.~Ma, and J.~Liu, ``Network characteristics of {LEO} satellite constellations: A {Starlink-based} measurement from end users,'' in {\em IEEE INFOCOM 2023-IEEE Conference on Computer Communications}, pp.~1--10, IEEE, 2023.

\bibitem{mohan2023multifaceted}
N.~Mohan, A.~Ferguson, H.~Cech, P.~R. Renatin, R.~Bose, M.~Marina, and J.~Ott, ``A multifaceted look at {Starlink} performance,'' in {\em Proceedings of the ACM Web Conference 2024 (WWW ’24)}, 2024.

\bibitem{NORAD}
T.~Kelso, ``{CelesTrak}: Current {NORAD} two-line element sets.'' \url{https: //www.celestrak.com/NORAD/elements/}, 2024.

\bibitem{StarlinkInsider2024}
S.~Insider, ``Starlink ground station locations: An overview.'' \url{https://starlinkinsider.com/starlink-gateway-locations}, 2024.
\newblock Online.

\bibitem{SpaceTrack2024}
{Space-Track Team}, ``Space-track.'' \url{https://www.space-track.org/}, 2024.
\newblock Online.

\bibitem{maxmind2024}
MaxMind, ``{GeoIP} accuracy comparison.'' \url{https://www.maxmind.com/en/geoip-accuracy-comparison}, 2024.

\bibitem{Stem}
M.~Maltsev, ``Stem.'' \url{https://github.com/torproject/stem}, 2023.

\bibitem{usersrouted-ccs13}
A.~Johnson, C.~Wacek, R.~Jansen, M.~Sherr, and P.~Syverson, ``Users get routed: Traffic correlation on {Tor} by realistic adversaries,'' in {\em Proceedings of the 20th ACM Conference on Computer and Communications Security}, 2013.

\bibitem{yen1971finding}
J.~Y. Yen, ``Finding the k shortest loopless paths in a network,'' {\em management Science}, vol.~17, no.~11, pp.~712--716, 1971.

\bibitem{kali-hping3}
K.~Linux, ``hping3.'' \url{https://www.kali.org/tools/hping3/}, 2024.

\bibitem{statistic-latency}
Medium, ``Statistics behind latency metrics: Understanding p90, p95, and p99.'' \url{https://medium.com/tuanhdotnet/statistics-behind-latency-metrics-understanding-p90-p95-and-p99-dc87420d505d/}, 2024.
\newblock Online.

\bibitem{cangialosi2015ting}
F.~Cangialosi, D.~Levin, and N.~Spring, ``Ting: Measuring and exploiting latencies between all tor nodes,'' in {\em Proceedings of the 2015 Internet Measurement Conference}, pp.~289--302, 2015.

\bibitem{izhikevich2024democratizing}
L.~Izhikevich, M.~Tran, K.~Izhikevich, G.~Akiwate, and Z.~Durumeric, ``Democratizing {LEO} satellite network measurement,'' {\em Proceedings of the ACM on Measurement and Analysis of Computing Systems}, vol.~8, no.~1, pp.~1--26, 2024.

\bibitem{pan2025stableroute}
T.~Pan, G.~Ruan, Q.~Fu, Z.~Luo, J.~Huang, X.~Luo, and T.~Huang, ``Stableroute: When dijkstra's algorithm meets topology-varying satellite networks,'' in {\em IEEE INFOCOM 2025-IEEE Conference on Computer Communications}, pp.~1--10, IEEE, 2025.

\bibitem{FiberPrice2025}
CableTV, ``Best fiber internet providers: Prices, speeds, and more.'' \url{https://www.cabletv.com/internet/best-fiber-internet-providers/}, 2025.
\newblock Online.

\bibitem{TorRequiement}
{The Tor Project}, ``Relay requirements.'' \url{https://community.torproject.org/relay/relays-requirements/}, 2024.
\newblock Online.

\bibitem{ipinfo2024}
{IPinfo}, ``{IPinfo}: Trusted {IP} data provider.'' \url{https://ipinfo.io/}, 2024.

\bibitem{oneterminalmulti}
R.~Gupta, ``Can starlink support multiple users simultaneously?.'' \url{https://spacetek.com.au/blogs/news/can-starlink-support-multiple-users-simultaneously}, 2024.
\newblock Online.

\bibitem{torbudget}
Alsmith, ``Transparency, openness, and our 2021-2022 financials.'' \url{https://blog.torproject.org/transparency-openness-and-our-2021-and-2022-financials/}, 2023.
\newblock Online.

\bibitem{torserver}
S.~Leibfarth, ``torservers.net.'' \url{https://torservers.net/}, 2025.
\newblock Online.

\bibitem{fanou2017investigating}
R.~Fanou, F.~Valera, and A.~Dhamdhere, ``Investigating the causes of congestion on the {African IXP} substrate,'' in {\em Proceedings of the 2017 Internet Measurement Conference}, pp.~57--63, 2017.

\bibitem{fontugne2020persistent}
R.~Fontugne, A.~Shah, and K.~Cho, ``Persistent last-mile congestion: Not so uncommon,'' in {\em Proceedings of the ACM internet measurement conference}, pp.~420--427, 2020.

\bibitem{sendnermirageflow}
C.~Sendner, J.~Stang, A.~Dmitrienko, R.~Wijewickrama, and M.~Jadliwala, ``{MirageFlow}: A new bandwidth inflation attack on {Tor},'' in {\em Network and Distributed System Security (NDSS) Symposium}, 2024.

\bibitem{mathews2023sok}
N.~Mathews, J.~K. Holland, S.~E. Oh, M.~S. Rahman, N.~Hopper, and M.~Wright, ``Sok: A critical evaluation of efficient website fingerprinting defenses,'' in {\em 2023 IEEE Symposium on Security and Privacy (SP)}, pp.~969--986, IEEE, 2023.

\bibitem{oh2022deepcoffea}
S.~E. Oh, T.~Yang, N.~Mathews, J.~K. Holland, M.~S. Rahman, N.~Hopper, and M.~Wright, ``{DeepCoFFEA}: Improved flow correlation attacks on {Tor} via metric learning and amplification,'' in {\em 2022 IEEE Symposium on Security and Privacy (SP)}, pp.~1915--1932, IEEE, 2022.

\bibitem{sun2017counter}
Y.~Sun, A.~Edmundson, N.~Feamster, M.~Chiang, and P.~Mittal, ``{Counter-RAPTOR}: Safeguarding {Tor} against active routing attacks,'' in {\em 2017 IEEE Symposium on Security and Privacy (SP)}, pp.~977--992, IEEE, 2017.

\bibitem{master2023worldwide}
A.~Master and C.~Garman, ``A worldwide view of nation-state {Internet} censorship,'' {\em Free and Open Communications on the Internet}, 2023.

\bibitem{AmazonKuiper2025}
Amazon, ``Project {Kuiper}.'' \url{https://www.aboutamazon.com/what-we-do/devices-services/project-kuiper/}, 2025.
\newblock Online.

\bibitem{singh2024connecting}
P.~Singh, D.~Barradas, T.~Elahi, and N.~Limam, ``Connecting the dots in the sky: Website fingerprinting in low earth orbit satellite {Internet},'' Network and Distributed System Security Symposium (NDSS), 2024.

\bibitem{nasr2018deepcorr}
M.~Nasr, A.~Bahramali, and A.~Houmansadr, ``{DeepCorr}: Strong flow correlation attacks on {Tor} using deep learning,'' in {\em Proceedings of the 2018 ACM SIGSAC Conference on Computer and Communications Security}, pp.~1962--1976, 2018.

\bibitem{rochet2018dropping}
F.~Rochet and O.~Pereira, ``Dropping on the edge: Flexibility and traffic confirmation in onion routing protocols.,'' {\em Proceedings on Privacy Enhancing Technology}, vol.~2018, no.~2, pp.~27--46, 2018.

\bibitem{yang2015mtor}
L.~Yang and F.~Li, ``{mTor}: A multipath {Tor} routing beyond bandwidth throttling,'' in {\em 2015 IEEE Conference on Communications and Network Security (CNS)}, pp.~479--487, IEEE, 2015.

\end{thebibliography}

\appendices

\section{Baseline Latency Measurement}
\label{appendix:baselin-latency-measurement}
\textbf{Baseline Latency Dataset.} Table \ref{tab:baseline-satellite-latency} shows the satellite latency data collected from LENS dataset. The measurements were conducted across five geographical locations: Alaska, Frankfurt, Seattle, Vancouver, and Victoria. Seattle has two dishes subscribing to Starlink's regular and high-priority service plan. We extract a total 1.624 billion data points of ICMP latencies from the LENS dataset, spanning a two-month period from April to May, 2024. As of terrestrial latency, as listed in Table \ref{tab:baseline-terrestrial-latency},  we extract over 104.64 million data points from RIPE dataset, covering a one-week period starting on Sep. 2, 2024. The data is divided based on the round-trip length, by a 2k kilometers interval. 

\begin{table}[ht]
  \centering
  \caption{Baseline Satellite Latency from LENS Dataset}
  \label{tab:baseline-satellite-latency}
  \setlength{\tabcolsep}{3pt}
  \begin{tabular}{lcccc}
    \toprule
    \textbf{Dish Loc.} & \textbf{PoP Loc.} & \textbf{Route Len. (km)} & \textbf{Duration} & \textbf{Data Num.} \\
    \midrule
    Alaska      & Seattle    & 5,119 & 26 days & 134.81M \\
    Frankfurt   & Frankfurt  & 2,200 & 40 days & 227.58M \\
    Seattle     & Seattle    & 2,200 & 46 days & 264.89M \\
    Seattle\_hp & Seattle    & 2,200 & 54 days & 327.21M \\
    Vancouver   & Seattle    & 2,234 & 59 days & 333.94M \\
    Victoria    & Seattle    & 2,212 & 61 days & 339.97M \\
    \bottomrule
  \end{tabular}
\end{table}

\begin{table}[ht]
  \centering
  \caption{Baseline Terrestrial Latency from RIPE Dataset}
  \label{tab:baseline-terrestrial-latency}
  \setlength{\tabcolsep}{7pt}
  \begin{tabular}{lcc}
    \toprule
    \textbf{Traffic Route Length (km)} & \textbf{Pairs Number} & \textbf{Data Number} \\
    \midrule
    0--2k   & 213,752 & 56.98M \\
    2k--4k  & 55,388  & 13.23M \\
    4k--6k  & 49,656  & 12.15M \\
    6k--8k  & 96,827  & 22.28M \\
    \rowcolor{gray!15} Total & 415,623 & 104.64M \\
    \bottomrule
  \end{tabular}
\end{table}

\begin{figure}[h]
    \centering
	\subfigure[LENS (Starlink)]{		
		\centering
		\includegraphics[width=0.46\linewidth]{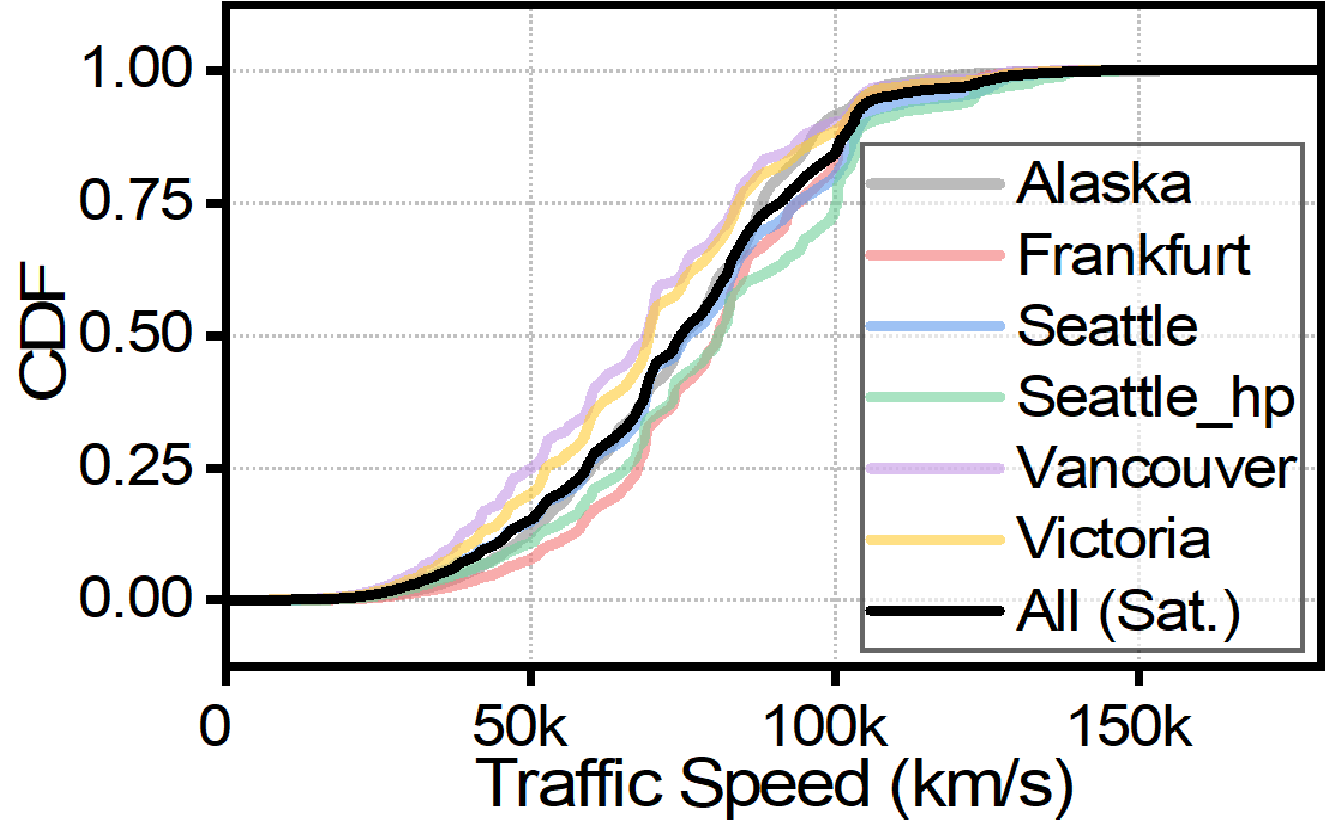}
		\label{fig:baseline-speed-starlink}
	}
	\subfigure[RIPE (Terrestrial)]{
		\centering
		\includegraphics[width=0.46\linewidth]{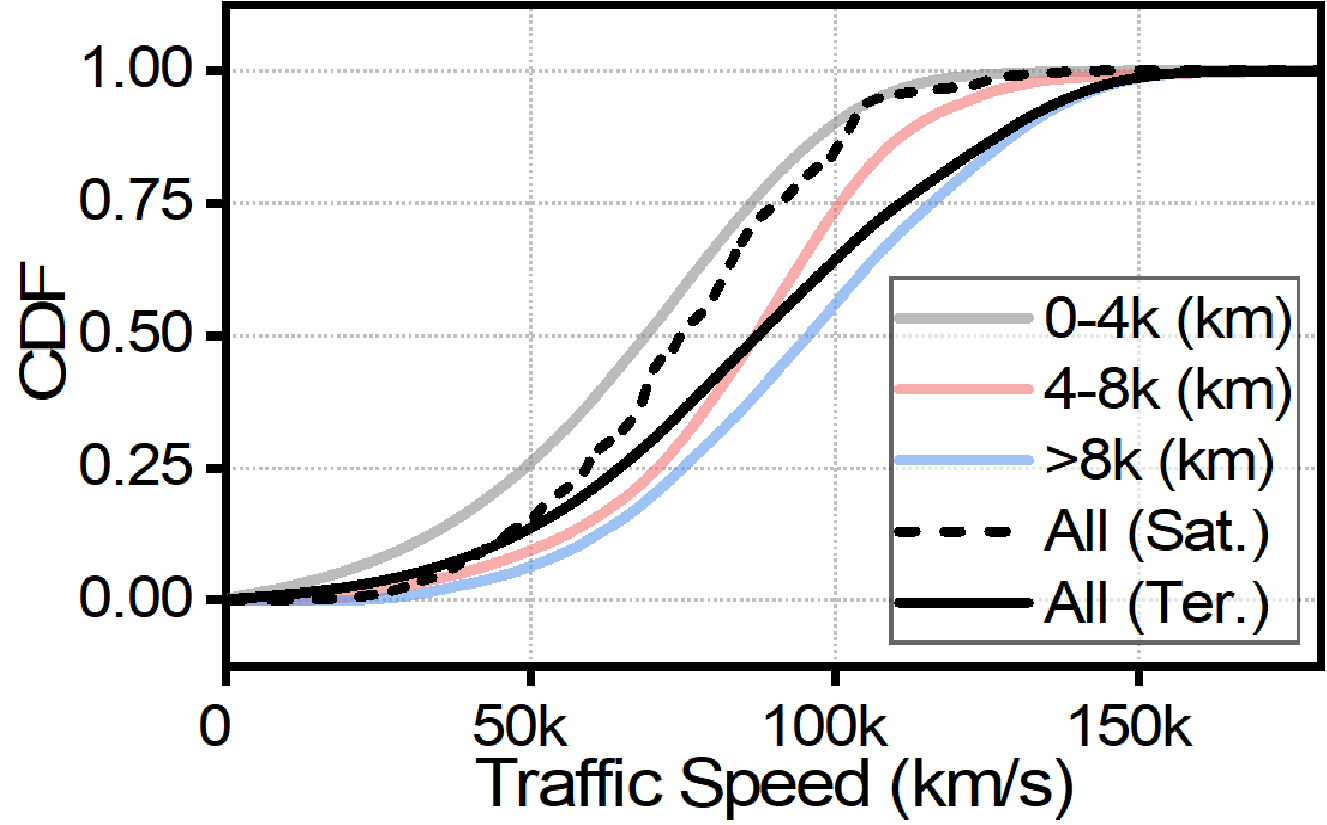}
		\label{fig:baseline-speed-terrestrial}
	}
	\caption{Traffic speed CDF of LENS and RIPE Dataset. }
	\label{fig:baseline-speed}
\end{figure}

\textbf{Baseline Traffic Speed.} The CDFs for the speed of satellite and terrestrial traffic are derived from LENS and RIPE datasets. The latencies are measured using ICMP transmissions, capturing all practical latency components. They also reflect the influence of factors such as satellite movement, weather conditions, and network reconfiguration \cite{mohan2023multifaceted}. Continuously sampling from this CDF over an extended duration can replicate the long-term latency characteristics of satellite and terrestrial routing at a global scale.

Fig. \ref{fig:baseline-speed-starlink} presents the satellite traffic speed across five locations. Seattle includes two dishes with regular and high-priority subscriptions, with the latter expected to show better latency performance. The round-trip length of satellite traffic is $\approx$ 2,200 km, except for Alaska, which is $\approx$5,119 km. Vancouver (light purple curve) exhibits the slowest traffic speed, while Frankfurt (light red) and Seattle's priority dish (light green) demonstrate the best speed performance. Traffic speed differences across global locations may result from the varying density of the satellite constellation, ground stations, and dish performance. The combinatorial CDF across all locations (black curve) shows that satellite traffic transmits slower than 50k km/s during $\approx$ 15\% of the time, and slower than 100k km/s during $\approx$ 85\% of the time.

Terrestrial traffic speeds are measured across $\approx$12,000 global probes on the RIPE Atlas platform. The distances between probe pairs range from 0.06 to 10k km (round-trip route lengths of 0.12 to 20k km). These figures are straight-line end-to-end distances, disregarding the routing detours through intermediate devices. Fig. \ref{fig:baseline-speed-terrestrial} indicates that shorter round-trip corresponds to slower traffic. When round-trip lengths are between 0-4k km (the transparent black curve) the speed CDF is above all other curves, meaning that a larger portion of the data is concentrated within the lower speed range. In contrast, transmissions $>$8k km (transparent blue curve) exhibit the fastest traffic speeds among all groups, probably because very long-distance transmissions, such as links between Europe and America, often utilize high-capacity backbone connections optimized for global communication, whereas shorter transmissions may experience suboptimal inter-domain routing. 


\section{Satellite Routing Strategy}
\label{appendix:satellite-routing-strategy}

A routing graph model is used for simulating satellite routing latency between two nodes, detailed below.

\textbf{Routing Graph Nodes.} \textit{User} means the sending and receiving party of satellite communication. In Tor, this could be a client, relay and destination server. \textit{Satellite} circles the Earth at a fixed altitude, acting as a ``mirror'' to reflect or forward signals without processing or caching. \textit{Ground Station}, equipped with antennas and electronic transmitters, directly communicates with satellites for tracking, controlling and data relaying. \textit{Point of Presence (PoP)} acts as a gateway between satellite and terrestrial networks. Traffic from satellites must pass through a ground station, and then a PoP before entering the terrestrial Internet.

\begin{figure}[h!]
    \centering
	\subfigure[Possible Link Types]{		
		\centering
		\includegraphics[width=0.46\linewidth]{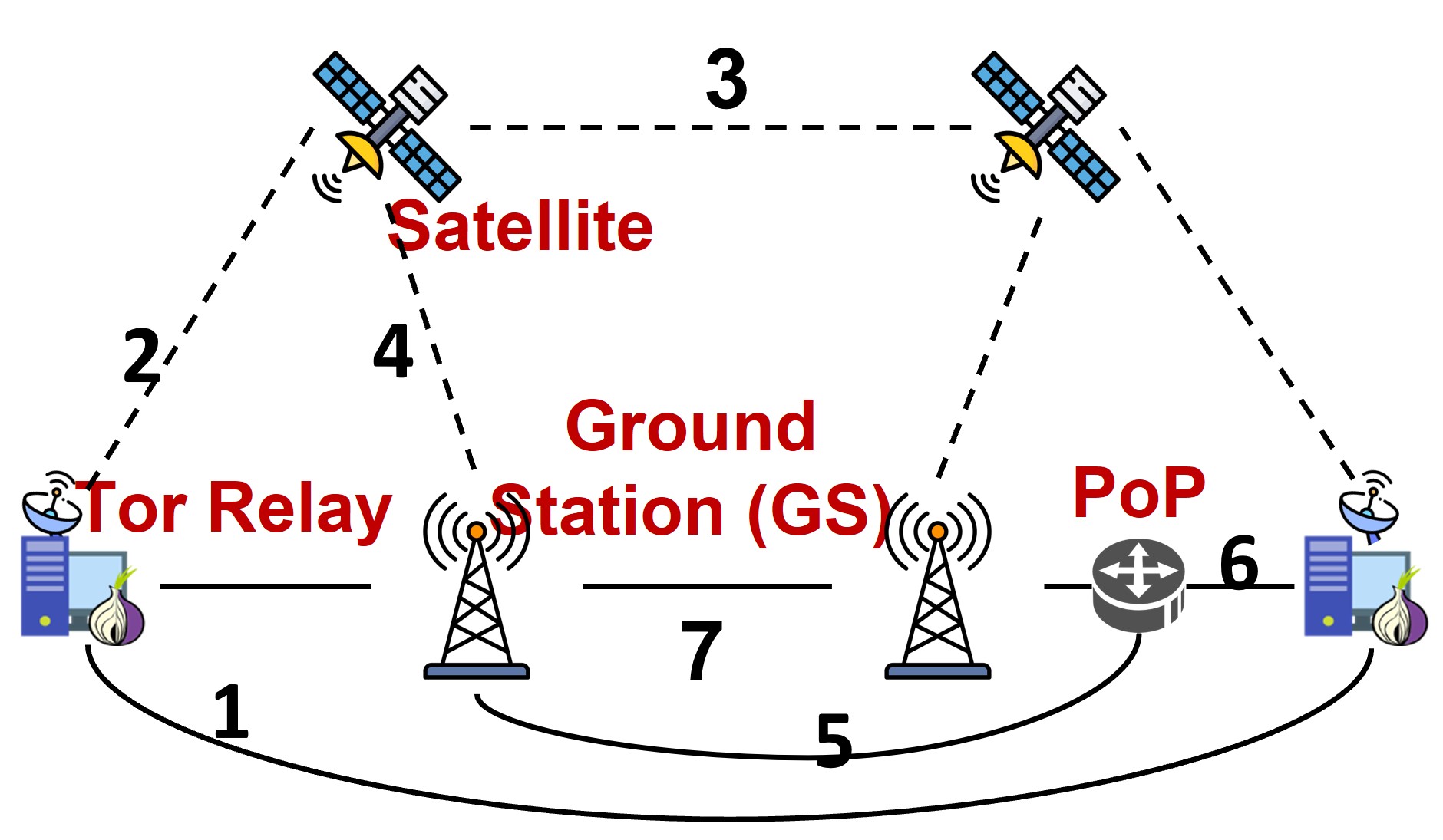}
		\label{fig:routing-strategy-link-type}
	}
    \subfigure[Traditional Terrestrial Routing]{
		\centering
		\includegraphics[width=0.46\linewidth]{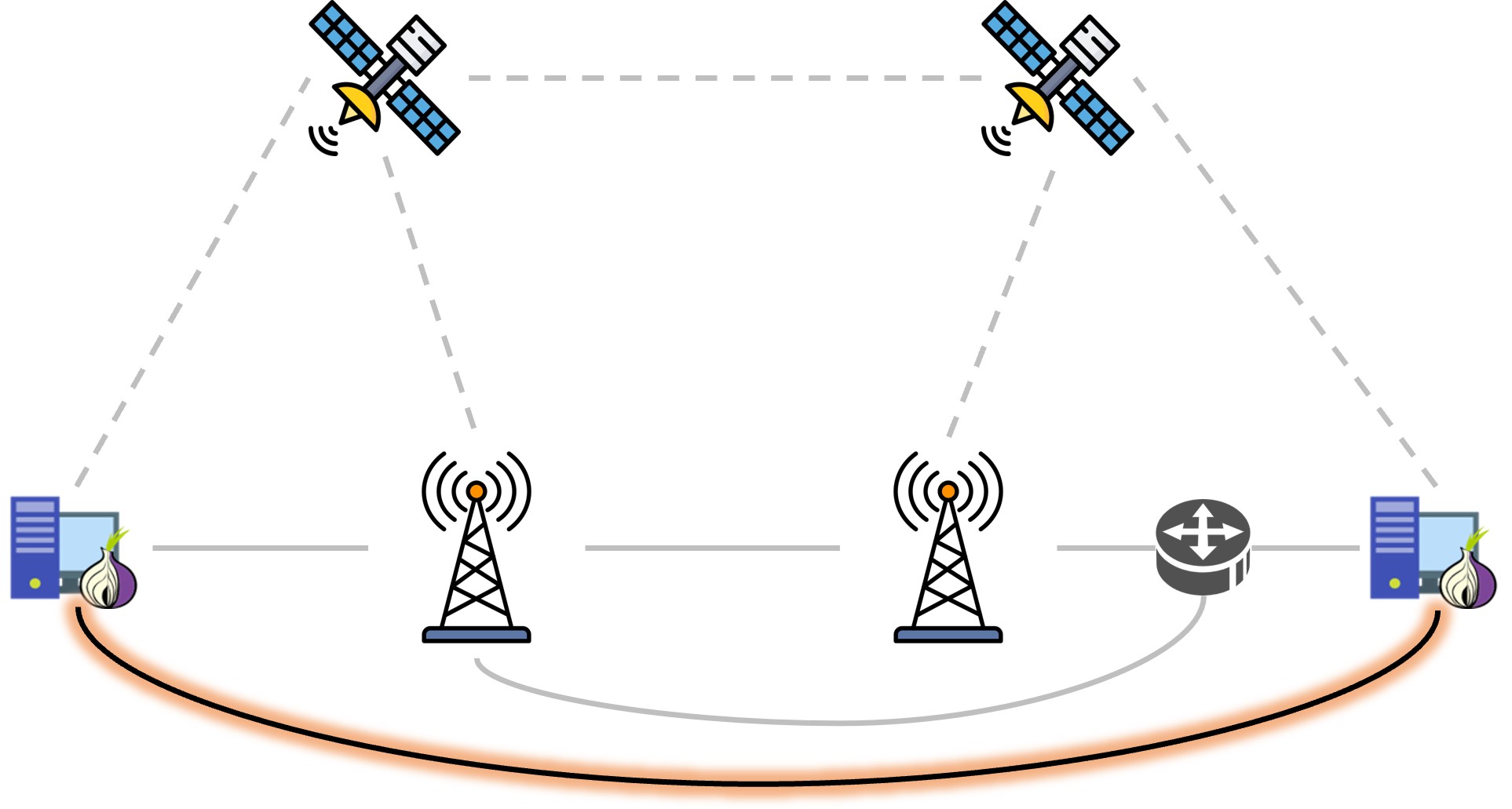}
		\label{fig:routing-strategy-terrestrial}
	}
	\subfigure[Single Bent-pipe Routing]{
		\centering
		\includegraphics[width=0.46\linewidth]{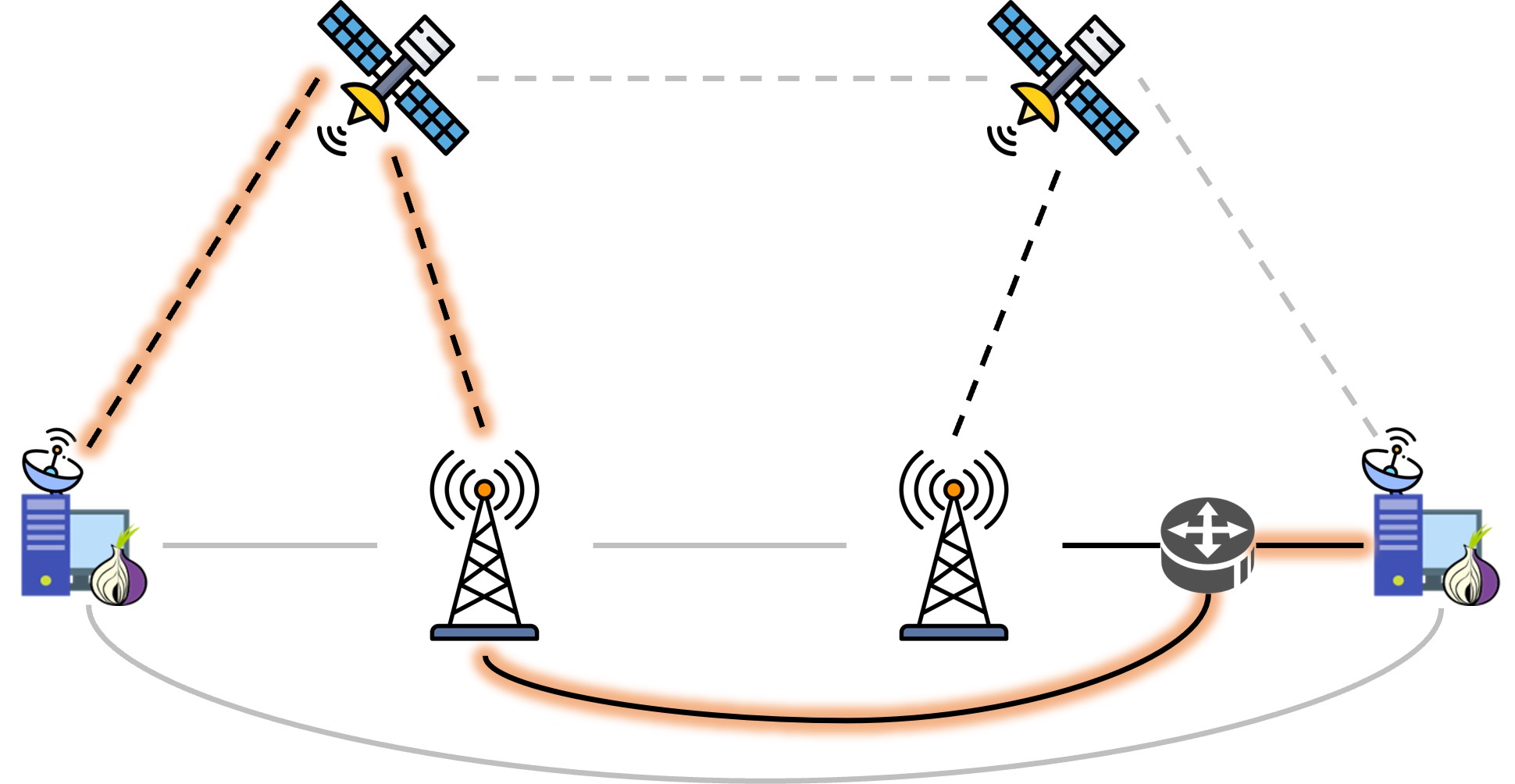}
		\label{fig:routing-strategy-one-bent-pipe}
	}
    \subfigure[ISL-enabled Routing]{		
		\centering
		\includegraphics[width=0.46\linewidth]{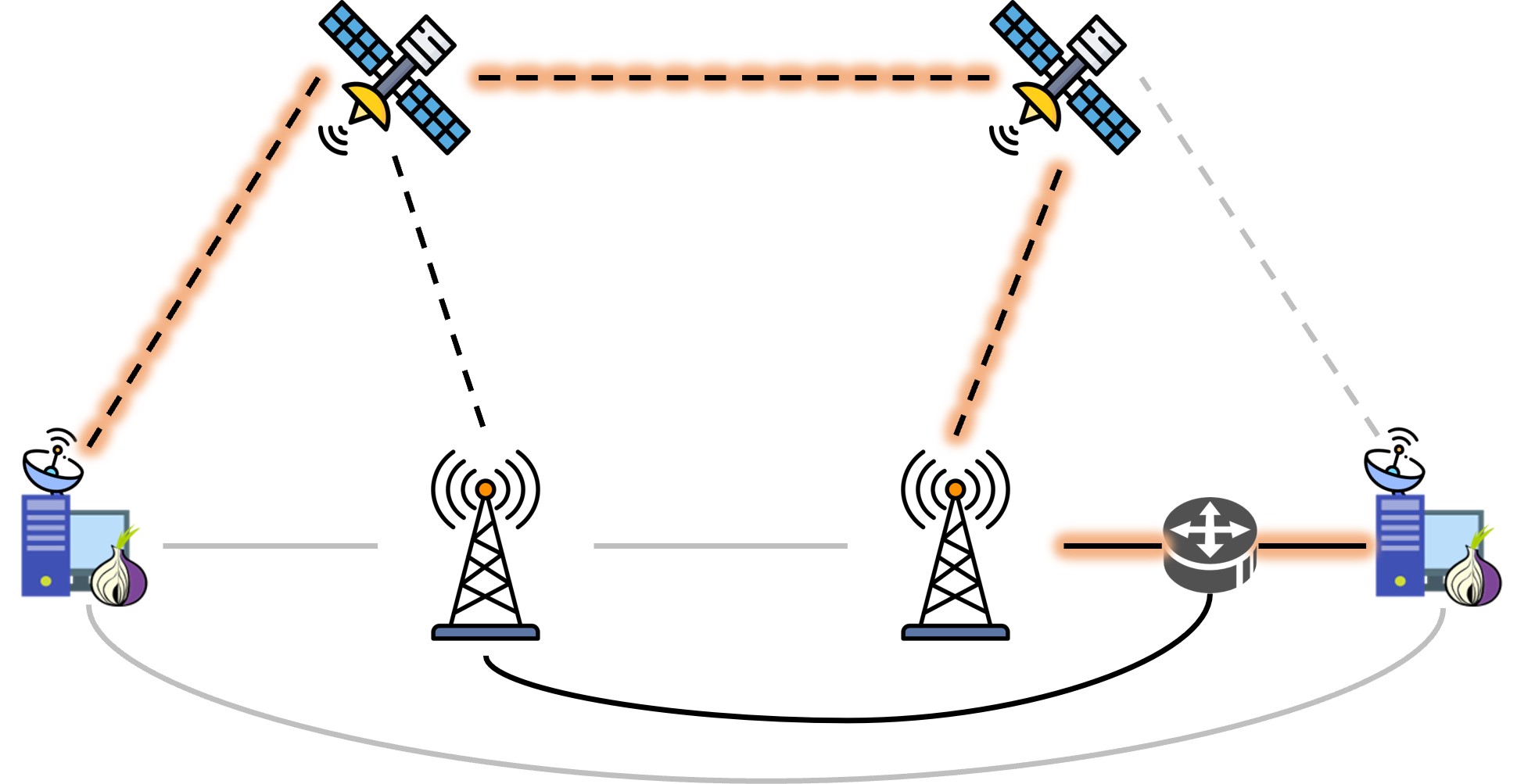}
		\label{fig:routing-strategy-ISL}
	}
	\caption{Routing graph between relays. Black line indicates available links while gray means unavailable ones.}
	\label{fig:routing-strategy}
\end{figure}

\textbf{Routing Graph Edges.} As shown in Fig.~\ref{fig:routing-strategy-link-type}, Inter-User Link (IUL, link~1) connects users through the terrestrial Internet. User-Satellite Link (USL, link~2) carries data from a user to a nearby satellite via spot-beams through the atmosphere. Inter-Satellite Link (ISL, link~3) is a laser-based connection between satellites, transmitting at the speed of light in vacuum. Groundstation-Satellite Link (GSL, link~4) exchanges traffic between satellites and ground stations using spot-beams. Groundstation-PoP Link (GPL, link~5) connects ground stations to nearby PoPs before traffic enters the terrestrial Internet, with a global latency of about 5 ms~\cite{mohan2023multifaceted}. Finally, User-PoP Link (UPL, link~6) delivers traffic between PoPs and users over the terrestrial Internet.

\textbf{Routing Strategies.} Fig. \ref{fig:routing-strategy-terrestrial} illustrates terrestrial routing, where neither relay connects to the satellite network and only IUL is used. In Fig. \ref{fig:routing-strategy-one-bent-pipe}, USL, GSL, GPL, and UPL are available, forming a ``bent-pipe'' structure: traffic from the source relay travels to satellite-1, down to a nearby ground station, then proceeds to the PoP and destination via the terrestrial network. When ISLs are available, as shown in Fig. \ref{fig:routing-strategy-ISL}, traffic is relayed across multiple satellites before descending to a ground station near the destination. This ``ISL-enabled'' routing keeps data in space for most of its journey, fully exploiting satellite acceleration potential.

\begin{figure}
    \centering
    \includegraphics[width=0.8\linewidth]{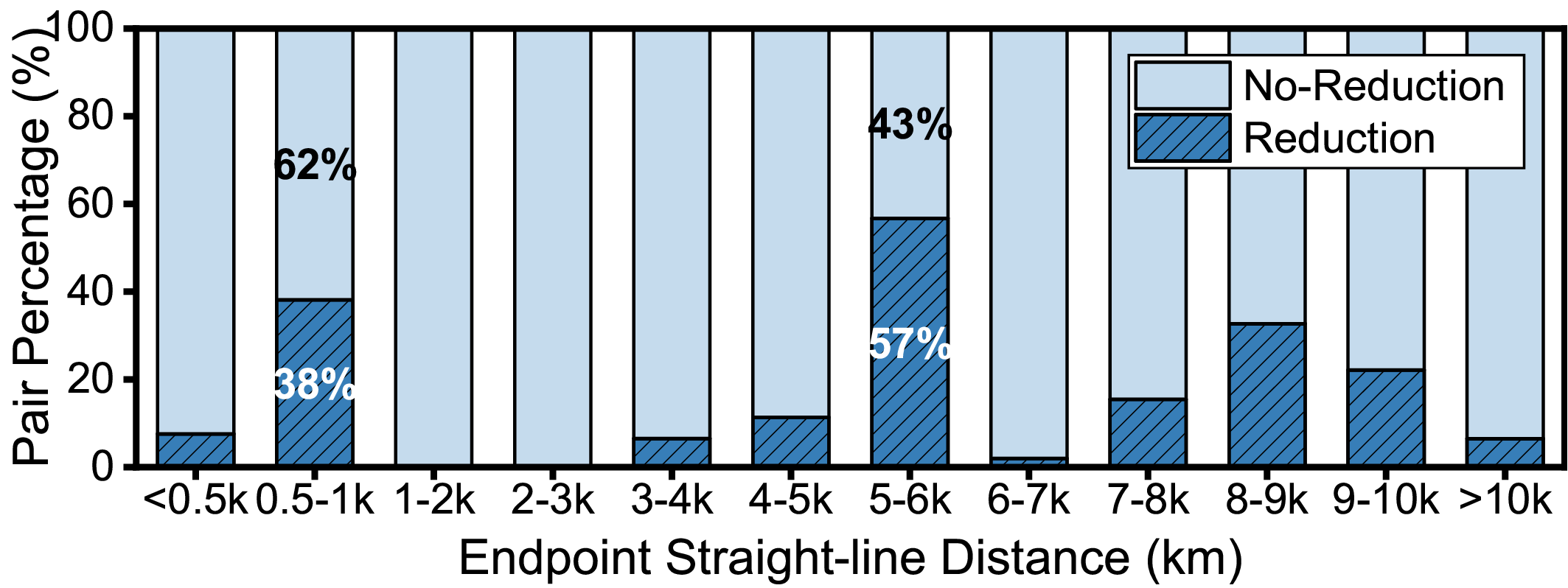}
    \caption{SaTor latency reduction efficacy across varying end-to-end distance between relay pairs.}
    \label{fig:reduction-increment-distance}
\end{figure}

\section{Satellite Routing across Varying Distance}

\label{Appendix:satellite-routing-across-varying-distance}

Fig. \ref{fig:reduction-increment-distance} shows SaTor's latency reduction grouped by end-to-end geographic distance. Satellite routing achieves the greatest gains in the 5–6k km (57\%) ranges. Moderate improvements (20–40\%) are observed at 8–10k km and 0.5-1k km. An intuitive explanation is that, for connections shorter than 0.5k km, terrestrial routing is already fast, while bent-pipe satellite links add propagation overhead. In the 0.5–1k km range, terrestrial delays are inflated due to local congestion or inefficient routing across regional ASes. Between 1k and 3k km, terrestrial traffic might be handed off to Tier-1 backbones, enabling optimized routing to reduce AS-path inflation. The 5–6k km range often involves transoceanic links, where terrestrial routes may experience congestion during suboptimal cable routing, while inter-satellite links can provide more stable and direct delivery. Beyond 6k km, the satellite advantage diminishes, possibly due to increased ISL hops and routing inefficiencies. 

\section{SaTor in the Real World}
\label{appendix:sator-in-the-real-world}
\begin{table}[t!]
  \centering
  \caption{AS Type of Tor Relays}
  \label{tab:as-type-tor-relay}
  \setlength{\tabcolsep}{2pt}
  \begin{tabular}{lllll}
    \toprule
    \textbf{AS Type} & \textbf{Top-2k} & \textbf{Top-4k} & \textbf{Top-6k} & \textbf{All (8,730)} \\
    \midrule
    Cloud Plat.   & 1,199 (60\%) & 2,886 (72\%) & 4,393 (73\%) & 6,149 (70\%) \\
    General ISP      & 422 (21\%)  & 571 (14\%)  & 887 (15\%)  & 1,771 (20\%) \\
    Business         & 66 (3\%)    & 97 (2\%)    & 233 (4\%)   & 249 (3\%) \\
    Education        & 313 (16\%)  & 445 (12\%)  & 485 (8\%)   & 559 (7\%) \\
    \bottomrule
  \end{tabular}
\end{table}

According to SpaceX's website, Starlink services are currently offered in over 65 countries. As of December 2024, out of 8,730 Tor relays in total, 8,280 (94\%) are situated in countries where Starlink is operational, demonstrating the practicality of SaTor under current conditions.

Based on Tor Metrics \cite{TorMetrics} and IPinfo database \cite{ipinfo2024}, all relay ASes are categorized into four types: \emph{cloud platform}, \emph{general ISP}, \emph{business network}, and \emph{education network}. As Table \ref{tab:as-type-tor-relay} shows, cloud platforms host 60\% of the top 2k relays and 70\% of all relays. Deploying SaTor on cloud-hosted relays would require the cloud providers to access their platform to satellite networks. Fortunately, major cloud providers are moving in that direction. Cloud providers like Amazon, Google, and Microsoft are cooperating with satellite companies to enhance their service \cite{SpaceCloud2021,HowSatellite2023}. About 30\% of relays are hosted on general ISP, business, or education networks, where SaTor deployment could be more straightforward. Around 40\% of the top 2k relays are non-cloud, which is sufficient to realize SaTor's potential.

\end{document}